\documentclass[nofootinbib,aps,twocolumn,showpacs,preprintnumbers,eqsecnum,pra]{revtex4}

\usepackage{graphicx}
\usepackage{dcolumn}
\usepackage{bm}
\usepackage{color}
\def\uu{\uparrow\uparrow}
\def\ud{\uparrow\downarrow}

\begin{document}

\preprint{paper1.tex}

\title{Theory of surface plasmons and surface-plasmon polaritons}
\author{J. M. Pitarke$^{1,2}$, V. M. Silkin$^{3}$, E. V. Chulkov$^{2,3}$, and
P. M. Echenique$^{2,3}$}
\affiliation{$^1$Materia
Kondentsatuaren Fisika Saila, Zientzi Fakultatea, Euskal Herriko
Unibertsitatea,\\ 644 Posta kutxatila, E-48080 Bilbo, Basque Country\\
$^2$Donostia International Physics Center (DIPC) and Unidad de F\'\i
sica de Materiales CSIC-UPV/EHU,\\ Manuel de Lardizabal Pasealekua,
E-20018
Donostia, Basque
Country\\
$^3$Materialen Fisika Saila, Kimika Fakultatea, Euskal Herriko
Unibertsitatea,\\ 1072 Posta kutxatila, E-20080 Donostia, Basque
Country}

\date{\today}

\begin{abstract}
Collective electronic excitations at metal surfaces are well known to
play a key role in a wide spectrum of science, ranging from physics
and materials science to biology. Here we focus on a
theoretical description of the many-body dynamical electronic
response of solids, which underlines the existence of various
collective electronic excitations at metal surfaces, such as the
conventional surface plasmon, multipole plasmons, and the recently
predicted acoustic surface plasmon. We also review existing
calculations, experimental
measurements, and applications.
\end{abstract}


\maketitle

\tableofcontents

\section{Introduction}

In his pioneering treatment of characteristic energy losses of fast
electrons passing through thin metal films, Ritchie predicted the
existence of self-sustained collective excitations at metal
surfaces~\cite{ritchie1}. It had already been pointed out by Pines
and Bohm~\cite{pines1,pines2} that the long-range nature of the
Coulomb interaction between valence electrons in metals yields
collective plasma oscillations similar to the
electron-density oscillations observed by Tonks and Langmuir in
electrical discharges in gases~\cite{tonks}, thereby explaining early
experiments by Ruthemann~\cite{ruth} and Lang~\cite{lang0} on the
bombardment of thin metallic films by fast electrons. Ritchie
investigated the impact of the film boundaries on the production of
collective excitations and found that the boundary effect is to cause
the appearance of a new {\it lowered} loss due
to the excitation of {\it surface} collective
oscillations~\cite{ritchie1}. Two years later, in a series of
electron energy-loss experiments Powell and Swan~\cite{powell}
demonstrated the existence of these collective excitations, the
quanta of which Stern and Ferrell called the {\it surface
plasmon}~\cite{stern1}.

Since then, there has been a significant advance in both
theoretical and experimental investigations of surface plasmons,
which for researches in the field of condensed matter and surface
physics have played a key role in the interpretation of a great
variety of experiments and the understanding of various
fundamental properties of solids. These include the nature of Van
der Waals forces~\cite{inglesvw,zk,ser}, the classical image
potential acting between a point classical charge and a metal
surface~\cite{feibelmani,ritchie2,mahan1,lucas1}, the energy
transfer in gas-surface interactions~\cite{gadzuk}, surface
energies~\cite{lucas2,ingles,langreth}, the damping of surface
vibrational modes~\cite{chabal,persson}, the energy loss of
charged particles moving outside a metal
surface~\cite{pendry1,inkson}, and the de-excitation of adsorbed
molecules~\cite{ueba}. Surface plasmons have also been employed in
a wide spectrum of studies ranging from
electrochemistry~\cite{knoll0}, wetting~\cite{wetting}, and
biosensing~\cite{bio0,bio1,bio2}, to scanning tunneling
microscopy~\cite{stm}, the ejection of ions from
surfaces~\cite{compton}, nanoparticle growth~\cite{nano1,nano2},
surface-plasmon microscopy~\cite{nature1,science0}, and
surface-plasmon resonance
technology~\cite{spr1,spr2,nature2,spr3,spr4,spr5,spr6}. Renewed
interest in surface plasmons has come from recent advances in the
investigation of the electromagnetic properties of nanostructured
materials~\cite{pendry2,halas}, one of the most attractive aspects
of these collective excitations now being their use to concentrate
light in subwavelength structures and to enhance transmission
through periodic arrays of subwavelength holes in optically thick
metallic films~\cite{ebbesen1,ebbesen3}.

The so-called field of plasmonics represents an exciting new area for the
application of surface and interface plasmons, and area in which
surface-plasmon based circuits merge the fields of photonics and electronics
at the nanoscale~\cite{ozsc06}. Indeed, surface-plasmon polaritons can serve
as a basis for constructing  nanoscale photonic circuits that will be able to
carry optical signals and electric currents~\cite{ebbesen2,noohapl05}. Surface
plasmons can also serve as a basis for the design, fabrication, and
characterization of subwavelength waveguide
components~\cite{quleol98,chbeol00,lakrapl01,nileapl03,krlael02,krwe04,makinm03,
muasprb04,mabaapl04,mafrapl05,bechjap05,nomaprb02,piogapl05,bovoprl05,bovon06}.
In the framework of plasmonics, modulators and switches have also been
investigated~\cite{krzhapl04,krzajo05}, as well as the use of surface plasmons
as mediators in the transfer of energy from donor to acceptors molecules on
opposite sides of metal films~\cite{barnes}. 

According to the work of Pines and Bohm, the quantum energy
collective plasma oscillations in a free electron gas with
equilibrium density $n$ is $\hbar\omega_p=\hbar(4\pi
ne^2/m_e)^{1/2}$, $\omega_p$ being the so-called plasmon
frequency~\cite{optical}. In the presence of a planar boundary, there
is a new mode (the {\it surface} plasmon), the frequency of which
equals in the nonretarded region (where the speed of light can be
taken to be infinitely large) Ritchie's frequency
$\omega_s=\omega_p/\sqrt{2}$ at wave vectors ${\bf q}$ in the range
$\omega_s/c<<q<<q_F$ ($q_F$ being the magnitude of the Fermi wave
vector) and exhibits
some dispersion as the wave vector is increased. In the retarded
region, where the phase velocity $\omega_s/q$ of the surface
plasmon is comparable to the velocity of light, surface plasmons
couple with the free electromagnetic field. These surface-plasmon
polaritons propagate along the metal surface with frequencies ranging
from zero (at $q=0$) towards the asymptotic value
$\omega_s=\omega_p/\sqrt{2}$, the dispersion relation $\omega(q)$
lying to the right of the light line and the propagating vector
being, therefore, larger than that of bare light waves of the same
energy. Hence, surface-plasmon polaritons in an ideal semi-infinite
medium are nonradiative in nature, i.e., cannot decay by emitting a
photon and, conversely, light incident on an ideal surface cannot
excite surface plasmons.

In the case of thin films, the electric fields of both
surfaces interact. As a result, there are (i) {\it tangential}
oscillations characterized by a symmetric disposition of charge
deficiency or excess at opposing points on the two surfaces and (ii)
{\it normal} oscillations in which an excess of charge density at a
point on one surface is accompanied by a deficiency at the point
directly across the thin film. The phase velocity of the {\it
tangential} surface plasmon is always less than the speed of light,
as occurs in the case of a semi-infinite electron system. However,
the phase velocity of {\it normal} oscillations may surpass that of
light, thereby becoming a radiative surface plasmon that should be
responsible for the emission of light~\cite{ferrell0}. This
radiation was detected using electron beam bombardment of thin films
of Ag, Mg, and Al with thicknesses ranging between $500$ and $1000$
\AA~\cite{radi1,radi2}. More recently, light emission was observed
in the ultraviolet from a metal-oxide-metal tunnel diode and was
attributed to the excitation of the radiative surface
plasmon~\cite{radi3}.

Nonradiative surface plasmons in both thin and thick films can couple
to electromagnetic radiation in the presence of surface roughness or
a grating, as suggested by Teng and Stern~\cite{stern2}.
Alternatively, prism coupling can be used to enhance the momentum of
incident light, as demonstrated by Otto~\cite{otto} and by Kretchmann
and Raether~\cite{raether}. Since then, this so-called attenuated
reflection (ATR) method and variations upon it have been used by
several workers in a large variety of
applications~\cite{c1,c2,c3,c4,c5,sambles,rev}.

During the last decades, there has also been a significant advance in
our understanding of surface plasmons in the {\it
nonretarded} regime. Ritchie~\cite{ritchie0} and
Kanazawa~\cite{kanazawa} were the first to attack the problem of
determining the dispersion $\omega(q)$ of the nonretarded surface
plasmon. Bennett~\cite{bennett} used a hydrodynamical model with a
continuous decrease of the electron density at the metal surface, and
found that a continuous electron-density variation yields
two collective electronic excitations: Ritchie's surface plasmon at
$\omega\sim\omega_s$, with a {\it negative} energy dispersion at low
wave vectors, and an upper surface plasmon at higher energies. In the
direction normal to the surface, the
distribution of Ritchie's surface plasmon consists of a single peak,
i.e., it has a {\it monopole} character; however, the charge
distribution of the upper mode has a node, i.e., it has a {\it
dipole} character and is usually called multipole surface plasmon.

Bennett's qualitative conclusions were generally confirmed by
microscopic descriptions of the electron gas. On the one hand,
Feibelman showed that in the long-wavelength limit the classical
result $\omega_s=\omega_p/\sqrt{2}$ is correct for a semi-infinite
plane-bounded electron gas, irrespective of the exact variation of
the electron density in the neighborhood of the surface~\cite{l1}. On
the other hand, explicit expressions for the {\it linear} momentum
dispersion of the conventional monopole surface plasmon that are
sensitive to the actual form of the electron-density fluctuation at
the surface were derived by Harris and Griffin~\cite{harris} using
the equation of motion for the Wigner distribution function in the
random-phase approximation (RPA) and by Flores and Garc\'\i
a-Moliner~\cite{flores} solving Maxwell's equations in combination
with
an integration of the field components over the surface region.
Quantitative RPA calculations of the linear dispersion of the
monopole surface plasmon were carried out by several authors by using
the infinite-barrier model (IBM) of the surface~\cite{beck1}, a step
potential~\cite{beck2,inglessp}, and the more realistic
Lang-Kohn~\cite{lk} self-consistent surface
potential~\cite{feibelman1}. Feibelman's calculations showed that for
the typical electron densities in metals ($2<r_s<6$) the
initial slope of the momentum dispersion of monopole surface plasmons
of jellium surfaces is negative~\cite{feibelman1}, as anticipated by
Bennett~\cite{bennett}.

Negative values of the momentum dispersion had been observed by
high-energy electron transmission on uncharacterized Mg
surfaces~\cite{kun} and later by inelastic low-energy electron
diffraction on the (100) and (111) surfaces of
Al~\cite{leed1,leed2}. Nevertheless, Klos and
Raether~\cite{raether1} and Krane and Raether~\cite{raether2} did
not observe a negative dispersion for Mg and Al films. Conclusive
experimental confirmation of the negative surface plasmon dispersion
of a variety of simple metals (Li, Na, K, Cs, Al, and Mg) did not
come until several years
later~\cite{plummer1,plummer2,plummer3,plummer5}, in a series of
experiments based on angle-resolved low-energy inelastic electron
scattering~\cite{note1}. These experiments showed good agreement
with self-consistent dynamical-response calculations carried out for
a jellium surface~\cite{note2} in a time-dependent adiabatic
extension of the density-functional theory (DFT) of Hohenberg, Kohn,
and Sham~\cite{dft}. Furthermore, these experiments also showed that
the multipole surface plasmon was observable, its energy and
dispersion being in quantitative agreement with the self-consistent
jellium calculations that had been reported by
Liebsch~\cite{liebsch1}.

Significant deviations from the dispersion of surface plasmons at
jellium surfaces occur on Ag~\cite{rocca1,rocca2,rocca3,rocca4} and
Hg~\cite{plummer6}, due to the presence of filled $4d$ and $5d$
bands, respectively, which in the case of Ag yields an anomalous
positive dispersion. In order to describe the observed features of Ag
surface plasmons, various simplified models for the screening of $d$
electrons have been developed~\cite{ag1,ag2,ag3,ag4,ag5}. Most
recently, calculations have been found to yield a qualitative
understanding of the existing electron energy-loss measurements by
combining a self-consistent jellium model for valence $5s$ electrons
with a so-called dipolium model in which the occupied $4d$ bands are
represented in terms of polarizable spheres located at the sites of a
semi-infinite face-cubic-centered (fcc) lattice~\cite{lopez}.

{\it Ab initio} {\it bulk} calculations of the dynamical response and
plasmon dispersions of {\it noble} metals with occupied $d$ bands have been
carried out recently~\cite{abi1,abi2,abi3}. However, first-principles
calculations of the surface-plasmon energy and linewidth dispersion
of real solids have been carried out only in the case of the
simple-metal prototype surfaces Mg(0001) and Al(111)~\cite{mg,vac}. These
calculations lead to an accurate description of the measured surface-plasmon
energy dispersion that is superior to that obtained in the jellium model,
and they show that the band structure is of paramount importance for a
correct description of the surface-plasmon linewidth.

The multipole surface plasmon, which is originated in the selvage
electronic structure at the surface, has been observed in a variety
of simple metals at $\omega\sim
0.8\omega_p$~\cite{plummer1,plummer2,plummer3,plummer5}, in
agreement with theoretical predictions. Nevertheless, electron
energy-loss spectroscopy (EELS) measurements of Ag, Hg, and Li revealed no
clear evidence of the multipole surface plasmon. In the case of Ag,
high-resolution energy-loss spectroscopy low-energy electron diffraction
(ELS-LEED) measurements indicated that a peak was obtained at $3.72\,{\rm eV}$
by subtracting the data for two different impact
energies~\cite{roccal}, which was interpreted to be the Ag multipole
plasmon. However, Liebsch argued that the frequency of the Ag multipole surface
plasmon should be in the $6-8\,{\rm eV}$ range {\it above} rather than {\it
below} the bulk plasma frequency, and suggested that the observed
peak at $3.72\,{\rm eV}$ might not be associated with a multipole
surface plasmon~\cite{liebschl}.

An alternative spectroscopy technique to investigate multipole surface
plasmons is provided by angle- and energy-resolved photoyield
experiments (AERPY)~\cite{levinson}. In fact, AERPY is more suitable
than electron
energy-loss spectroscopy to identify the multipole surface plasmon,
since the monopole surface plasmon of clean flat surfaces (which is
the dominant feature in
electron-energy loss spectra) is not excited by photons and thus the
weaker multipole surface mode (which intersects the radiation line in
the retardation regime) can be observed. A large increase in the
surface photoyield was observed at $\omega=0.8\omega_p$ from
Al(100)~\cite{levinson} and Al(111)~\cite{barman1}. Recently, the
surface electronic structure and optical response of Ag has been
studied using this technique~\cite{barman2}. In these experiments,
the Ag multipole surface plasmon is observed at $3.7\,{\rm eV}$,
while no signature of the multipole surface plasmon is observed above
the plasma frequency ($\omega_p=3.8\,{\rm eV}$) in disagreement with
the existing theoretical prediction~\cite{liebschl}. Hence, further
theoretical work is needed on the surface electronic response of Ag
that go beyond the $s$-$d$ polarization model described in
Ref.~\cite{liebschl}.

Another collective electronic excitation at metal surfaces is the
so-called acoustic surface plasmon that has been predicted to exist
at solid surfaces where a partially occupied quasi-two-dimensional
surface-state band coexists with the underlying three-dimensional
continuum~\cite{sigael04,sipiprb05}. This {\it new} low-energy collective
excitation exhibits linear dispersion at low wave vectors, and might
therefore affect electron-hole (e-h) and phonon dynamics near the
Fermi level~\cite{note3}. It has been demonstrated that it is a
combination of the nonlocality of the 3D dynamical screening and the
spill out of the 3D electron density into the vacuum which allows
the formation of 2D electron-density acoustic oscillations at metal
surfaces, since these oscillations would otherwise be completely
screened by the surrounding 3D substrate~\cite{pitarke}. This {\it
novel} surface-plasmon mode has been observed recently at the (0001)
surface of Be, showing a linear energy dispersion that is in very
good agreement with first-principles calculations~\cite{acexp}.

Finally, we note that metal-dielectric interfaces of arbitrary
geometries also support charge density oscillations similar to the
surface plasmons characteristic of planar interfaces. These are
{\it localized} Mie plasmons occurring at frequencies which are
characteristic of the interface geometry~\cite{bohren}. The
excitation of localized plasmons on small particles has attracted
great interest over the years in scanning transmission electron
microscopy~\cite{batson,howie1,ferrell,rivacoba1,howie2,rivacoba2}
and near-field optical spectroscopy~\cite{klar}. Recently, new
advances in structuring and manipulating on the nanometer scale
have rekindled interest this field~\cite{nano}. In nanostructured metals and
carbon-based structures, such as fullerenes and carbon nanotubes, localized
plasmons can be excited
by light and can therefore be easily detected as pronounced
optical resonances~\cite{lucas,heer,fj1}. Furthermore, very
localized dipole and multipole modes in the vicinity of highly
coupled structures are responsible for surface-enhanced Raman
scattering~\cite{fj2,fj3} and other striking properties like,
e.g., the blackness of colloidal silver~\cite{ingles1}.

Collective electronic excitations in thin adsorbed overlayers,
semiconductor heterostructures, and parabolic quantum wells have also
attracted attention over the last years. The adsorption of thin films
is important, because of the drastic changes that they produce in the
electronic properties of the substrate and also because of related
phenomena such as catalytic promotion~\cite{aruga}; however, the
understanding of adsorbate-induced collective excitations is still
incomplete~\cite{over1,over2,over3,over4,over5,over6,plummerl,over7}. The
excitation spectrum of collective
modes in semiconductor quantum wells has been described by several
authors~\cite{dobson,schaich1,zaremba,guo,qw}. These systems, which
have been grown in semiconductor heterostructures with the aid
of Molecular Beam Epitaxy~\cite{sm1}, form a nearly ideal
free-electron gas and have been, therefore, a playground on which to
test existing many-body theories~\cite{halp1,halp2}.

Major reviews on the theory of collective electronic excitations at
metal surfaces have been given by Ritchie~\cite{ritchier},
Feibelman~\cite{feibelman0}, and Liebsch~\cite{liebsch0}.
Experimental reviews are also available, which focus on high-energy
EELS experiments~\cite{raetherr1}, surface plasmons on smooth and
rough surfaces and on gratings~\cite{raetherr2}, and angle-resolved
low-energy EELS investigations~\cite{plummer0,rocca}.
An extensive review on plasmons and magnetoplasmons in semiconductor
heterostructures has been given recently by Kushwaha~\cite{magneto}.

This review will focus on a unified theoretical description of the
many-body dynamical electronic response of solids, which underlines the
existence of various collective electronic excitations at metal
surfaces, such as the conventional surface plasmon, multipole
plasmons, and the acoustic surface plasmon. We also review existing
calculations, experimental measurements, and some of the most recent
applications including particle-solid interactions, scanning transmission
electron microscopy, and surface-plasmon based photonics, i.e., plasmonics.

\section{Surface-plasmon polariton: Classical approach}\label{spp1}

\subsection{Semi-infinite system}\label{IIA}

\subsubsection{The surface-plasmon condition}

\begin{figure}
\includegraphics[width=0.95\linewidth]
{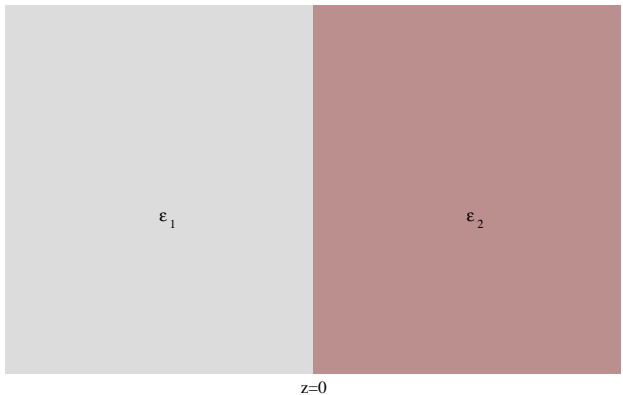}
\caption{Two semi-infinite media with dielectric functions
$\epsilon_1$ and $\epsilon_2$ separated by a planar interface at
$z=0$.}\label{fig1}
\end{figure}

We consider a classical model consisting of two semi-infinite
nonmagnetic media with local (frequency-dependent) dielectric
functions $\epsilon_1$ and $\epsilon_2$ separated by a planar
interface at $z=0$ (see Fig. 1). The full set of Maxwell's equations
in the absence of external sources can be expressed as
follows~\cite{jackson}
\begin{equation}\label{m1}
{\bf\nabla}\times{\bf H}_i=\epsilon_i\,{1\over c}\
{\partial\over\partial t}\,{\bf E}_i,
\end{equation}
\begin{equation}\label{m2}
{\bf\nabla}\times{\bf E}_i=-{1\over c}\,{\partial\over\partial t}\,
{\bf H}_i,
\end{equation}
\begin{equation}\label{m3}
{\bf\nabla}\cdot(\epsilon_i\,{\bf E}_i)=0,
\end{equation}
and
\begin{equation}\label{m4}
{\bf\nabla}\cdot{\bf H}_i=0,
\end{equation}
where the index $i$ describes the media: $i=1$ at $z<0$, and $i=2$ at
$z>0$.

Solutions of Eqs.~(\ref{m1})-(\ref{m4}) can generally be classified
into $s$-polarized and $p$-polarized electromagnetic modes, the
electric field ${\bf E}$ and the magnetic field ${\bf H}$ being
parallel to the interface, respectively. For an ideal surface, if
waves are to be formed that propagate along the interface there must
necessarily be a component of the electric field normal to the
surface. Hence, $s$-polarized surface oscillations (whose electric
field ${\bf E}$ is parallel to the interface) do not exist; instead,
we seek conditions under which a
traveling wave with the magnetic field ${\bf H}$ parallel to the
interface ($p$-polarized wave) may propagate along the surface
($z=0$), with the fields tailing off into the positive ($z>0$) and
negative ($z<0$) directions. Choosing the $x$-axis along the
propagating direction, we write
\begin{equation}\label{e}
{\bf E}_i=(E_{i_x},0,E_{i_z})\,{\rm e}^{-\kappa_i|z|}\,{\rm
e}^{i(q_ix-\omega t)}
\end{equation}
and
\begin{equation}\label{h}
{\bf H}_i=(0,E_{i_y},0)\,{\rm e}^{-\kappa_i|z|}\,{\rm
e}^{i(q_ix-\omega t)},
\end{equation}
where $q_i$ represents the magnitude of a wave vector that is parallel
to the surface. Introducing Eqs.~(\ref{e}) and (\ref{h}) into
Eqs.~(\ref{m1})-(\ref{m4}), one finds
\begin{equation}\label{h1}
i\,\kappa_1\,H_{1_{y}}=+{\omega\over c}\,\epsilon_1\,E_{1_x},
\end{equation}
\begin{equation}\label{h2}
i\,\kappa_2\,H_{2_{y}}=-{\omega\over c}\,\epsilon_2\,E_{2_x},
\end{equation}
and
\begin{equation}\label{retarded1}
\kappa_i=\sqrt{q_i^2-\epsilon_i\,{\omega^2\over c^2}}.
\end{equation}

The boundary conditions imply that the component of the electric and
magnetic fields parallel to the surface must be
continuous. Using Eqs.~(\ref{h1}) and (\ref{h2}), one writes the
following system of equations:
\begin{equation}
{\kappa_1\over\epsilon_1}\,H_{1_y}+{\kappa_2\over\epsilon_2}\,H_{2_y}=0
\end{equation}
and
\begin{equation}
H_{1_y}-H_{2_y}=0,
\end{equation}
which has a solution only if the determinant is zero, i.e.,
\begin{equation}\label{retarded}
{\epsilon_1\over\kappa_1}+{\epsilon_2\over\kappa_2}=0.
\end{equation}
This is the surface-plasmon condition.

From the boundary conditions also follows the continuity of the 2D
wave vector ${\bf q}$ entering Eq.~(\ref{retarded1}), i.e.,
$q_1=q_2=q$.
Hence, the surface-plasmon
condition [Eq.~(\ref{retarded})] can also be expressed as
follows~\cite{ritchiep},
\begin{equation}\label{eqr}
q(\omega)={\omega\over c}\sqrt{\epsilon_1\,
\epsilon_2\over\epsilon_1+\epsilon_2},
\end{equation}
where $\omega/c$ represents the magnitude of the light wave
vector. For a metal-dielectric interface with the dielectric
characterized by $\epsilon_2$, the solution $\omega(q)$ of
Eq.~(\ref{eqr}) has slope equal to $c/\sqrt{\epsilon_2}$ at the
point $q=0$ and is a monotonic increasing function of $q$, which
is always smaller than $c\,q/\sqrt{\epsilon_2}$ and for large $q$
is asymptotic to the value given by the solution of
\begin{equation}\label{nonretarded}
\epsilon_1+\epsilon_2=0.
\end{equation}
This is the {\it nonretarded} surface-plasmon condition
[Eq.~(\ref{retarded}) with $\kappa_1=\kappa_2=q$], which is valid as
long as the phase velocity $\omega/q$ is much smaller than the speed
of light.

\subsubsection{Energy dispersion}

In the case of a Drude semi-infinite metal in vacuum, one has
$\epsilon_2=1$ and~\cite{ashcroft}
\begin{equation}\label{drude}
\epsilon_1=1-{\omega_p^2\over\omega(\omega+i\eta)},
\end{equation}
$\eta$ being a positive infinitesimal. Hence, in this case
Eq.~(\ref{eqr}) yields
\begin{equation}\label{retarded5}
q(\omega)={\omega\over c}\,\sqrt{\omega^2-\omega_p^2\over
2\omega^2-\omega_p^2}.
\end{equation}

\begin{figure}
\includegraphics[width=0.45\textwidth,height=0.3375\textwidth]
{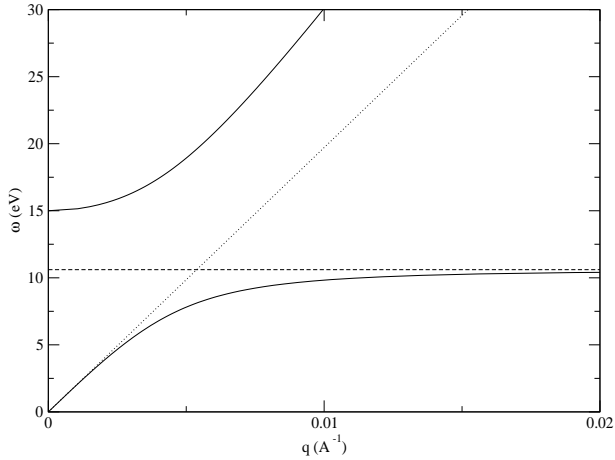}
\caption{The solid lines represent the solutions of
Eq.~(\ref{retarded5}) with $\omega_p=15\,{\rm eV}$: the dispersion of
light in the solid (upper
line) and the surface-plasmon polariton (lower line). In the retarded
region ($q<\omega_s/c$), the surface-plasmon polariton dispersion
curve approaches the light line $\omega=cq$ (dotted line). At short
wave lengths ($q>>\omega_s/c$), the surface-plasmon polariton
approaches asymptotically the nonretarded surface-plasmon frequency
$\omega_s=\omega_p/\sqrt{2}$ (dashed line).}\label{fig2}
\end{figure}

We have represented in Fig. 2 by solid lines the dispersion relation
of Eq.~(\ref{retarded5}), together with the light line $\omega=c\,q$
(dotted line). The upper solid line represents the dispersion of
light in the solid. The lower solid line is the surface-plasmon
polariton
\begin{equation}\label{retarded4}
\omega^2(q)=\omega_p^2/2+c^2q^2-\sqrt{\omega_p^4/4+c^4q^4},
\end{equation}
which in the retarded region (where $q<\omega_s/c$)
couples with the free electromagnetic field and in the non-retarded
limit ($q>>\omega_s/c$) yields the {\it classical} nondispersive
surface-plasmon frequency $\omega_s=\omega_p/\sqrt{2}$.

\begin{figure}
\includegraphics[width=0.95\linewidth]
{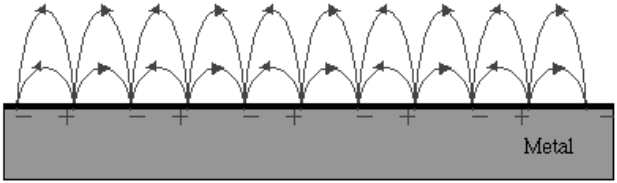}
\caption{Schematic representation of the electromagnetic field
associated with
a surface-plasmon polariton propagating along a metal-dielectric
interface.
The field strength ${\bf E}_i$ [see Eq.~(\ref{e})] decreases
exponentially
with the distance $|z|$ from the surface, the decay constant
$\kappa_i$ being
given by Eq.~(\ref{kappa}). + and - represent the regions with lower
and
higher electron density, respectively.}\label{fig0}
\end{figure}

We note that the wave vector ${\bf q}$ entering the dispersion
relation of Eq.~(\ref{retarded4}) (lower solid line of Fig. 2) is a
2D wave vector in the plane of the surface. Hence, if light hits the
surface in an arbitrary direction the external radiation dispersion
line will always lie somewhere between the light line $c\,q$ and the
vertical line, in such a way that it will not intersect the
surface-plasmon polariton line, i.e., light incident on an ideal
surface cannot excite surface plasmons. Nevertheless, there are two
mechanisms that allow external radiation to be coupled to
surface-plasmon polaritons: surface roughness or gratings, which can
provide the requisite momentum via {\it umklapp}
processes~\cite{stern2}, and attenuated total reflection (ATR) which
provides the external radiation with an imaginary wave vector in the
direction perpendicular to the surface~\cite{otto,raether}.

\subsubsection{Skin depth}

\begin{figure}
\includegraphics[width=0.95\linewidth]
{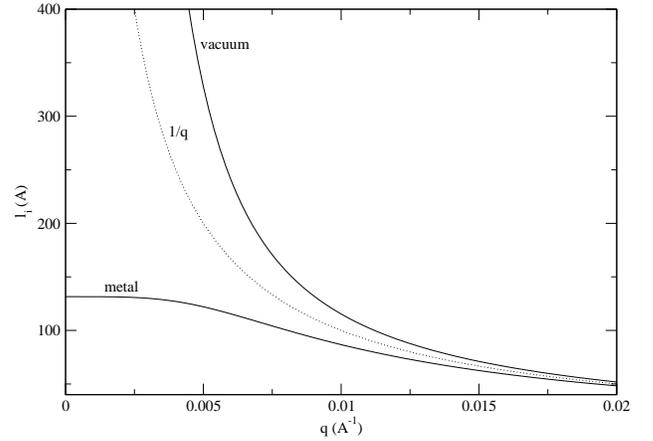} \caption{Attenuation length $l_i=1/\kappa_i$, versus
$q$, as obtained from Eq.~(\ref{kappa}) at the surface-plasmon
polariton condition [Eq.~(\ref{retarded4})] for a Drude metal in
vacuum. $\epsilon_1$ has been taken to be of the form of
Eq.~(\ref{drude}) with $\omega_p=15\,{\rm eV}$ and $\epsilon_2$ has
been set up to unity. The dotted line represents the large-$q$ limit
of both $l_1$ and $l_2$, i.e., $1/q$.}\label{figl0}
\end{figure}

Finally, we look at the spatial extension of the electromagnetic field
associated with the surface-plasmon polariton (see Fig.~\ref{fig0}).
Introducing the
surface-plasmon condition of Eq.~(\ref{eqr}) into
Eq.~(\ref{retarded1}) (with $q_1=q_2=q$), one
finds the following expression for the surface-plasmon decay
constant $\kappa_i$ perpendicular to the interface:
\begin{equation}\label{kappa}
\kappa_i={\omega\over c}\,
\sqrt{-\epsilon_i^2\over\epsilon_1+\epsilon_2},
\end{equation}
which allows to define the attenuation length $l_i=1/\kappa_i$ at which the
electromagnetic field falls to $1/{\rm e}$. Fig.~\ref{figl0} shows $l_i$ as a
function of the magnitude $q$ of the surface-plasmon polariton wave vector for
a Drude metal [$\epsilon_1$ of Eq.~(\ref{drude})] in vacuum ($\epsilon_2=0$).
In the vacuum side of the interface, the attenuation length is over the
wavelength involved ($l_2>1/q$), whereas the attenuation length into the metal
is determined at long-wavelengths ($q\to 0$) by the so-called skin-depth. At
large $q$ [where the nonretarded surface-plasmon condition of
Eq.~(\ref{nonretarded}) is fulfilled], the skin depth is $l_i\sim 1/q$ thereby
leading to a strong concentration of the electromagnetic surface-plasmon field
near the interface.

\subsection{Thin films}

Thin films are also known to support surface collective oscillations.
For this geometry, the electromagnetic fields of both surfaces
interact in such a way that the {\it retarded} surface-plasmon
condition of Eq.~(\ref{retarded}) splits into two new conditions (we
only consider nonradiative surface plasmons), depending on whether
electrons in the two surfaces oscillate in phase or
not. In the case of a thin film of thickness $a$ and dielectric
function $\epsilon_1$ in a medium of dielectric function
$\epsilon_2$, one finds~\cite{raetherr2}:
\begin{equation}\label{couplingr}
{\epsilon_1\over\kappa_1\tanh(\kappa_1
a/2)}+{\epsilon_2\over\kappa_2}=0
\end{equation}
and
\begin{equation}\label{coupling}
{\epsilon_1\over\kappa_1\coth(\kappa_1
a/2)}+{\epsilon_2\over\kappa_2}=0.
\end{equation}
Instead, if the film is surrounded by dielectric layers of dielectric
constant $\epsilon_0$ and equal thickness $t$ on either side, one
finds
\begin{equation}\label{couplingr2}
{\epsilon_1\over\kappa_1\nu\tanh(\kappa_1
a/2)}+{\epsilon_0\over\kappa_0}=0
\end{equation}
and
\begin{equation}\label{coupling2}
{\epsilon_1\over\kappa_1\nu\coth(\kappa_1
a/2)}+{\epsilon_0\over\kappa_0}=0,
\end{equation}
where
\begin{equation}
\nu={1-\Delta\,{\rm e}^{-2\kappa_0 t}\over
1+\Delta\,{\rm e}^{-2\kappa_0 t}},
\end{equation}
with
\begin{equation}
\Delta={\kappa_2\epsilon_0-\kappa_0\epsilon_2\over
\kappa_2\epsilon_0+\kappa_0\epsilon_2}
\end{equation}
and
\begin{equation}\label{kappa0}
\kappa_0=\sqrt{q^2-\epsilon_0\,{\omega^2\over c^2}}.
\end{equation}

\begin{figure}
\includegraphics[width=0.45\textwidth,height=0.3375\textwidth]
{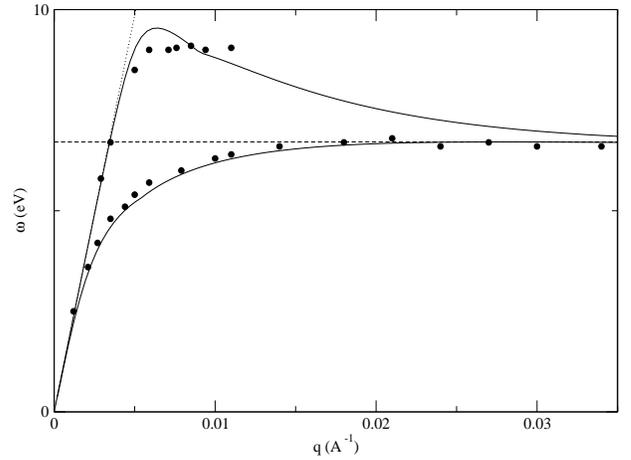}
\caption{Dispersion $\omega(q)$ of the surface-plasmon polariton of an
Al film of thickness $a=120\,{\rm\AA}$ surrounded by dielectric layers
of equal thickness $t=40\,{\rm\AA}$. The solid lines represent the
result
obtained from Eqs.~(\ref{couplingr2})-(\ref{kappa0}) with
$\epsilon_2=1$, $\epsilon_0=4$, and a frequency-dependent Drude
dielectric function $\epsilon_1$ [see Eq.~(\ref{drude})] with
$\omega_p=15\,{\rm eV}$ and $\eta=0.75\,{\rm eV}$~\cite{eta}. The
solid
circles represent the electron
spectrometry measurements reported by Petit, Silcox, and
Vincent~\cite{silcox}. The dashed line represents the nonretarded
surface-plasmon frequency $\omega_p/\sqrt{5}$, which is the
solution of Eq.~(\ref{nonretarded}) with $\epsilon_2=4$ and a Drude
dielectric function $\epsilon_1$. The dotted
line represents the light line $\omega=cq$.}\label{fig3}
\end{figure}

Electron spectrometry measurements of the dispersion of the
surface-plasmon polariton in oxidized Al films were reported by
Pettit, Silcox, and Vincent~\cite{silcox}, spanning the energy range
from the short-wavelength limit where $\omega\sim\omega_p/\sqrt{2}$
all the way to the long-wavelength limit where $\omega\sim c\,q$. The
agreement between the experimental measurements and the prediction of
Eqs.~(\ref{couplingr2})-(\ref{kappa0}) (with a Drude dielectric
function for the Al film and a dielectric constant $\epsilon_0=4$ for
the surrounding oxide) is found to be very good, as shown in Fig.~\ref{fig3}.

In the nonretarded regime ($q>>\omega_s/c$), where
$\kappa_1=\kappa_2=q$, Eqs.~(\ref{couplingr}) and (\ref{coupling})
take the form
\begin{equation}\label{thin1}
{\epsilon_1+\epsilon_2\over\epsilon_1-\epsilon_2}=
\mp{\rm e}^{-qa},
\end{equation}
which for a Drude thin slab [$\epsilon_1$ of Eq.~(\ref{drude})]
in vacuum ($\epsilon_2=1$) yields~\cite{ritchie1}:
\begin{equation}\label{tfilm}
\omega={\omega_p\over\sqrt{2}}\,\left(1\pm{\rm e}^{-qa}\right)^{1/2}.
\end{equation}
This equation has two limiting cases, as discussed by
Ferrell~\cite{ferrell0}. At short wavelengths ($qa>>1$), the surface
waves become decoupled and each surface sustains independent
oscillations at the reduced frequency $\omega_s=\omega_p/\sqrt{2}$
characteristic of a semi-infinite
electron gas with a single plane boundary. At long wavelengths
($qa<<1$), there are
{\it normal} oscillations at $\omega_p$ and {\it tangential} 2D
oscillations at \begin{equation}\label{2D}
\omega_{2D}=(2\pi naq)^{1/2},
\end{equation}
which were later discussed by Stern~\cite{stern} and observed in
artificially
structured semiconductors~\cite{allen} and more recently in a metallic
surface-state band on a silicon surface~\cite{nagao}.

\section{Nonretarded surface plasmon: Simplified models}

The classical picture leading to the retarded Eq.~(\ref{retarded}) and
nonretarded Eq.~(\ref{nonretarded}) ignores both the nonlocality of
the electronic response of the system and the microscopic spatial
distribution of the electron density near the surface. This
microscopic effects can generally be ignored at long wavelengths
where $q<<q_F$; however, as the excitation wavelength approaches
atomic dimensions nonlocal effects can be important.

As nonlocal effects can generally be ignored in the retarded region
where $q<\omega_s/c$ (since $\omega_s/c<<q_F$), here we focus our
attention on the nonretarded regime where $\omega_s/c<q$. In
this regime and in the absence of external sources, the
$\omega$-components of the time-dependent electric and displacement
fields associated with
collective oscillations at a metal surface satisfy the
quasi-static Maxwell's equations
\begin{equation}\label{laplace1}
{\bf\nabla}\cdot{\bf E}({\bf r},\omega)=-4\pi\,
\delta n({\bf r},\omega),
\end{equation}
or, equivalently,
\begin{equation}\label{poisson}
\nabla^2\phi({\bf r},\omega)=4\pi\,\delta n({\bf r},\omega),
\end{equation}
and
\begin{equation}\label{laplace2}
{\bf\nabla}\cdot{\bf D}({\bf r},\omega)=0,
\end{equation}
$\delta n({\bf r},\omega)$ being the fluctuating electron density
associated with the surface plasmon, and $\phi({\bf r},\omega)$ being
the $\omega$ component of the time-dependent scalar potential.

\subsection{Planar surface plasmon}

\subsubsection{Classical model}

In the classical limit, we consider two semi-infinite media with local
(frequency-dependent) dielectric functions $\epsilon_1$ and
$\epsilon_2$ separated by a planar interface at $z=0$, as in
Section~\ref{IIA} (see Fig. 1). In this case, the fluctuating
electron density
$\delta n({\bf r},\omega)$ corresponds to a delta-function sheet at
$z=0$:
\begin{equation}\label{sheet}
\delta n({\bf r},\omega)=\delta n({\bf r}_\parallel,\omega)\
\delta(z),
\end{equation}
where ${\bf r}_\parallel$ defines the position vector in the surface
plane, and the displacement field ${\bf D}({\bf r},\omega)$ takes the
following form:
\begin{equation}\label{classical}
{\bf D}({\bf r},\omega)=\cases{\epsilon_1\,{\bf E}({\bf r},
{\omega}),&$z<0$,\cr\cr
\epsilon_2\,{\bf E}({\bf r},{\omega}),&$z>0$.}
\end{equation}

Introducing Eq.~(\ref{sheet}) into Eq.~(\ref{poisson}), one finds that
self-sustained solutions of Poisson's equation take the form
\begin{equation}\label{nabla1}
\phi({\bf r},\omega)=\phi_0\,{\rm e}^{{\bf q}\cdot{\bf r}_\parallel}\,
{\rm e}^{-q|z|},
\end{equation}
where ${\bf q}$ is a 2D wave vector in the plane of the surface, and
$q=|{\bf q}|$. A combination of Eqs.~(\ref{laplace2}),
(\ref{classical}), and (\ref{nabla1}) with
${\bf E}({\bf r},\omega)=-{\bf\nabla}\phi({\bf r},\omega)$
yields the {\it nonretarded}
surface-plasmon condition of Eq.~(\ref{nonretarded}), i.e.,
\begin{equation}\label{nonretardedbis}
\epsilon_1+\epsilon_2=0.
\end{equation}

\subsubsection{Nonlocal corrections}

Now we consider a more realistic {\it jellium} model of the solid
surface
consisting of a fixed semi-infinite uniform positive background at
$z\leq 0$ plus a neutralizing {\it nonuniform} cloud of interacting
electrons. Within this
model, there is translational invariance in the plane of the surface;
hence, we can define 2D Fourier transforms ${\bf E}(z;q,\omega)$ and
${\bf D}(z;q,\omega)$, the most general linear relation between them
being
\begin{equation}\label{general}
{\bf D}(z;q,\omega)=\int dz'{\bf\epsilon}(z,z';q,\omega)\cdot{\bf
E} (z';q,\omega),
\end{equation}
where the tensor ${\bf\epsilon}(z,z';q,\omega)$ represents the
dielectric function of the medium.

In order to avoid an explicit calculation of ${\bf\epsilon}(z,z';q,
\omega)$, one can assume that far from the surface and at low wave
vectors (but still in the nonretarded regime, i.e.,
$\omega_s/c<q<q_F$) Eq.~(\ref{general}) reduces to an
expression of
the form of Eq.~(\ref{classical}):
\begin{equation}
{\bf D}(z;q,\omega)=\cases{\epsilon_1\,{\bf E}(z;q,
\omega),&$z<z_1$,\cr\cr
\epsilon_2\,{\bf E}(z;q,\omega),&$z>z_2$,}
\end{equation}
where $z_1<<0$ and $z_2>>0$. Eqs.~(\ref{poisson}) and (\ref{laplace2})
with ${\bf E}({\bf r},\omega)=-{\bf\nabla}\phi({\bf r},\omega)$
then yield the following integration of the field components $E_z$
and $D_x$ in terms of the potential $\phi(z)$ at $z_1$ and $z_2$
[where it reduces to the classical potential of Eq.~(\ref{nabla1})]:
\begin{equation}\label{one}
\int_{z_1}^{z_2}dz\,E_z(z;q,\omega)=\phi(z_2;q,\omega)-\phi(z_1;q,
\omega)
\end{equation}
and
\begin{equation}\label{two}
-i\int_{z_1}^{z_2} dz\,D_x(z;q,\omega)=\epsilon_2\,
\phi(z_2;q,\omega)-\epsilon_1\,\phi(z_1;q,\omega).
\end{equation}

Neglecting quadratic and higher-order terms in the wave vector,
Eqs.~(\ref{one}) and (\ref{two}) are found to be compatible
under the surface-plasmon condition~\cite{flores}
\begin{equation}\label{ds}
{\epsilon_1+\epsilon_2\over\epsilon_1-\epsilon_2}=q\left[d_\perp(\omega)-
d_\parallel(\omega)\right],
\end{equation}
$d_\perp(\omega)$ and $d_\parallel(\omega)$ being the so-called
$d$-parameters introduced by Feibelman~\cite{feibelman0}:
\begin{eqnarray}\label{dperp}
d_\perp(\omega)&=&\int dz\,z\,{d\over dz}E_z(z,\omega)\,/\,\int dz\
{d\over dz}E_z(z,\omega)\cr\cr
&=&\int dz\,z\,\delta n(z,\omega)\,/\,\int dz\,\delta n(z,\omega)
\end{eqnarray}
and
\begin{equation}
d_\parallel(\omega)=\int dz\,z\,{d\over dz}D_x(z,\omega)\,/\,\int dz\
{d\over dz}D_x(z,\omega),
\end{equation}
where $E_z(z,\omega)$, $D_x(z,\omega)$, and $\delta n(z,\omega)$
represent the fields and the induced density evaluated in
the $q\to 0$ limit.

For a Drude semi-infinite metal in vacuum [$\epsilon_2=1$ and
Eq.~(\ref{drude}) for $\epsilon_1$], the nonretarded
surface-plasmon condition of Eq.~(\ref{ds})
yields the nonretarded dispersion relation
\begin{equation}\label{sdr}
\omega=\omega_s\left\{1-q{\rm
Re}\left[d_\perp(\omega_s)-d_\parallel(\omega_s)\right]/2+\dots\right\},
\end{equation}
where $\omega_s$ is Ritchie's frequency: $\omega_s=\omega_p/\sqrt{2}$.
For neutral jellium surfaces, $d_\parallel(\omega)$ coincides with the
jellium edge and the linear coefficient of the surface-plasmon
dispersion $\omega(q)$, therefore, only depends on the
position $d_\perp(\omega_s)$ of the centroid of the induced electron
density at $\omega_s$ [see Eq.~(\ref{dperp})] with respect to the
jellium edge.

\subsubsection{Hydrodynamic approximation}\label{hydro}

In a hydrodynamic model, the collective motion of electrons in an
arbitrary inhomogeneous system is expressed in terms of the electron
density $n({\bf r},t)$ and the hydrodynamical velocity ${\bf v}({\bf
r},t)$, which assuming irrotational flow we express as the gradient
of a velocity potential $\psi({\bf r},t)$ such that ${\bf v}({\bf
r},t)=-{\bf\nabla}\psi({\bf r},t)$. First of all, one writes the
basic hydrodynamic Bloch's equations (the continuity equation and the
Bernoulli's equation) in the absence of external
sources~\cite{lundqvist}:
\begin{equation}\label{hd1}
{d\over dt}n({\bf r},t)={\bf\nabla}\cdot\left[n({\bf r},t)\
{\bf\nabla}\psi({\bf r},t)\right]
\end{equation}
and
\begin{equation}\label{hd2}
{d\over dt}\psi({\bf r},t)={1\over 2}\,
\left|{\bf\nabla}\psi({\bf r},t)\right|^2+{\delta G[n]\over\delta
n}+\phi({\bf r},t),
\end{equation}
and Poisson's equation:
\begin{equation}\label{pois}
\nabla^2\phi({\bf r},t)=4\pi\,n({\bf r},t),
\end{equation}
where $G[n]$ is the internal kinetic energy, which is typically
approximated by the Thomas-Fermi functional
\begin{equation}\label{tf}
G[n]={3\over{10}}\,(3\pi^2)^{2/3}\,\left[n({\bf r},t)\right]^{5/3}.
\end{equation}

The hydrodynamic equations [Eqs.~(\ref{hd1})-(\ref{pois})] are
nonlinear equations, difficult to solve. Therefore, one typically uses
perturbation theory to expand the electron density and the velocity
potential as follows
\begin{equation}
n({\bf r},t)=n_0({\bf r})+n_1({\bf r},t)+\dots
\end{equation}
and
\begin{equation}
\psi({\bf r},t)=0+\psi_1({\bf r},t)\dots,
\end{equation}
so that Eqs.~(\ref{hd1})-(\ref{pois}) yield the linearized
hydrodynamic equations
\begin{equation}\label{hd1p}
{d\over dt}n_1({\bf r},t)={\bf\nabla}\cdot\left[n_0({\bf r})\
{\bf\nabla}\psi_1({\bf r},t)\right],
\end{equation}
\begin{equation}\label{hd2p}
{d\over dt}\psi_1({\bf r},t)=\left[\beta({\bf r})\right]^2\,{n_1({\bf
r},t)\over n_0({\bf r})}+\phi_1({\bf r},t),
\end{equation}
and
\begin{equation}\label{hd3p}
\nabla^2\phi_1({\bf r},t)=4\pi\,n_1({\bf r},t),
\end{equation}
where $n_0({\bf r})$ is the unperturbed electron density and
$\beta({\bf r})=\sqrt{1/3}\,\left[3\pi^2n_0({\bf r})\right]^{1/3}$
represents the speed of
propagation of hydrodynamic disturbances in the electron
system~\cite{noteh}.

We now consider a semi-infinite metal in vacuum consisting of an
abrupt step of the unperturbed electron density at the interface,
which we choose to be located at $z=0$:
\begin{equation}\label{density0}
n_0(z)=\cases{\bar n,&$z\leq 0$,\cr\cr
0,&$z>0$.}
\end{equation}
Hence, within this model $n_0({\bf r})$ and $\beta({\bf r})$ are
constant at $z\leq 0$ and vanish at $z>0$.

Introducing Fourier transforms, Eqs.~(\ref{hd1p})-(\ref{density0})
yield the basic differential equation
for the plasma normal modes at $z\leq 0$:
\begin{equation}\label{normal}
\nabla^2(\omega^2-\omega_p^2+\beta^2\nabla^2)\psi_1({\bf r},\omega)
\qquad
(z\leq 0),
\end{equation}
and Laplace's equation at $z>0$:
\begin{equation}
\nabla^2\phi_1({\bf r},\omega)=0\qquad (z>0),
\end{equation}
where both $n_1({\bf r},\omega)$ and $\psi_1({\bf r},\omega)$ vanish.
Furthermore, translational invariance in the plane of the surface
allows to introduce the 2D Fourier transform
$\psi_1(z;{\bf q},\omega)$, which according to Eq.~(\ref{normal}) must
satisfy the
following equation at $z\leq 0$:
\begin{equation}\label{fourier2}
(-q^2+d^2/dz^2)\left[\omega^2-\omega_p^2-\beta
(-q^2+d^2/dz^2)\right]\psi_1(z;{\bf q},\omega),
\end{equation}
where ${\bf q}$ represents a 2D wave vector in the plane of the
surface.

Now we need to specify the boundary conditions. Ruling out exponential
increase at $z\to\infty$ and noting that the normal component of the
hydrodynamical velocity should vanish at the interface, for each value
of $q$ one finds
solutions to Eq.~(\ref{fourier2}) with
frequencies~\cite{wilems,barton}
\begin{equation}\label{plasmonb}
\omega^2\geq\omega_p^2+\beta^2\,q^2
\end{equation}
and
\begin{equation}\label{plasmons}
\omega^2={1\over 2}\left[\omega_p^2+\beta^2q^2+
\beta q\sqrt{2\omega_p^2+\beta^2q^2}\right].
\end{equation}
Eqs.~(\ref{plasmonb}) and (\ref{plasmons}) represent a continuum of
bulk normal modes and a surface normal mode, respectively. At long
wavelengths, where $\beta\,q/\omega_p<<1$ (but still in the
nonretarded regime where $\omega_s/c<q$), Eq.~(\ref{plasmons})
yields the surface-plasmon dispersion relation
\begin{equation}\label{hdr}
\omega=\omega_p/\sqrt{2}+\beta\,q/2,
\end{equation}
which was first derived by Ritchie~\cite{ritchie0} using Bloch's
equations, and later by Wagner~\cite{wagner} and by Ritchie and
Marusak~\cite{marusak} by assuming, within a Boltzmann
transport-equation approach, specular reflection at the surface.

\subsection{Localized surface plasmons: classical approach}

Metal-dielectric interfaces of arbitrary geometries also support
charge density oscillations similar to the surface plasmons
characteristic of planar interfaces. In the long-wavelength (or
classical) limit, in which the interface separates two media with
local (frequency-dependent) dielectric functions $\epsilon_1$ and
$\epsilon_2$, one writes
\begin{equation}\label{locall}
{\bf D}_i({\bf r},\omega)=\epsilon_i\,{\bf E}_i({\bf r},\omega),
\end{equation}
where the index $i$ refers to the media 1 and 2 separated by the
interface. In the case of simple geometries, such as spherical and
cylindrical interfaces, Eqs.~(\ref{laplace1})-(\ref{laplace2}) can be
solved explicitly with the aid of Eq.~(\ref{locall}) to find explicit
expressions for the nonretarded surface-plasmon condition.

\subsubsection{Simple geometries}

\paragraph{Spherical interface.} In the case of a sphere of dielectric
function $\epsilon_1$ in a host medium of dielectric function
$\epsilon_2$, the classical (long-wavelength) planar surface-plasmon
condition of Eq.~(\ref{nonretardedbis}) is easily found to be
replaced by~\cite{bohren}
\begin{equation}\label{spheres}
l\,\epsilon_1+(l+1)\,\epsilon_2=0,\qquad l=1,2,\dots,
\end{equation}
which in the case of a Drude metal sphere [$\epsilon_1$ of
Eq.~(\ref{drude})] in vacuum ($\epsilon_2=1$) yields the Mie
plasmons at frequencies
\begin{equation}\label{l2}
\omega_l=\omega_p\,\sqrt{l\over2l+1}.
\end{equation}

\paragraph{Cylindrical interface.} In the case of an infinitely long
cylinder of dielectric function $\epsilon_1$ in a host medium of
dielectric function $\epsilon_2$, the classical (long-wavelength)
surface-plasmon condition depends on the direction
of the electric field. For electromagnetic waves with the electric
field normal to the interface ($p$-polarization), the corresponding
long-wavelength (and nonretarded) surface-plasmon condition coincides
with that of a planar surface, i.e.,~\cite{sch,pitarke2,pitarke3}
\begin{equation}\label{cyl}
\epsilon_1+\epsilon_2=0,
\end{equation}
which for Drude cylinders [$\epsilon_1$ of Eq.~(\ref{drude})] in
vacuum ($\epsilon_2=1$) yields the planar surface-plasmon frequency
$\omega_s=\omega_p/\sqrt{2}$.

For electromagnetic waves with the electric field parallel to the axis
of the cylinder ($s$-polarization), the presence of the interface
does not modify the electric field and one easily finds that only the
bulk mode of the host medium is present, i.e., one finds the plasmon
condition
\begin{equation}\label{bulk}
\epsilon_2=0.
\end{equation}
In some situations, instead of having one single cylinder in a host
medium, an array of parallel cylinders may be present with a filling
fraction $f$. In this case and for electromagnetic waves polarized
along the cylinders ($s$-polarization), the plasmon condition of
Eq.~(\ref{bulk}) must be replaced by~\cite{berg0}
\begin{equation}\label{berg}
f\,\epsilon_1+(1-f)\,\epsilon_2=0,
\end{equation}
which for Drude cylinders [$\epsilon_1$ of Eq.~(\ref{drude})] in
vacuum ($\epsilon_2=1$) yields the reduced plasmon frequency
$\omega=\sqrt{f}\,\omega_p$.

\subsubsection{Boundary-charge method}

In the case of more complex interfaces, a so-called boundary-charge
method (BCM) has been used by several authors to determine
numerically the classical (long-wavelength) frequencies of localized
surface
plasmons. In this approach, one first considers the $\omega$-component
of the time-dependent surface charge
density arising from the difference between the normal components of
the electric fields inside and outside the surface:
\begin{equation}
\sigma_s({\bf r},\omega)={1\over 4\pi}
\left[\left.{\bf E}({\bf r},\omega)\cdot{\bf n}
\right|_{{\bf r}={\bf r}^-}+\left.{\bf E}({\bf r},\omega)\cdot{\bf n}
\right|_{{\bf r}={\bf r}^+}\right],
\end{equation}
which noting that the normal component of the displacement vector [see
Eq.~(\ref{locall})] must be continuous yields the following
expression:
\begin{equation}\label{sigma}
\sigma_s({\bf r},\omega)={1\over 4\pi}\
{\epsilon_1-\epsilon_2\over\epsilon_1}\,
\left.{\bf E}({\bf r},\omega)\cdot{\bf n}
\right|_{{\bf r}={\bf r}^+},
\end{equation}
where ${\bf n}$ represents a unit vector in the direction
perpendicular to the interface.

An explicit expression for the normal component of the electric field
at a point of medium 2 that is infinitely close to the interface
(${\bf r}={\bf r}^+$) can be obtained with the use of Gauss' theorem. One
finds:
\begin{equation}\label{normal2}
\left.{\bf E}({\bf r},\omega)\cdot{\bf n}
\right|_{{\bf r}={\bf r}^+}=-{\bf n}\cdot
{\bf\nabla}\phi({\bf r},\omega)+2\pi\sigma_s({\bf r},\omega),
\end{equation}
where $\phi({\bf r},\omega)$ represents the scalar potential. In the
absence of external sources, this potential is entirely due to the
surface charge density itself:
\begin{equation}\label{phi}
\phi({\bf r},\omega)=\int d^2{\bf r}'\,{\sigma_s({\bf r}',\omega)\over
|{\bf r}-{\bf r}'|}.
\end{equation}
Combining Eqs.~(\ref{sigma})-(\ref{phi}), one finds the following
integral equation:
\begin{equation}
2\pi\,{\epsilon_1+\epsilon_2\over\epsilon_1-\epsilon_2}\,\sigma_s({\bf
r},\omega)-\int d^2{\bf r}'\,{{\bf r}-{\bf r}'\over|{\bf r}-{\bf r}'|
^3}\cdot{\bf n}\,\sigma_s({\bf r}',\omega)=0,
\end{equation}
which describes the self-sustained oscillations of the system.

The boundary-charge method has been used by several authors to
determine the normal-mode frequencies of a cube~\cite{fuchs1,fuchs2}
and of bodies of arbitrary shape~\cite{ouy1,ouy2}. More recent
applications of this method include investigations of the surface
modes of channels cut on planar surfaces~\cite{ouy3}, the surface
modes of coupled parallel wires~\cite{ouy4}, and the electron energy
loss near inhomogeneous dielectrics~\cite{javi1,javi2}. A
generalization of this procedure that includes relativistic
corrections has been reported as well~\cite{javihowie}.

\subsubsection{Composite systems: effective-medium
approach}\label{classicals}

Composite systems with a large number of interfaces can often be
replaced by an effective homogeneous medium that in the
long-wavelength
limit is characterized by a local
effective dielectric function $\epsilon_{eff}(\omega)$.
Bergman~\cite{berg1} and Milton~\cite{berg2} showed that in the case
of a two-component system with local
(frequency-dependent) dielectric functions $\epsilon_1$ and
$\epsilon_2$ and volume fractions $f$ and $1-f$, respectively, the
long-wavelength effective dielectric function of the system can be
expressed as a sum of simple poles that only depend on the
microgeometry of the composite material and not on the dielectric
functions of the components:
\begin{equation}\label{eff1}
\epsilon_{eff}(\omega)=\epsilon_2\left[1-f\sum_\nu{B_\nu\over
u-m_\nu}\right],
\end{equation}
where $u$ is the spectral variable
\begin{equation}\label{spectral}
u=\left[1-\epsilon_1/\epsilon_2\right]^{-1},
\end{equation}
$m_\nu$ are depolarization factors, and $B_\nu$ are the strengths of
the corresponding normal modes, which all add up to unity:
\begin{equation}\label{sum1}
\sum_\nu B_\nu=1.
\end{equation}
Similarly,
\begin{equation}\label{eff2}
\epsilon_{eff}^{-1}
(\omega)=\epsilon_2^{-1}\left[1+f\sum_\nu{C_\nu\over
u-n_\nu}\right],
\end{equation}
with
\begin{equation}\label{sum2}
\sum_\nu C_\nu=1.
\end{equation}
The optical absorption and the long-wavelength energy
loss of moving charged particles
are known to be dictated by the poles of the {\it local} effective
dielectric function $\epsilon_{eff}(\omega)$ and inverse dielectric
function $\epsilon_{eff}^{-1}(\omega)$, respectively. If there is one
single interface, these poles are known to coincide.

In particular, in the case of a two-component isotropic system
composed of {\it identical} inclusions of dielectric function
$\epsilon_1$ in a host medium of dielectric function $\epsilon_2$,
the effective dielectric function $\epsilon_{eff}(\omega)$ can be
obtained from the following relation:
\begin{equation}\label{macro}
(\epsilon_{eff}-\epsilon_2)\,{\bf E}=f(\epsilon_1-\epsilon_2)\,{\bf
E}_{in},
\end{equation}
where ${\bf E}$ is the macroscopic electric field averaged over the
composite:
\begin{equation}\label{average}
{\bf E}=f{\bf E}_{in}+(1-f){\bf E}_{out},
\end{equation}
${\bf E}_{in}$ and ${\bf E}_{out}$ representing the average electric
field inside and outside the inclusions, respectively~\cite{noteani}.

\paragraph{Simple geometries.}\label{cylinder} If there is only one
mode with strength different from zero, as occurs (in the
long-wavelength limit) in the case of one
single sphere or cylinder in a host medium, Eqs.~(\ref{eff1}) and
(\ref{eff2}) yield
\begin{equation}\label{efone}
\epsilon_{eff}(\omega)=\epsilon_2\left[1-f{1\over u-m}\right]
\end{equation}
and
\begin{equation}\label{eftwo}
\epsilon_{eff}^{-1}(\omega)=\epsilon_2^{-1}\left[1+f{1\over
u-n}\right],
\end{equation}
normal modes occurring, therefore, at the frequencies dictated by the
following conditions:
\begin{equation}\label{ef1}
m\,\epsilon_1+(1-m)\,\epsilon_2=0
\end{equation}
and
\begin{equation}\label{ef2}
n\,\epsilon_1+(1-n)\,\epsilon_2=0.
\end{equation}
For Drude particles [$\epsilon_1$ of Eq.~(\ref{drude})] in vacuum
($\epsilon_2=1$), these frequencies are easily found to
be $\omega=\sqrt{m}\,\omega_p$ and $\omega=\sqrt{n}\,\omega_p$,
respectively.

Indeed, for a {\it single} 3D spherical or 2D circular~\cite{p}
inclusion in a host medium, an
elementary
analysis shows that the electric field ${\bf E}_{in}$ in the interior
of the inclusion is
\begin{equation}\label{d}
{\bf E}_{in}={u\over u-m}\,{\bf E},
\end{equation}
where $m=1/D$, $D$ representing the dimensionality of the inclusions,
i.e., $D=3$ for spheres and $D=2$ for cylinders.
Introduction of Eq.~(\ref{d}) into Eq.~(\ref{macro}) leads to an
effective dielectric function of the form of Eq.~(\ref{efone}) with
$m=1/D$, which
yields [see Eq.~(\ref{ef1})] the surface-plasmon condition dictated by
Eq.~(\ref{spheres}) with $l=1$ in the case of
spheres ($D=3$) and the surface-plasmon condition of Eq.~(\ref{cyl})
in the case of cylinders ($D=2$). This result indicates that in the
nonretarded long-wavelength limit (which holds for wave vectors ${\bf
q}$ such
that $\omega_sa/c<q\,a<<1$, $a$ being the radius of the inclusions)
both the absorption of light and the energy-loss spectrum of a single
3D spherical or 2D circular inclusion exhibit one single strong
maximum at the dipole resonance where $\epsilon_1+2\epsilon_2=0$ and
$\epsilon_1+\epsilon_2=0$, respectively, which for a Drude sphere and
cylinder [$\epsilon_1$ of Eq.~(\ref{drude})] in vacuum
($\epsilon_2=1$) yield $\omega=\omega_p/\sqrt{3}$ and
$\omega=\omega_p/\sqrt{2}$.

In the case of electromagnetic waves polarized along one single
cylinder or array of parallel cylinders ($s$-polarization), the
effective dielectric function of the composite is simply the average
of the dielectric functions of its constituents, i.e.:
\begin{equation}
\epsilon_{eff}=f\,\epsilon_1+(1-f)\,\epsilon_2,
\end{equation}
which can also be written in the form of Eqs.~(\ref{efone}) and
(\ref{eftwo}), but now with $m=0$ and $n=f$, respectively. Hence, for
this
polarization the absorption of light exhibits no maxima (in the case
of a
dielectric host medium with constant dielectric function $\epsilon_2$)
and the
long-wavelength energy-loss spectrum exhibits a strong
maximum at
frequencies dictated by the plasmon condition of Eq.~(\ref{berg}),
which in the case of Drude cylinders [$\epsilon_1$ of
Eq.~(\ref{drude})] in vacuum ($\epsilon_2=1$) yields the reduced
plasmon frequency $\omega=\sqrt{f}\,\omega_p$.

\paragraph{Maxwell-Garnett approximation.}

The interaction among spherical (or circular) inclusions in a host
medium can be introduced approximately in the framework of the
well-known
Maxwell-Garnett (MG) approximation~\cite{bohren}.

The basic assumption of this approach is that the average electric
field ${\bf E}_{in}$ within a particle located in a system of
identical particles is
related to the average field ${\bf E}_{out}$ in the medium {\it
outside} as in the case of a single isolated (noninteracting)
particle, thereby only dipole interactions being taken into account.
Hence, in this approach the electric field ${\bf E}_{in}$ is taken to
be of the form of Eq.~(\ref{d}) but with the macroscopic electric
field ${\bf E}$ replaced by the electric field ${\bf E}_{out}$
outside, which together with Eqs.~(\ref{macro}) and (\ref{average})
yields the effective dielectric function and effective inverse
dielectric function of Eqs.~(\ref{efone}) and (\ref{eftwo}) with the
depolarization factors $m=n=1/D$ (corresponding to the dilute limit,
where $f\to 0$) replaced by
\begin{equation}\label{mg1}
m={1\over D}\,(1-f)
\end{equation}
and
\begin{equation}\label{mg2}
n={1\over D}\,\left[1+(D-1)f\right].
\end{equation}

\subsubsection{Periodic structures}

Over the years, theoretical studies of the normal modes of complex
composite systems had been generally restricted to mean-field
theories of the Maxwell-Garnett type, which approximately account
for the behavior of localized dipole plasmons~\cite{bohren}.
Nevertheless, a number of methods have been developed recently for a
full solution of Maxwell's equations in periodic
structures~\cite{bands1,bands2,bands3,bands4,bands5,english}. The
transfer matrix method has been used to determine the normal-mode
frequencies of a lattice of metallic cylinders~\cite{pitarken1} and
rods~\cite{pitarken2}, a so-called on-shell method has been employed
by Yannopapas {\it et al.} to investigate the plasmon modes of a
lattice of metallic spheres in the low filling fraction
regime~\cite{modinos}, and a finite difference time domain (FDTD)
scheme has been adapted to extract the effective response of
metallic structures~\cite{english}.

Most recently, an embedding method~\cite{bands4} has been employed to
solve Maxwell's equations, which has allowed to calculate the
photonic band structure of three- and two-dimensional lattices of
nanoscale metal spheres and cylinders in the frequency range of the
Mie plasmons~\cite{ingles1}. For small filling fractions, there is a
surface-plasmon polariton which in the
non-retarded region yields the non-dispersive Mie plasmon with
frequency $\omega_p/\sqrt{D}$. As the filling fraction
increases, a continuum of plasmon modes is found to exist
between zero frequency and the bulk metal plasmon
frequency~\cite{ingles1}, which yield strong absorption of incident
light and whose energies can be tuned according to the
particle-particle separation~\cite{javinew}.

\subsubsection{Sum rules}

\begin{figure}
\includegraphics[width=0.95\linewidth]{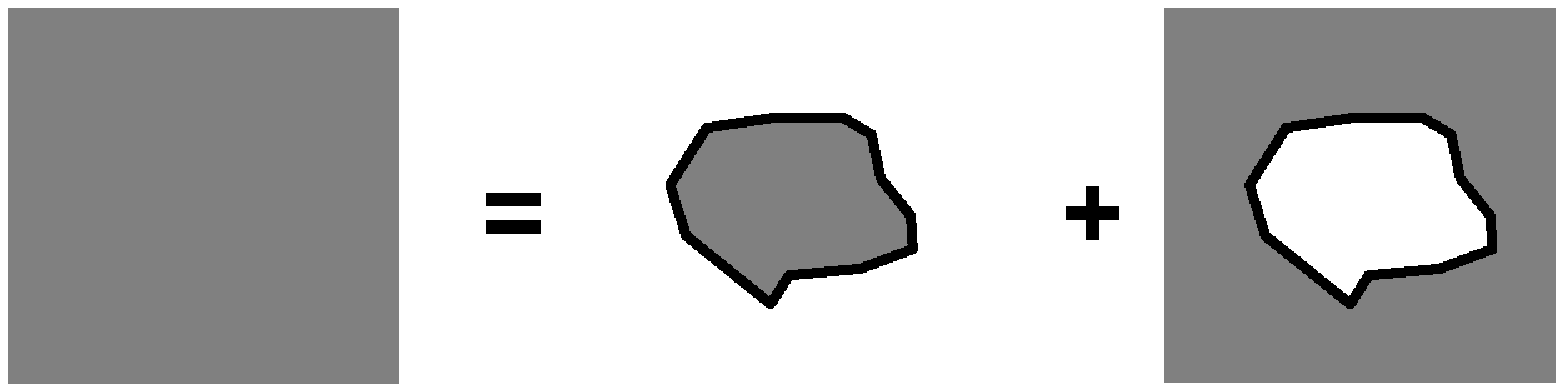}\\
\vspace{1cm}
\includegraphics[width=0.95\linewidth]{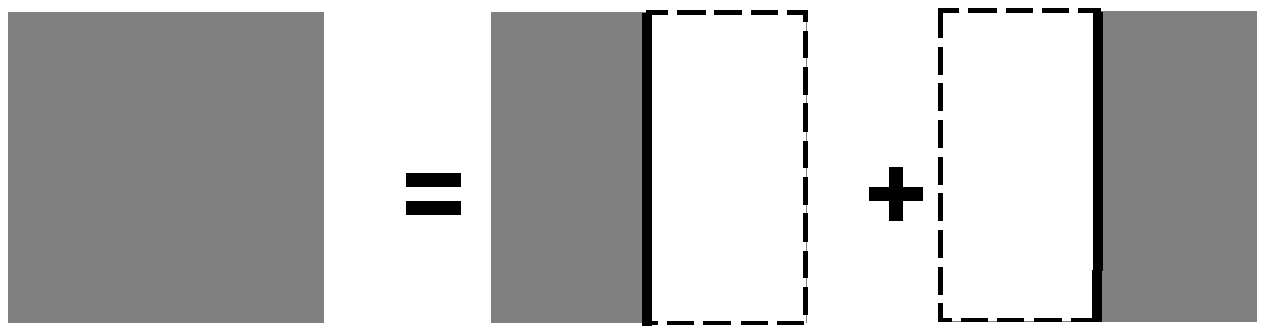}\\
\caption{Complementary systems in which the regions of plasma and
vacuum are interchanged. The top panel represents the general
situation. The bottom panel represents a half-space filled with metal
and interfaced with vacuum. The surface-mode frequencies
$\omega_{s_1}$ and $\omega_{s_2}$ of these systems fulfill the sum
rule of Eq.~(\ref{sumr1}).}\label{fig4}
\end{figure}

Sum rules have played a key role in providing insight in the
investigation of a variety of physical situations. A useful sum rule
for the surface modes in complementary media with arbitrary geometry
was introduced by Apell {\it et al.}~\cite{apell}, which in the
special case of a
metal/vacuum interface implies that~\cite{ronve}
\begin{equation}\label{sumr1}
\omega_{s_1}^2+\omega_{s_2}^2=\omega_p^2,
\end{equation}
where $\omega_{s_1}$ is the surface-mode frequency of a given system,
and $\omega_{s_2}$ represents the surface mode of a second
complementary system in which the regions of plasma and vacuum are
interchanged (see Fig.~\ref{fig4}).

For example, a half-space filled with a metal of bulk plasma frequency
$\omega_p$ and interfaced with vacuum maps into itself (see
bottom panel of Fig.~\ref{fig4}), and therefore Eq.~(\ref{sumr1})
yields
\begin{equation}\label{srule1}
\omega_{s_1}=\omega_{s_2}=\omega_p/\sqrt{2},
\end{equation}
which is Ritchie's frequency of plasma oscillations at a metal/vacuum
planar
interface.

Other examples are a Drude metal sphere in vacuum, which sustains {\it
localized} Mie plasmons at frequencies given by Eq.~(\ref{l2}), and a
spherical void in a Drude metal, which shows Mie plasmons
at frequencies
\begin{equation}\label{l3}
\omega_l=\omega_p\,\sqrt{l+1\over2l+1}.
\end{equation}
The squared surface-mode frequencies of the sphere [Eq.~(\ref{l2})]
and the void [(Eq.~(\ref{l3})] add up to $\omega_p^2$ for all $l$, as
required by Eq.~(\ref{sumr1}).

The splitting of surface modes that occurs in thin films
due to the coupling of the electromagnetic fields in the two surfaces
[see Eq.~(\ref{thin1})]
also occurs in the case of localized modes. Apell {\it et
al.}~\cite{apell} proved a second sum rule, which relates the
surface modes corresponding to the in-phase and out-of-phase linear
combinations of the screening charge densities at the interfaces. In
the case of metal/vacuum interfaces this sum rule takes the form of
Eq.~(\ref{sumr1}), but now $\omega_{s_1}$ and $\omega_{s_2}$ being
in-phase and out-of-phase modes of the same system.

For a Drude metal film with equal and abrupt planar surfaces, the
actual values of the {\it nonretarded} $\omega_{s_1}$ and
$\omega_{s_2}$ are those given by
Eq.~(\ref{tfilm}), which fulfill the sum rule dictated by
Eq.~(\ref{sumr1}). For a spherical fullerene molecule described by
assigning a Drude dielectric function to every point between the inner
and outer surfaces of radii $r_1$ and $r_2$, one finds the
following frequencies for the in-phase and out-of-phase surface
modes~\cite{apell2}:
\begin{equation}
\omega_s^2={\omega_p^2\over 2}\left[1\pm{1\over
2l+1}\sqrt{1+4l(l+1)(r_1/r_2)^{2l+1}}\right],
\end{equation}
also fulfilling the sum rule of Eq.~(\ref{sumr1}).

Another sum rule has been reported
recently~\cite{pitarken1,pitarken2}, which relates the frequencies of
the modes that can be excited by light [as dictated by the poles of
the effective dielectric function of
Eq.~(\ref{eff1})] and those modes that can be excited by moving
charged particles [as dictated by the poles of the effective inverse
dielectric
function of Eq.~(\ref{eff2})]. Numerical calculations for various
geometries have shown that the depolarization factors $m_\nu$ and
$n_\nu$ entering
Eqs.~(\ref{eff1}) and (\ref{eff2}) satisfy the
relation~\cite{pitarken1,pitarken2}
\begin{equation}\label{srulen}
n_\nu=1-(D-1)\,m_\nu,
\end{equation}
where $D$ represents the dimensionality of the inclusions.

Furthermore, combining Eqs.~(\ref{efone}) and (\ref{eftwo})
(and assuming, therefore, that only dipole interactions are present)
with the sum rule of Eq.~(\ref{srulen}) yields
Eqs.~(\ref{mg1}) and (\ref{mg2}), i.e., the MG approximation.
Conversely, as long as multipolar modes contribute to the spectral
representation of the effective response [see Eqs.~(\ref{eff1}) and
(\ref{eff2})], the strength of the dipolar modes decreases [see
Eqs.~(\ref{sum1}) and (\ref{sum2})] and a
combination of Eqs.~(\ref{eff1}) and (\ref{eff2}) with
Eq.~(\ref{srulen}) leads to the conclusion that the dipolar
resonances must necessarily deviate from their MG counterparts
dictated by Eqs.~(\ref{mg1}) and (\ref{mg2}). That a nonvanishing
contribution from multipolar modes appears together with a deviation
of the frequencies of the dipolar modes with respect to their MG
counterparts was shown explicitly in Ref.~\cite{pitarken1}.

\section{Dynamical structure factor}

The dynamical structure factor $S({\bf r},{\bf r}';\omega)$
represents a key quantity in the description of both
single-particle and collective electronic excitations in a
many-electron system~\cite{pines}. The rate for the generation of
electronic excitations by an external potential, the inelastic
differential cross section for external particles to scatter in a
given direction, the inelastic lifetime of excited hot electrons,
the so-called stopping power of a many-electron system for moving
charged particles, and the ground-state energy of an arbitrary
many-electron system (which is involved in, e.g., the surface
energy and the understanding of Van der Waals interactions) are
all related to the dynamical structure factor of the system.

The dynamical structure factor, which accounts for the
particle-density fluctuations of the system, is defined as follows
\begin{equation}
S({\bf r},{\bf r}';\omega)=\sum_n\,\delta\hat\rho_{0n}({\bf r}_1)\,
\delta\hat\rho_{n0}({\bf r}_2)\,\delta(\omega-E_n+E_0).
\end{equation}
Here, $\delta\hat\rho_{n0}({\bf r})$ represent
matrix elements, taken
between the many-particle ground state $|\Psi_{0}\rangle$
of energy $E_0$ and the many-particle excited state $|\Psi_{n}\rangle$
of energy $E_n$, of the operator $\hat\rho({\bf r})-n_0({\bf r})$,
where $\hat\rho({\bf r})$ is the electron-density operator~\cite{fetter}
\begin{equation}\label{eqa4}
\hat\rho({\bf r})=\sum_{i=1}^N\hat\delta({\bf r}-{\bf r}_i),
\end{equation}
with $\hat\delta$ and ${\bf r}_i$ describing the Dirac-delta operator and
electron coordinates, respectively, and $n_0({\bf r})$ represents
the ground-state electron density, i.e.:
\begin{equation}\label{groundd}
n_0({\bf r})=<\Psi_0|\hat\rho({\bf r})|\Psi_0>.
\end{equation}

The many-body ground and excited states of a many-electron system
are {\it unknown} and the dynamical structure factor is, therefore,
difficult to calculate. Nevertheless, one can use the
zero-temperature limit of the fluctuation-dissipation
theorem~\cite{callen}, which relates the dynamical structure factor
$S({\bf r},{\bf r}';\omega)$ to the dynamical density-response
function $\chi({\bf r},{\bf r}';\omega)$ of linear-response theory.
One writes,
\begin{equation}
S({\bf r},{\bf r}';\omega)=-{\Omega\over\pi}\,{\rm Im}
\chi({\bf r},{\bf r}';\omega)\,\theta(\omega),
\end{equation}
where $\Omega$ represents the normalization volume and $\theta(x)$ is
the Heaviside step function.

\section{Density-response function}

Take a system of $N$ interacting electrons exposed to a
frequency-dependent
external potential $\phi^{ext}({\bf r},\omega)$. Keeping terms of
first order
in the external perturbation and neglecting {\it retardation} effects,
time-dependent perturbation theory yields
the following expression for the induced electron density~\cite{fetter}:
\begin{equation}\label{eqa1}
\delta n({\bf r},\omega)=\int{\rm d}{\bf r}'\,
\chi({\bf r},{\bf  r'};\omega)\,\phi^{ext}({\bf r'},\omega),
\end{equation}
where $\chi({\bf r},{\bf  r'};\omega)$ represents the so-called
density-response function of the many-electron system:
\begin{eqnarray}\label{eqa2}
\chi({\bf r}, {\bf r'};\omega)&=&\sum_n
\rho^{*}_{n0}({\bf r}) \rho_{n0}({\bf r}')
\left[{1\over E_0-E_n+\hbar(\omega+{\rm i}\eta)}\right.\cr\cr
&&\left.-{1\over E_0+E_n+\hbar(\omega+{\rm i}\eta)}\right],
\end{eqnarray}
$\eta$ being a positive infinitesimal.

The imaginary part of the true density-response function of
Eq.~(\ref{eqa2}), which accounts for the creation of both collective
and single-particle excitations in the many-electron system, is known
to satisfy the so-called
$f$-sum rule:
\begin{equation}\label{fsumrule}
\int_{-\infty}^\infty d\omega\,\omega\,{\rm Im}
\chi({\bf r},{\bf r}';\omega)=
-\pi{\bf\nabla}\cdot{\bf\nabla}'\left[n_0({\bf r})\delta({\bf r},{\bf
r}')\right],
\end{equation}
with $n_0({\bf r})$ being the unperturbed ground-state electron
density of
Eq.~(\ref{groundd}).

\subsection{Random-phase approximation (RPA)}

In the so-called random-phase or, equivalently, time-dependent Hartree
approximation, the electron density
$\delta n({\bf r},\omega)$ induced in an {\it interacting} electron
system by a small external potential $\phi^{ext}({\bf r},\omega)$ is
obtained as the electron density induced in a {\it noninteracting}
Hartree system (of electrons moving in a self-consistent Hartree
potential) by both the external potential
$\phi^{ext}({\bf r},\omega)$ and the induced potential
\begin{equation}\label{ext}
\delta\phi^H({\bf r},\omega)=\int d{\bf r}'\,v({\bf r},{\bf r}')\
\delta
n({\bf r}',\omega),
\end{equation}
with $v({\bf r},{\bf r}')$ representing the bare Coulomb interaction.
Hence, in this approximation one writes
\begin{eqnarray}\label{eqa1new}
&&\delta n({\bf r},\omega)=\int{\rm d}{\bf r}'\,
\chi^0({\bf r},{\bf  r'};\omega)\,\cr\cr
&&\times\left[\phi^{ext}({\bf r'},\omega)
+\int d{\bf r}''\,v({\bf r}',{\bf r}'')\,\delta
n({\bf r}'',\omega)\right],
\end{eqnarray}
which together with Eq.~(\ref{eqa1}) yields the following Dyson-type
equation for the {\it interacting} density-response function:
\begin{eqnarray}\label{eq:Xalda}
\chi({\bf r},{\bf r}';\omega)&=&\chi^0({\bf r},{\bf
r}';\omega)+\int{\rm d}{\bf
r}_1\int{\rm d}{\bf r}_2\,\chi^0({\bf r},{\bf r}_1;\omega)\cr
\cr&\times&v({\bf r}_1,{\bf r}_2)\,\chi({\bf r}_2, {\bf r}';\omega),
\end{eqnarray}
where $\chi^0({\bf r},{\bf r}';\omega)$ denotes the density-response
function of noninteracting Hartree electrons:
\begin{eqnarray}\label{chi0}
\chi^0({\bf r},{\bf r}';\omega)
&=&{2\over\Omega}\sum_{i,j}(f_i-f_j)\cr\cr
&\times&{\psi_i({\bf r})\psi_j^*({\bf r})
\psi_j({\bf r}')\psi_i^*({\bf r}')\over
\omega-\varepsilon_j+\varepsilon_i+{\rm i}\eta}.
\end{eqnarray}
Here, $f_i$ are Fermi-Dirac occupation factors,
which at zero temperature take the form
$f_i=\theta(\varepsilon_F-\varepsilon_i)$, $\varepsilon_F$ being the
Fermi energy, and the single-particle states and energies
$\psi_i({\bf r})$ and
$\varepsilon_i$ are the eigenfunctions and eigenvalues of a Hartree
Hamiltonian, i.e.:
\begin{equation}\label{hartree}
\left[-{1\over 2}\nabla^2+v_H[n_0]({\bf r})\right]\,\psi_i({\bf
r})=\varepsilon_i\,\psi_i({\bf r}),
\end{equation}
where
\begin{equation}
v_H[n_0]({\bf r})=v_0({\bf r})+\int d{\bf r}'\,v({\bf r},{\bf r}')
n_0({\bf r}'),
\end{equation}
with $v_0({\bf r})$ denoting a static external potential and
$n_0({\bf r})$ being the unperturbed Hartree electron density:
\begin{equation}\label{groundh}
n_0({\bf r})=\sum_{i=1}^N|\psi_i({\bf r})|^2.
\end{equation}

\subsection{Time-dependent density-functional theory}

In the framework of time-dependent density-functional theory
(TDDFT)~\cite{tddft}, the {\it exact} density-response function of an
interacting many-electron system is found to obey the following
Dyson-type equation:
\begin{eqnarray}\label{eq:Xalda2}
&&\chi({\bf r},{\bf r}';\omega)=\chi^0({\bf r},{\bf
r}';\omega)+\int{\rm d}{\bf
r}_1\int{\rm d}{\bf r}_2\,\chi^0({\bf r},{\bf r}_1;\omega)\cr
\cr&&\times\left\{v({\bf r}_1,{\bf r}_2)+f_{xc}[n_0]({\bf r}_1,{\bf
r}_2;\omega)\right\}\chi({\bf r}_2, {\bf r}';\omega).
\end{eqnarray}
Here, the noninteracting density-response function
$\chi^0({\bf r},{\bf r}';\omega)$ is of the form of Eq.~(\ref{chi0}),
but with the single-particle states and energies $\psi_i({\bf r})$
and $\varepsilon_i$ being now the eigenfunctions and eigenvalues of
the Kohn-Sham Hamiltonian of DFT, i.e:
\begin{equation}\label{dft}
\left[-{1\over 2}\nabla^2+v_{KS}[n_0]({\bf r})\right]\,\psi_i({\bf
r})=\varepsilon_i\,\psi_i({\bf r}),
\end{equation}
where
\begin{equation}\label{vks}
v_{KS}[n_0]({\bf r})=v_H[n_0]({\bf r})+v_{xc}[n_0]({\bf r}),
\end{equation}
with
\begin{equation}\label{vxc}
v_{xc}[n_0]({\bf r})=\left.{\delta E_{xc}[n]\over\delta n({\bf
r})}\right|_{n=n_0}.
\end{equation}
$E_{xc}[n]$ represents the {\it unknown} XC energy functional
and $n_0({\bf r})$ denotes the exact unperturbed electron density of
Eq.~(\ref{groundd}), which DFT shows to coincide with that of
Eq.~(\ref{groundh}) but with the Hartree eigenfunctions $\psi_i({\bf r})$ of
Eq.~(\ref{hartree}) being replaced by their Kohn-Sham counterparts of
Eq.~(\ref{dft}). The xc kernel $f_{xc}[n_0]({\bf r},{\bf r}';\omega)$ denotes
the Fourier transform of
\begin{equation}\label{kernel}
f_{xc}[n_0]({\bf r},t;{\bf r}';t')=\left.
{\delta v_{xc}[n]({\bf r},t)\over\delta n({\bf r}',t')}\right|_{n=n_0},
\end{equation}
with $v_{xc}[n]({\bf r},t)$ being the exact time-dependent xc potential of
TDDFT.

If short-range XC effects are ignored altogether by setting the {\it
unknown} XC potential $v_{xc}[n_0]({\bf r})$ and XC kernel
$f_{xc}[n_0]({\bf r},{\bf r}';\omega)$ equal to zero, the TDDFT
density-response function of
Eq.~(\ref{eq:Xalda2}) reduces to the RPA Eq.~(\ref{eq:Xalda}).

\subsubsection{The XC kernel}

Along the years, several approximations have been used to evaluate the
{\it unknown} XC kernel of Eq.~(\ref{kernel}).

\paragraph{Random-phase approximation (RPA).}  Nowadays, one usually
refers to the RPA as the result of simply setting the XC kernel
$f_{xc}[n_0]({\bf r},{\bf r}';\omega)$ equal to zero:
\begin{equation}\label{zero}
f_{xc}^{RPA}[n_0]({\bf r},{\bf r}';\omega)=0,
\end{equation}
but still using in Eqs.~(\ref{chi0}) and (\ref{groundh}) the full
single-particle states
and energies $\psi_i({\bf r})$ and $\varepsilon_i$ of DFT [i.e., the
solutions of Eq.~(\ref{dft})] with $v_{xc}[n_0]({\bf r})$ set
different from zero. This is sometimes called DFT-based RPA.

\paragraph{Adiabatic local-density approximation (ALDA).} In this
approximation, also called
time-dependent local-density approximation (TDLDA)~\cite{soven}, one
assumes that both the unperturbed $n_0({\bf r})$ and the induced
$\delta n({\bf r},\omega)$ electron densities
vary slowly in space and time and, therefore, one replaces the
dynamical XC kernel by the long-wavelength ($Q\to 0$) limit of the
static XC kernel of a homogeneous electron gas at the local density:
\begin{equation}\label{alda}
f_{xc}^{ALDA}[n_0]({\bf r},{\bf r}';\omega)=\left.
{d^2\left[n\varepsilon_{xc}(n)\right]\over
dn^2}\right|_{n=n_0({\bf r})}\delta({\bf r}-{\bf r}'),
\end{equation}
where $\varepsilon_{xc}(n)$ is the XC energy per particle of a
homogeneous electron gas of density $n$.

\paragraph{PGG and BPG.}In the spirit of the optimized
effective-potential method~\cite{oep},
Petersilka, Gossmann, and Gross (PGG)~\cite{peter} derived the
following frequency-independent exchange-only approximation for
inhomogeneous systems:
\begin{equation}
f_{x}^{PGG}[n_0]({\bf r},{\bf r}';\omega)=-{2\over|{\bf r}-{\bf r}'|}
\,{\left|\sum_i\,f_i\,\psi_i({\bf r})\,
\psi_i^*({\bf r}')\right|^2\over n_0({\bf r})n_0({\bf r}')},
\end{equation}
where $\psi_i({\bf r})$ denote the solutions of the Kohn-Sham
Eq.~(\ref{dft}).

More recently, Burke, Petersilka, and Gross (BPG)~\cite{burke} devised
a hybrid formula for the XC kernel, which combines expressions for
symmetric and antisymmetric spin orientations from the exchange-only
PGG scheme and the ALDA. For an unpolarized many-electron system, one
writes~\cite{burke}:
\begin{equation}
f_{xc}^{BPG}[n_0]({\bf r},{\bf r}';\omega)={1\over 2}\,
\left[f_{xc}^{\uu,PGG}+f_{xc}^{\ud,LDA}\right],
\end{equation}
where $f_{xc}^{\uu}$ and $f_{xc}^{\ud}$ represent the XC kernel for
electrons with parallel and antiparallel spin, respectively.

\paragraph{Average approximation} The investigation of short-range XC
effects in solids has been focused in a great extent onto the simplest
possible many-electron system, which is the homogeneous electron gas.
Hence, recent attempts to account for XC effects in inhomogeneous
systems
have adopted the following approximation~\cite{pp,gg}:
\begin{equation}\label{av1}
f_{xc}^{av}[n_0]({\bf r},{\bf r}';\omega)=f_{xc}^{hom}(\tilde n;
|{\bf r}-{\bf r}'|;\omega),
\end{equation}
where $\tilde n$ represents a function of the electron densities at
points ${\bf r}$ and ${\bf r}'$, typically the arithmetical average
\begin{equation}\label{av2}
\tilde n={1\over 2}\left[n_0({\bf r})+n_0({\bf r}')\right],
\end{equation}
and $f_{xc}^{hom}(\tilde n;|{\bf r}-{\bf r}'|;\omega)$
denotes the XC kernel of a homogeneous electron gas
of density $\tilde n$, whose 3D Fourier transform
$f_{xc}^{hom}(\tilde n;Q,\omega)$ is
directly connected to the so-called local-field factor
$G(\tilde n;Q,\omega)$:
\begin{equation}\label{av3}
f_{xc}^{hom}(\tilde n;Q,\omega)=-{4\pi\over Q^2}\,
G(\tilde n;Q,\omega).
\end{equation}

In the ALDA, one writes
\begin{eqnarray}
G^{ALDA}(\tilde n;Q,\omega)&=&G(\tilde n;Q\to 0,\omega=0)\cr\cr
&=&-{Q^2\over 4\pi}\,\left.
{d^2\left[n\varepsilon_{xc}(n)\right]\over
dn^2}\right|_{n=\tilde n},
\end{eqnarray}
which in combination with Eqs.~(\ref{av1})-(\ref{av3}) yields the ALDA
XC kernel of Eq.~(\ref{alda}). However, more
accurate nonlocal dynamical expressions for the local-field factor
$G(\tilde n;Q,\omega)$ are available nowdays, which together with
Eqs.~(\ref{av1})-(\ref{av3}) should yield an accurate (beyond the
ALDA) representation of the XC kernel of inhomogeneous systems.

During the last decades, much effort has gone into the determination
of the static local-field factor
$G^{static}(\tilde n;Q)=
G(\tilde n;Q,\omega=0)$~\cite{s1,s2,s3,s4,s5,s6,s7,s8}, the most
recent works
including diffusion Monte Carlo (DMC)
calculations~\cite{bowen,moroni} and the parametrization of the DMC
data of Ref.~\cite{moroni} given by Corradini {\it et
al.}~\cite{cdop}:
\begin{equation}\label{corradini}
G^{static}(\tilde n;Q)=C\hat Q^2+B\hat Q^2/(g+\hat Q^2)+
\alpha\,\hat Q^4\,{\rm e}^{-\beta\,\hat Q^2},
\end{equation}
where $\hat Q=Q/q_F$, and the parameters $B$, $C$, $g$, $\alpha$, and
$\beta$ are the dimensionless functions of $\tilde n$
listed in Ref.~\cite{cdop}.

Calculations of the frequency dependence of the local-field factor
$G(\tilde n;Q,\omega)$ have been carried out mainly in the limit of
long wavelengths ($Q\to 0$)~\cite{v1,v2,v3,v4,v5,v6}, but work has
also been done for finite wave vectors~\cite{z1,z2,z3,z4}.

\section{Inverse dielectric function}

In the presence of a many-electron system, the total potential
$\phi({\bf r},\omega)$ of a unit test charge at point ${\bf r}$ that
is exposed to the external potential $\phi^{ext}({\bf r},\omega)$
can be expressed in the following form:
\begin{equation}\label{tpot}
\phi({\bf r},\omega)=\phi^{ext}({\bf r},\omega)+
\delta\phi^H({\bf r},\omega),
\end{equation}
where $\delta\phi^H({\bf r},\omega)$ represents the induced potential
of
Eq.~(\ref{ext}). Using Eqs.~(\ref{eqa1}) and (\ref{ext}), the total
potential
$\phi({\bf r},\omega)$ of Eq.~(\ref{tpot}) is easily found to take the
following form:
\begin{equation}\label{tpot2}
\phi({\bf r},\omega)=\int d{\bf r}'\,\epsilon^{-1}({\bf r},{\bf r}';
\omega)\
\phi^{ext}({\bf r},\omega),
\end{equation}
where
\begin{equation}\label{epsilon}
\epsilon^{-1}({\bf r},{\bf r}';\omega)=\delta({\bf r}-{\bf
r}')+\int{\rm d}{\bf r}''\,
v({\bf r}-{\bf r}'')\,\chi({\bf r}'',{\bf r}';\omega).
\end{equation}
This is the so-called inverse {\it longitudinal} dielectric function
of the
many-electron system, whose poles dictate the occurrence of
collective electronic
excitations and which can be evaluated in the RPA or in the framework
of TDDFT from the knowledge of the density-response function of
Eqs.~(\ref{eq:Xalda}) and
(\ref{eq:Xalda2}), respectively.

Some quantities, such as the optical absorption and the electron energy loss of
charged particles moving in arbitrary inhomogeneous media, can be described by
the so-called effective inverse dielectric function
$\epsilon_{eff}^{-1}({\bf Q},\omega)$, which is defined as a 3D
Fourier transform of the inverse dielectric function
$\epsilon^{-1}({\bf r},{\bf r}';\omega)$:
\begin{equation}\label{effective1}
\epsilon_{eff}^{-1}({\bf Q},\omega)={1\over\Omega}\,\int d{\bf r}\
\int d{\bf r}'\,
{\rm e}^{-i{\bf Q}\cdot({\bf r}-{\bf r}')}\,\epsilon^{-1}({\bf r},{\bf
r}';\omega),
\end{equation}
and which at long wavelengths ($Q\to 0$) should take the form of
Eq.~(\ref{eff2})~\cite{lt}.

In particular, in the case of a homogeneous system and in the {\it classical}
long-wavelength limit, where the total potential $\phi({\bf r},\omega)$ of a
unit test charge at point ${\bf r}$ only depends on the external potential
$\phi^{ext}({\bf r},\omega)$ at that point, the inverse dielectric function
takes the following form:
\begin{equation}
\epsilon^{-1}({\bf r},{\bf r}';\omega)=\epsilon^{-1}(\omega)\,
\delta({\bf r}-{\bf r}'),
\end{equation}
which in combination with Eq.~(\ref{tpot2}) yields the classical
formula
\begin{equation}\label{local}
\phi({\bf r},\omega)=\phi^{ext}({\bf r},\omega)/\epsilon(\omega),
\end{equation}
$\epsilon(\omega)$ representing the so-called {\it local} dielectric
function of the medium.

\section{Screened interaction}

Another key quantity in the description of electronic excitations in a
many-electron system, which also dictates the occurrence of collective
electronic excitations, is the frequency-dependent complex screened
interaction $W({\bf r},{\bf r}';\omega)$. This quantity yields the
total potential $\phi({\bf r},\omega)$ of a unit test charge at point
${\bf
r}$ in the presence of an external test charge of density
$n^{ext}({\bf r}',\omega)$ at point ${\bf r}'$:
\begin{equation}\label{tpotnew}
\phi({\bf r},\omega)=\int d{\bf r}'\,W({\bf r},{\bf r}';
\omega)\,n^{ext}({\bf r}',\omega).
\end{equation}
The potential $\phi^{ext}({\bf r},\omega)$ due to the external test
charge density $n^{ext}({\bf r},\omega)$ is simply
\begin{equation}\label{external}
\phi^{ext}({\bf r},\omega)=\int d{\bf r}'\,
v({\bf r},{\bf r}')\,n^{ext}({\bf r}',\omega).
\end{equation}
Hence, a comparison of Eqs.~(\ref{tpot2}) and (\ref{tpotnew}) yields
\begin{equation}\label{screened0}
W({\bf r},{\bf r}';\omega)=\int d{\bf r}''\,
\epsilon^{-1}({\bf r},{\bf r}'';\omega)\,v({\bf r}'',{\bf r}'),
\end{equation}
and using Eq.~(\ref{epsilon}):
\begin{eqnarray}\label{screened}
W({\bf r},{\bf r}';\omega)&=&v({\bf r},{\bf r}')
+\int d{\bf r}_1\int d{\bf r}_2\cr\cr
&\times& v({\bf r},{\bf r}_1)\,\chi({\bf r}_1,{\bf
r}_2,\omega)\,v({\bf r}_2,{\bf r}').
\end{eqnarray}

From Eqs.~(\ref{effective1}) and (\ref{screened0}) one easily finds
the following representation of the effective inverse dielectric
function:
\begin{equation}\label{effective2}
\epsilon_{eff}^{-1}({\bf Q},\omega)={1\over\Omega v_Q}\int d{\bf
r}\,\int d{\bf r}'\,
{\rm e}^{-i{\bf Q}\cdot({\bf r}-{\bf r}')}\,
W({\bf r},{\bf r}';\omega),
\end{equation}
where $v_Q=4\pi/Q^2$ denotes the 3D Fourier transform of the bare
Coulomb interaction $v({\bf r},{\bf r}')$.

\subsection{Classical model}

In a classical model consisting of two homogeneous media characterized
by local (frequency-dependent) dielectric functions $\epsilon_1$ and
$\epsilon_2$ and separated by an interface of arbitrary geometry, the
total potential at each medium is simply given by Eq.~(\ref{local})
and is, therefore, a solution of Poisson's equation
\begin{equation}
\nabla^2\phi({\bf r},\omega)=-{4\pi\over\epsilon_i(\omega)}\,
n^{ext}({\bf r},\omega),
\end{equation}
$\epsilon_i(\omega)$ being $\epsilon_1$ or $\epsilon_2$ depending on
whether the point ${\bf r}$ is located in medium 1 or in medium 2,
respectively. Hence, the screened interaction $W({\bf r},{\bf r}';
\omega)$ entering
Eq.~(\ref{tpotnew}) is a solution of the following equation:
\begin{equation}\label{poissonl}
\nabla^2 W({\bf r},{\bf r}';\omega)=-{4\pi\over\epsilon_i(\omega)}
\,\delta({\bf r}-{\bf r}').
\end{equation}

For simple geometries, such as the planar, spherical, and cylindrical
interfaces, Eq.~(\ref{poissonl}) can be solved explicitly by imposing
the ordinary boundary conditions of continuity of the potential and
the normal component of the displacement vector at the interface.

\subsubsection{Planar surface}

In the case of two semi-infinite media with local
(frequency-dependent) dielectric functions $\epsilon_1$ (at $z<0$) and
$\epsilon_2$ (at $z>0$) separated by a planar interface at $z=0$ (see
Fig.~\ref{fig1}),
there is translational invariance in two directions, which we take to
be normal to the $z$ axis. Hence, one can define the Fourier
transform $W(z,z';q,\omega)$, $q$ being the magnitude of a 2D wave
vector
in the plane of the interface, and imposing the ordinary boundary
conditions of continuity of the potential and the normal component of
the displacement vector at the interface, one finds:
\begin{widetext}
\begin{equation}\label{clas1}
W(z,z';q,\omega)={2\pi\over q}\cases{
\left[{\rm e}^{-q|z-z'|}+g\,{\rm e}^{-q(|z|+|z'|)}\right]/\epsilon_1,
&$z<0,z'<0$,\cr\cr
2\,g\,{\rm e}^{-q|z-z'|}/(\epsilon_1-\epsilon_2),&$z^<<0,z^>>0$,\cr\cr
\left[{\rm e}^{-q|z-z'|}-g\,
{\rm e}^{-q(|z|+|z'|)}\right]/\epsilon_2,&$z>0,z'>0$,}
\end{equation}
\end{widetext}
where $z^<$ ($z^>$) is the smallest (largest) of $z$ and $z'$, and $g$
is the classical surface-response function:
\begin{equation}\label{g1}
g(\omega)={\epsilon_1(\omega)-\epsilon_2(\omega)\over
\epsilon_1(\omega)+\epsilon_2(\omega)},
\end{equation}
or, equivalently,
\begin{equation}
g(\omega)=-{n\over u-n},
\end{equation}
where $u$ is the spectral variable of Eq.~(\ref{spectral}) and
$n=1/2$.

An inspection of Eqs.~(\ref{clas1}) and (\ref{g1}) shows that the
screened interaction
$W(z,z';q,\omega)$ has poles at the {\it classical} bulk- and
surface-plasmon conditions dictated by $\epsilon_i=0$ and by
Eq.~(\ref{nonretardedbis}), respectively.

\subsubsection{Spheres}

In the case of a sphere of radius $a$ and local (frequency-dependent)
dielectric function $\epsilon_1$ embedded in a host medium of local
(frequency-dependent) dielectric function $\epsilon_2$, we first
expand the
screened interaction $W({\bf r},{\bf r}';\omega)$ in spherical harmonics:
\begin{equation}\label{scr1}
W({\bf r},{\bf r}';\omega)=\sum_{l,m}\,{4\pi\over 2l+1}\,
W_l(r,r';\omega)\,Y_{l,m}^*(\Omega)\,Y_{l,m}(\Omega'),
\end{equation}
and we then derive the coefficients of this expansion by imposing the
boundary conditions. One finds~\cite{bausells}:
\begin{widetext}
\begin{equation}\label{scr3}
W_l(r,r';\omega)=
\cases{\displaystyle
\left[\displaystyle{(r^<)^l\over(r^>)^{l+1}}
+(l+1)\,g_l\,\displaystyle{(r\,r')^l\over a^{2l+1}}\right]/\epsilon_1,
&$r,r'<a$,\cr\cr
(l+1)\,g_l
\displaystyle{(r^<)^l\over(r^>)^{l+1}}/(\epsilon_1-\epsilon_2),
&$r^<<a,r^>>a$,\cr\cr
\left[\displaystyle{(r^<)^l\over
(r^>)^{l+1}}-l\,g_l\,\displaystyle{a^{2l+1}\over
(r\,r')^{l+1}}\right]/\epsilon_2,&$r,r'>a$,}
\end{equation}
\end{widetext}
where $r_<$ ($r_>$) is the smallest (largest) of $r$ and $r'$, and
\begin{equation}\label{scr4}
g_l(\omega)={\epsilon_1(\omega)-\epsilon_2(\omega)\over
l\,\epsilon_1(\omega)+(l+1)\,\epsilon_2(\omega)},
\end{equation}
or, equivalently,
\begin{equation}
g_l(\omega)=-{n_l\over u-n_l},
\end{equation}
with $u$ being the spectral variable of Eq.~(\ref{spectral}) and
\begin{equation}
n_l={l\over 2l+1}.
\end{equation}

As in the case of the planar surface, the screened interaction of
Eqs.~(\ref{scr1})-(\ref{scr4}) has poles at the {\it classical} bulk-
and surface-plasmon conditions, which in the case of a single sphere
in a host medium are dictated by $\epsilon_i=0$ and by
Eq.~(\ref{spheres}), respectively.

Introducing Eqs.~(\ref{scr1})-(\ref{scr4}) into
Eq.~(\ref{effective2}), one finds the following expression for the
effective inverse dielectric function~\cite{pitarke2}:
\begin{eqnarray}\label{effsphere}
&&\epsilon^{-1}_{eff}(Q,\omega)=\epsilon_2^{-1}+
f(\epsilon_1^{-1}-\epsilon_2^{-1})\cr\cr
&&\times\left[1+{3\over x}\,\sum_{l=0}^\infty\,(2l+1)\,g_l\,j_l(x)\
\Theta_l(x)\right],
\end{eqnarray}
where
\begin{equation}
\Theta_l(x)={l\,j_{l-1}(x)\,\epsilon_1-(l+1)\,j_{l+1}(x)\,\epsilon_2
\over\epsilon_1-\epsilon_2}
\end{equation}
and $x=Qa$. Here, $f$ represents the volume fraction filled by the
sphere
and $j_l(x)$ are spherical Bessel functions of the first
kind~\cite{abra}. This equation represents the dilute ($f\to 0$)
limit of the effective inverse dielectric function derived by Barrera
and Fuchs for a system composed by identical interacting spheres in a
host
medium~\cite{barrera}.

In the limit as $Qa<<1$, an expansion of Eq.~(\ref{effsphere}) yields
\begin{equation}
\epsilon_{eff}^{-1}(Q,\omega)=\epsilon_2^{-1}\left[1-
3f\,{\epsilon_1-\epsilon_2\over\epsilon_1+2\epsilon_2}\right],
\end{equation}
which is precisely the long-wavelength effective inverse dielectric
function obtained in Section~\ref{classicals} from Eqs.~(\ref{macro})
and
(\ref{d}) with $D=3$ and which admits the spectral representation of
Eq.~(\ref{eftwo}) with $n=1/3$. This result demonstrates the expected
result that in the limit as $Qa<<1$ a broad beam of charged particles
interacting with a single sphere of dielectric function $\epsilon_1$
in a host medium of dielectric function $\epsilon_2$ can only
create collective excitations at the {\it dipole} resonance where
$\epsilon_1+2\epsilon_2=0$ [Eq.~(\ref{spheres}) with $l=1$], which for a Drude
sphere in vacuum yields $\omega=\omega_p/\sqrt{3}$.

\subsubsection{Cylinders}

In the case of an infinitely long cylinder of radius $a$ and local
(frequency-dependent) dielectric function $\epsilon_1$ embedded in a
host medium of local (frequency-dependent) dielectric function
$\epsilon_2$, we expand the screened interaction
$W({\bf r},{\bf r}';\omega)$ in terms of the modified Bessel functions
$I_m(x)$ and $K_m(x)$~\cite{abra}, as follows
\begin{eqnarray}\label{cyl1}
&&W({\bf r},{\bf r}';\omega)={2\over\pi}\int_0^\infty dq_z\,
\cos\left[q_z(z-z')\right]\cr\cr
&&\times\sum_{m=0}^\infty \mu_m\,
W_m(\rho,\rho';\omega)\cos\left[m(\phi-\phi')\right],
\end{eqnarray}
where $z$ and $\rho$ represent the projections of the position vector
along the axis of the cylinder and in a plane perpendicular to the
cylinder, respectively, $q_z$ denotes the magnitude of
a wave vector along the axis of the cylinder, and $m_m$ are Neumann
numbers
\begin{equation}\label{cyl2}
\mu_m=\cases{1,&$m=0$,\cr\cr 2,&$m\ge 1$.}
\end{equation}
The coefficients $W_m(\rho,\rho';\omega)$ are then derived by imposing
the boundary conditions, i.e., by requiring that the total scalar
potential and the normal component of the displacement vectors be
continuous at the interface. One finds:
\begin{widetext}
\begin{equation}\label{cyl3}
 W_m(\rho,\rho';\omega)=\cases{
I_m(q_z\rho^<)\,\left[K_m(q_z\rho^>)
-K_m'(x)\,g_m\,I_m(q_z\rho^>)/I_m'(x)\right]
/\epsilon_1,
&$\rho,\rho'<a$,\cr\cr
\left[x\,I_m'(x)\,K_m(x)\right]^{-1}\,g_m\,I_m(q_z\rho^<)\,K_m(q_z\rho^>)/(\epsilon_1-\epsilon_2),&
$\rho^<<a,\rho^>>a$,\cr\cr
\left[I_m(q_z\rho^<)
-I_m(x)\,g_m\,K_m(q_z\rho^<)/K_m(x)\right]\,K_m(q_z\rho^>)
/\epsilon_2,&$\rho,\rho'>a$,}
\end{equation}
\end{widetext}
where $x=q_za$ and
\begin{equation}\label{cyl4}
g_m(x,\omega)={I_m'(x)\,K_m(x)\left[\epsilon_1(\omega)-\epsilon_
(\omega)\right]\over
I_m'(x)\,K_m(x)\,\epsilon_1(\omega)-I_m(x)\,K_m'(x)\,
\epsilon_2(\omega)},
\end{equation}
or, equivalently,
\begin{equation}
g_m(x,\omega)=-{n_m\over u-n_m},
\end{equation}
with $u$ being the spectral variable of Eq.~(\ref{spectral}) and
\begin{equation}\label{nm}
n_m=x\,I_m'(x)\,K_m(x).
\end{equation}

\begin{figure}
\includegraphics[width=0.45\textwidth,height=0.3375\textwidth]
{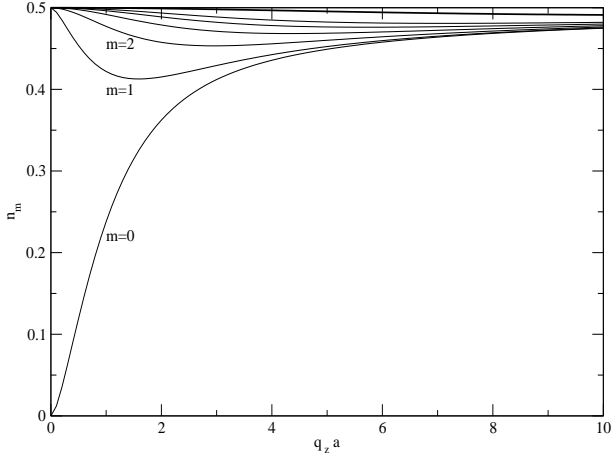}
\caption{Depolarization factors $n_m=x\,I_m'(x)\,K_m(x)$, as a
function of $x=q_za$, for $m=0,1,2,3,4,5$ (thin solid lines), and
$m=10$ (thick solid line). As $m\to\infty$, the depolarization factor
$n_m$ equals the planar surface-plasmon value $n_m=1/2$ for all
values of $x=q_z a$.}\label{fig5}
\end{figure}

In the limit as $x\to 0$ ($p$-polarization)~\cite{long}, the
depolarization factors of Eq.~(\ref{nm}) are easily found to be
$n_0=0$ (corresponding to the plasmon condition
$\epsilon_2=0$)~\cite{notep} and $n_m=1/2$ (corresponding to the
planar surface-plasmon condition $\epsilon_1+\epsilon_2=0$) for all
$m\neq 0$; in the limit as $x\to\infty$, Eq.~(\ref{nm}) yields
$n_m=1/2$ for all $m$. For the behavior of the depolarization
factors $n_m$ of Eq.~(\ref{nm}) as a function of $x$ see
Fig.~\ref{fig5}. This figure shows that the energies of all modes
are rather close to the planar surface-plasmon energy (corresponding
to $n_m=1/2$), except for $m=0$. The $m=0$ mode, which corresponds
to a homogeneous charge distribution around the cylindrical surface,
shifts downwards from the planar surface-plasmon energy ($n_0=1/2$)
as the adimensional quantity $x=q_za$ decreases, as occurs with the
symmetric low-energy mode in thin films [see Eqs.~(\ref{thin1}) and
(\ref{tfilm})].

Introducing Eqs.~(\ref{cyl1})-(\ref{cyl4}) into
Eq.~(\ref{effective2}), one finds the following expression for the
effective inverse dielectric
function~\cite{pitarke2}:
\begin{eqnarray}\label{effcyl1}
&&\epsilon^{-1}_{eff}({\bf Q},\omega)=\epsilon_2^{-1}+
f\,(\epsilon_1^{-1}-\epsilon_2^{-1})\cr\cr
&\times&\left[1+{2\over x^2+y^2}\,
\sum_{m=0}^\infty\mu_mJ_m(y)g_m\Theta_m(x,y)\right],
\end{eqnarray}
where
\begin{equation}\label{effcyl2}
\Theta_m(x,y)={I_m'(x)\,f_m^{(1)}(x,y)\,\epsilon_1+K_m'(x)\,
f_m^{(2)}(x,y)\,\epsilon_2\over I_m'(x)\,K_m(x)
\left[\epsilon_1-\epsilon_2\right]},
\end{equation}
\begin{equation}\label{effcyl3}
f_m^{(1)}(x,y)=x\,J_m(y)\,K_{m-1}(x)+y\,J_{m-1}(y)\,K_m(x),
\end{equation}
\begin{equation}\label{effcyl4}
f_m^{(2)}(x,y)=x\,J_m(y)\,I_{m-1}(x)-y\,J_{m-1}(y)\,I_m(x),
\end{equation}
$x=q_za$, and $y=qa$, $q_z$ and $q$ representing the components of the
total wave vector ${\bf Q}$ along the axis of the cylinder and in a
plane perpendicular to the cylinder, respectively. The volume
fraction filled by the cylinder is denoted by $f$, and
$J_m(x)$ are cylindrical Bessel functions of the first
kind~\cite{abra}. A spectral representation of the effective inverse
dielectric function of Eqs.~(\ref{effcyl1})-(\ref{effcyl4}) was
reported in Ref.~\cite{pitarke3}.

In the limit as $Qa<<1$, an expansion of
Eqs.~(\ref{effcyl1})-(\ref{effcyl4}) yields
\begin{equation}\label{cylong}
\epsilon^{-1}_{eff}({\bf Q},\omega)=\epsilon_2^{-1}\left\{
1-f\,{\epsilon_1-\epsilon_2\over x^2+y^2}\,\left[{x^2\over\epsilon_2}+
2\,{y^2\over\epsilon_1+\epsilon_2}\right]\right\},
\end{equation}
which admits the spectral representation of Eq.~(\ref{eff2}) with two
nonvanishing spectra strengths: $C_0=x^2/(x^2+y^2)$ and
$C_1=y^2/(x^2+y^2)$, the corresponding depolarization factors being
$n_0=0$ and $n_1=1/2$, respectively.

Equation~(\ref{cylong}) demonstrates that in the limit as $Qa<<1$
and for a
wave vector normal to the cylinder ($x=0$), moving charged particles
can only create collective excitations at the dipole resonance where
$n_1=1/2$, i.e., $\epsilon_1+\epsilon_2=0$, which for a Drude
cylinder in vacuum
yields Ritchie's frequency $\omega_s=\omega_p/\sqrt{2}$. Conversely,
still in the limit as $Qa<<1$ but for a wave vector along the axis of
the cylinder ($y=0$), moving charged particles can only excite the bulk
mode of the host medium dictated by the condition $u=0$ (corresponding
to $n_0=0$), i.e., $\epsilon_2=0$, in agreement with the discussion of
Section~\ref{cylinder}.

\subsection{Nonlocal models: planar surface}

{\it Nonlocal} effects that are absent in the classical model
described above can be incorporated in a variety of semiclassical and
quantal approaches, which we here only describe for a {\it planar
surface}.

\subsubsection{Hydrodynamic model}\label{hydrog}

\paragraph{Semiclassical hydrodynamic approach.} Within a
semiclassical hydrodynamic approach, the screened interaction
$W({\bf r},{\bf r}';\omega)$ [as defined in Eq.~(\ref{tpotnew})] can
be obtained from the linearized hydrodynamic
Eqs.~(\ref{hd1p})-(\ref{hd3p}). For a semi-infinite metal in vacuum
consisting of an abrupt step of the unperturbed electron density
$n_0(z)$ [see Eq.~(\ref{density0})], we can assume translational
invariance in the plane of the surface, and noting that the normal
component of the hydrodynamical velocity should vanish at the
interface Eqs.~(\ref{hd1p})-(\ref{hd3p}) yield the following
expression for the 2D Fourier transform $W(z,z';q,\omega)$:
\begin{widetext}
\begin{equation}\label{gh1}
W(z,z';q,\omega)={2\pi\over q}\cases{
\epsilon_s(z-z')+\epsilon_s(z+z')-2\,g\,
\displaystyle{\epsilon_s(z)\,\epsilon_s(z')\over 1-\epsilon_s^0},
&$z<0,z'<0$,\cr\cr
2\,g\,\displaystyle{\epsilon_s(z^<)\over 1-\epsilon_s^0}
\,{\rm e}^{-qz^>},&$z^<<0,z^>>0$,\cr\cr
{\rm e}^{-q|z-z'|}-g\,
{\rm e}^{-q(z+z')},&$z>0,z'>0$,}
\end{equation}
\end{widetext}
where $z^<$ ($z^>$) is the smallest (largest) of $z$ and $z'$,
\begin{equation}\label{gh2}
\epsilon_s(z;q,\omega)={\Lambda\,\omega\,(\omega+i\eta)\, {\rm
e}^{-q|z|}-q\,\omega_p^2\, {\rm e}^{-\Lambda|z|}
\over\Lambda\left[\omega(\omega+i\eta)-\omega_p^2\right]},
\end{equation}
\begin{equation}\label{gh3}
g(q,\omega)={\omega_p^2\over 2\beta^2\,\Lambda(\Lambda+q)-\omega_p^2},
\end{equation}
\begin{equation}\label{gh4}
\Lambda={1\over\beta}\sqrt{\omega_p^2+\beta^2\,q^2-\omega(\omega+i\eta)},
\end{equation}
$\beta=\sqrt{1/3}(3\pi^2n_0)^{1/3}$ [as in Eqs.~(\ref{plasmonb}) and
(\ref{plasmons})],
and
\begin{equation}\label{gh5}
\epsilon_s^0(q,\omega)=\epsilon_s(z=0;q,\omega).
\end{equation}

An inspection of Eqs.~(\ref{gh1})-(\ref{gh5}) shows that the
hydrodynamic
surface-response function $g(q,\omega)$ and, therefore, the
hydrodynamic screened interaction $W(z,z';q,\omega)$ become
singular at the hydrodynamic surface-plasmon condition dictated by
Eq.~(\ref{plasmons}). We also note that the second moment of the
imaginary
part of the hydrodynamic surface-response function $g(q,\omega)$ is
found to
be
\begin{equation}\label{sumrule}
\int_\infty^\infty d\omega\,\omega\,{\rm Im}g(q,\omega)=2\pi^2\,\bar
n,
\end{equation}
where $\bar n$ represents the electron density: $\bar
n=\omega_p^2/4\pi$.

Finally, we note that in the long-wavelength ($q\to 0$) limit the
hydrodynamic
screened interaction of Eqs.~(\ref{gh1})-(\ref{gh5}) reduces to the
{\it classical} screened interaction of Eqs.~(\ref{clas1}) and
(\ref{g1}) with the dielectric functions $\epsilon_1$ and $\epsilon_2$
being
replaced by the Drude dielectric function [Eq.~(\ref{drude})] and
unity,
respectively. The same result is also obtained by simply assuming
that the electron gas is nondispersive, i.e., by taking the
hydrodynamic speed  $\beta$ equal to zero.

\paragraph{Quantum hydrodynamic approach.} Within a quantized
hydrodynamic model of a many-electron system, one
first linearizes the hydrodynamic Hamiltonian with respect to the
induced electron density, and then quantizes this Hamiltonian on the
basis of the normal modes of oscillation (bulk and surface plasmons)
corresponding to Eqs.~(\ref{plasmonb}) and (\ref{plasmons}). One
finds:
\begin{equation}
H=H_G+H_0^B+H_0^S,
\end{equation}
where $H_G$ represents the Thomas-Fermi ground state of the static
unperturbed electron system~\cite{lundqvist}, and $H_0^B$ and $H_0^S$
are
free bulk and surface plasmon Hamiltonians, respectively:
\begin{equation}
H_0^B={1\over\Omega}\sum_{{\bf q},q_z}\left[1/2+\omega_Q^B\right]
a_{{\bf Q}}^\dagger(t)a_{{\bf Q}}(t)
\end{equation}
and
\begin{equation}
H_0^S={1\over A}\sum_{\bf q}\left[1/2+\omega_q^S\right]b_{{\bf
q}}^\dagger(t)b_{\bf q}(t).
\end{equation}
Here, $\Omega$ and $A$ represent the normalization volume and the
normalization area of the surface, respectively, $a_{\bf Q}(t)$
and $b_{\bf q}(t)$ are Bose-Einstein operators that
annihilate bulk and surface plasmons with wave vectors ${\bf
Q}=({\bf q},q_z)$ and ${\bf q}$, respectively, and $\omega_Q^B$
and $\omega_q^S$ represent the dispersion of bulk and surface
plasmons:
\begin{equation}
\left(\omega_Q^B\right)^2=\omega_p^2+\beta^2\,Q^2
\end{equation}
and
\begin{equation}\label{wqshd}
\left(\omega_q^S\right)^2={1\over 2}\left[\omega_p^2+
\beta^2\,q^2+\beta\,q\,
\sqrt{2\omega_p^2+\beta^2\,q^2}\right].
\end{equation}

Hence, within this approach one can distinguish the separate
contributions to the imaginary part of the hydrodynamic
surface-response function $g(q,\omega)$ of  Eq.~(\ref{gh3}) coming
from the excitation of
either bulk or surface plasmons. One finds~\cite{aitor}:
\begin{equation}\label{hydrot}
{\rm Im}g(q,\omega)={\rm Im}g^B(q,\omega)+{\rm Im}g^S(q,\omega),
\end{equation}
where
\begin{eqnarray}\label{hdp}
&&{\rm Im}g^B(q,\omega)={1\over 2}\,q\,\int_0^\infty dq_z
\,\delta(\omega-\omega_Q^B)\cr\cr
&&\times{(\omega_p^2/\omega_{Q}^B)\,q_z^2\over
q_z^4+q_z^2(q^2+\omega_p^2/\beta^2)+\omega_p^4/(4\beta^4)}
\end{eqnarray}
and
\begin{equation}\label{hds}
{\rm Im}g^S(q,\omega)={\pi\over 2}\,{\gamma_{q}\over
q+2\gamma_{q}}\,{\omega_p^2\over\omega_{q}^S}\,
\delta(\omega-\omega_{q}^S),
\end{equation}
with $Q=\sqrt{q^2+q_z^2}$ and
\begin{equation}\label{hydrot2}
\gamma_q={1\over 2\beta}\left(-\beta
q+\sqrt{2\omega_p^2+\beta^2q^2}\right).
\end{equation}

For the second moments of ${\rm Im}g^B(q,\omega)$ and
${\rm Im}g^S(q,\omega)$, one finds:
\begin{equation}\label{sumruleb}
\int_0^\infty d\omega\,\omega\,{\rm
Im}g^B(q,\omega)={\pi\over 4}\,{q\over
q+2\gamma_q}\,\omega_p^2
\end{equation}
and
\begin{equation}\label{sumrules}
\int_0^\infty d\omega\,\omega\,{\rm
Im}g^S(q,\omega)={\pi\over 4}\,{2\gamma_{q}\over
q+2\gamma_q}\,\omega_p^2,
\end{equation}
which add up to the second moment of Eq.~(\ref{sumrule}).

In the limit as $q\to 0$ the bulk contribution to the so-called
energy-loss
function ${\rm Im}g(q,\omega)$ [see
Eqs.~(\ref{hydrot})-(\ref{hydrot2})]
vanishes, and the imaginary part of
both Eqs.~(\ref{gh3}) and (\ref{hds}) yields the classical
result:
\begin{equation}\label{long}
{\rm Im}g(q,\omega)\to{\pi\over 2}\,\omega_s\,\delta(\omega-\omega_s),
\end{equation}
which can also be obtained from Eq.~(\ref{g1}) with $\epsilon_1$
replaced by the Drude dielectric function of Eq.~(\ref{drude}) and
$\epsilon_2$ set equal to unity. Equation~(\ref{long}) shows that in
the classical (long-wavelength) limit the energy loss is dominated by
the excitation of surface plasmons of energy
$\omega_s=\omega_p/\sqrt{2}$, as predicted by Ritchie.

\subsubsection{Specular-reflection model (SRM)}\label{srm}

An alternative scheme to incorporate {\it nonlocal} effects, which
has the virtue of expressing the screened interaction
$W(z,z';q,\omega)$ in terms of the dielectric function
$\epsilon(Q,\omega)$ of a homogeneous electron gas representing the
bulk material, is the so-called specular-reflection model reported
independently by Wagner~\cite{wagner} and by Ritchie and
Marusak~\cite{marusak}. In this model, the medium is described by an
electron gas in which all electrons are considered to be specularly
reflected at the surface, thereby the electron density vanishing
outside.

For a semi-infinite metal in vacuum, the unperturbed electron
density $n_0(z)$ is taken to be of the form of Eq.~(\ref{density0}),
and the SRM yields a screened interaction of the form of
Eq.~(\ref{gh1}) but with the quantities $\epsilon_s(z;q,\omega)$ and
$g(q,\omega)$ being replaced by the more general expressions:
\begin{equation}\label{epsilons}
\epsilon_s(z;q,\omega)={q\over\pi}\int_{-\infty}^{+\infty}
{dq_z\over Q^2}\,{\rm e}^{iq_zz}\epsilon^{-1}(Q,\omega)
\end{equation}
and
\begin{equation}\label{gsrm}
g(q,\omega)={1-\epsilon_s^0(q,\omega)\over 1+\epsilon_s^0(q,\omega)},
\end{equation}
with $\epsilon_s^0(q,\omega)$ defined as in Eq.~(\ref{gh5}), and
$Q=\sqrt{q^2+q_z^2}$.

The inverse dielectric function $\epsilon^{-1}(Q,\omega)$ entering
Eq.~(\ref{epsilons}) represents the 3D Fourier transform of the
inverse dielectric function $\epsilon^{-1}({\bf r},{\bf r}',\omega)$
of a homogeneous electron gas. From Eq.~(\ref{epsilon}), one finds:
\begin{equation}\label{epsilonq}
\epsilon^{-1}(Q,\omega)=1+v_Q\,\chi(Q,\omega),
\end{equation}
where $\chi(Q,\omega)$ represents the 3D Fourier transform of the
density-response function $\chi({\bf r},{\bf r}';\omega)$.

In the framework of TDDFT, one uses Eq.~(\ref{eq:Xalda2}) to find
\begin{eqnarray}\label{chiq}
&&\chi(Q,\omega)=\chi^0(Q,\omega)+
\chi^0(Q,\omega)\cr\cr
&&\times\left\{v_Q+f_{xc}(\bar n;Q,\omega)\right\}\
\chi(Q,\omega),
\end{eqnarray}
with $\chi^0(Q,\omega)$ and $f_{xc}(\bar n;Q,\omega)$ being the 3D
Fourier transforms of the noninteracting density-response function
and the XC kernel of Eqs.~(\ref{chi0}) and (\ref{kernel}),
respectively. For a homogeneous electron gas, the eigenfunctions
$\psi_i({\bf r})$ entering Eq.~(\ref{chi0}) are all plane waves; thus,
the integrations can be carried out analytically to yield
the well-known Lindhard function
$\chi^0(Q,\omega)$~\cite{lindhard}.
If one sets the XC kernel $f_{xc}(\bar n;Q,\omega)$ equal to zero,
introduction of
Eq.~(\ref{chiq}) into Eq.~(\ref{epsilonq}) yields the RPA dielectric
function
\begin{equation}\label{chiqrpa}
\epsilon^{RPA}(Q,\omega)=1-v_Q\,\chi^0(Q,\omega),
\end{equation}
which is easy to evaluate.

The RPA dielectric function $\epsilon^{RPA}(Q,\omega)$ of a
homogeneous electron gas can be further approximated in the framework
of the hydrodynamic scheme described in Section~\ref{hydro}. One
finds,
\begin{equation}\label{chihydro}
\epsilon^{hydro}(Q,\omega)=
1+{\omega_p^2\over\beta^2Q^2-\omega(\omega+i\eta)},
\end{equation}
which in the classical (long-wavelength) limit
 yields the {\it local} Drude dielectric function of
Eq.~(\ref{drude}). Introduction of Eq.~(\ref{chihydro}) into
Eq.~(\ref{epsilons}) yields the hydrodynamic screened interaction of
Eqs.~(\ref{gh1})-(\ref{gh5}).

We know from Eq.~(\ref{tpotnew}) that collective excitations are
dictated by singularities in the screened interaction or,
equivalently, maxima in the imaginary part of this quantity. For $z$
and
$z'$ coordinates well inside the solid ($z,z'\to-\infty$), one finds
\begin{equation}\label{in}
W^{in}(z,z';q,\omega)=\int_{-\infty}^\infty{dq_z\over 2\pi}\,
{\rm e}^{iq_z(z-z')}v_Q\,
\epsilon^{-1}(Q,\omega),
\end{equation}
which in the case of the Drude dielectric function
$\epsilon(Q,\omega)$
of Eq.~(\ref{drude}) and for positive frequencies ($\omega>0$) yields
\begin{equation}\label{inclas}
{\rm Im}W^{in}(z,z';q,\omega)\to-{\pi^2\over q}\,\omega_p\
\delta(\omega-\omega_p)
\,{\rm e}^{-q|z-z'|}.
\end{equation}

For $z$ and $z'$ coordinates both outside the solid ($z,z'>0$), one
finds
\begin{equation}\label{out}
W^{out}(z,z';q,\omega)={2\pi\over q}\left[{\rm e}^{-q|z-z'|}
-g(q,\omega)\,{\rm e}^{-q(z+z')}\right],
\end{equation}
which in the classical ($q\to 0$) limit and for positive frequencies
($\omega>0$) yields
\begin{equation}\label{outclas}
{\rm Im}W^{out}(z,z';q,\omega)\to-{\pi^2\over q}\,\omega_s\
\delta(\omega-\omega_s)
\,{\rm e}^{-q(z+z')}.
\end{equation}

\begin{figure}
\includegraphics[width=0.95\linewidth]
{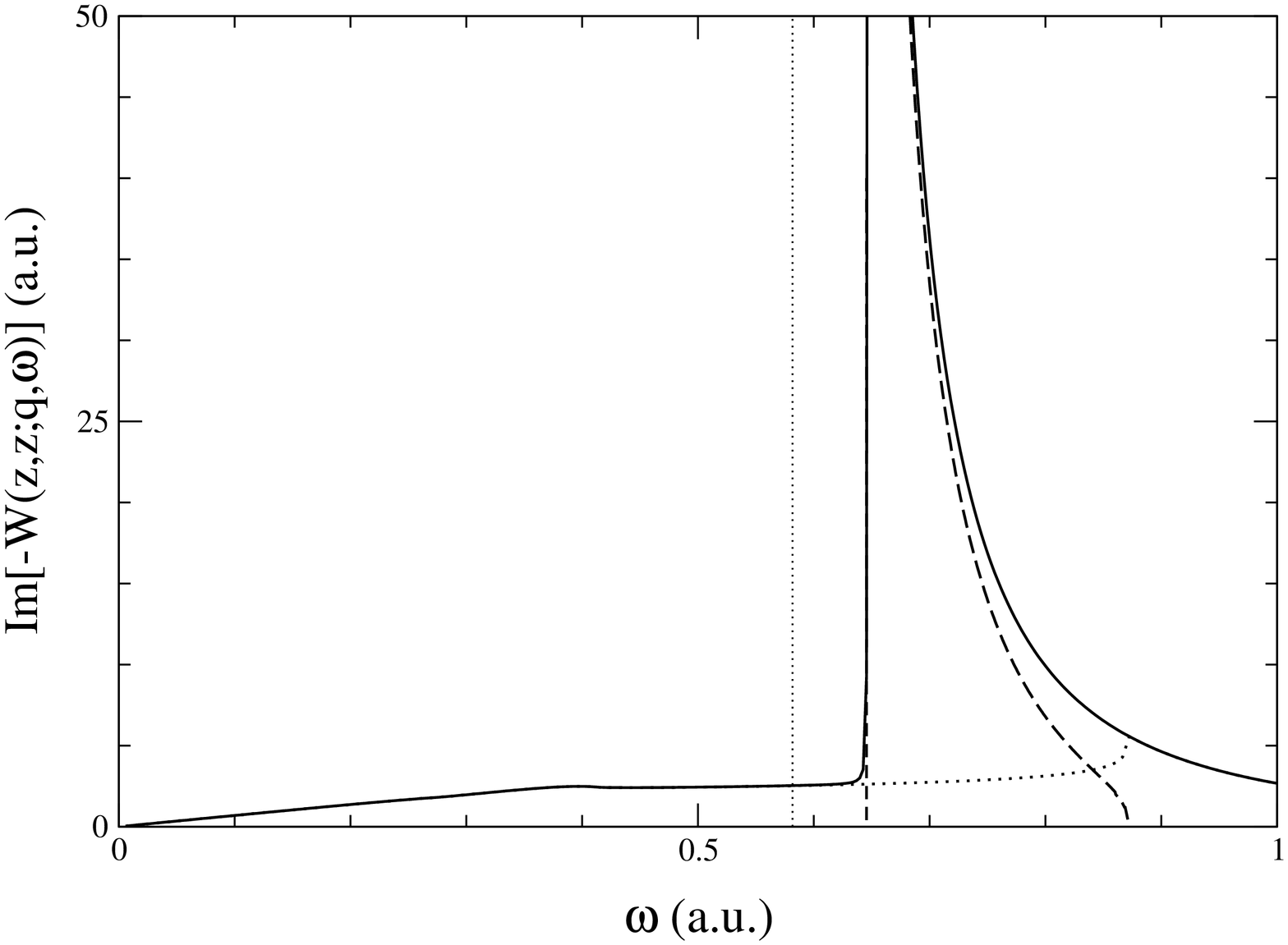}
\caption{The solid line represents the energy-loss function,
${\rm Im}[-W(z,z';q,\omega)]$, versus $\omega$, as obtained at $z=z'$,
$q=0.4q_F$, and $r_s=2.07$ from Eq.~(\ref{in}) by using the full RPA
dielectric function $\epsilon^{RPA}(Q,\omega)$. The thick dashed and
dotted
lines represent separate contributions from the excitation of
{\it bulk} collective modes and e-h pairs occurring at energies
$\omega_{Q=q}^B<\omega<\omega_{Q=Q_c}^B$ and
$\omega\leq qq_F+q^2/2$, respectively. The vertical dotted line
represents the energy $\omega_p=15.8\,{\rm eV}$ at which collective
oscillations
would occur in a Drude
metal.}\label{fig6}
\end{figure}

\begin{figure}
\includegraphics[width=0.95\linewidth]
{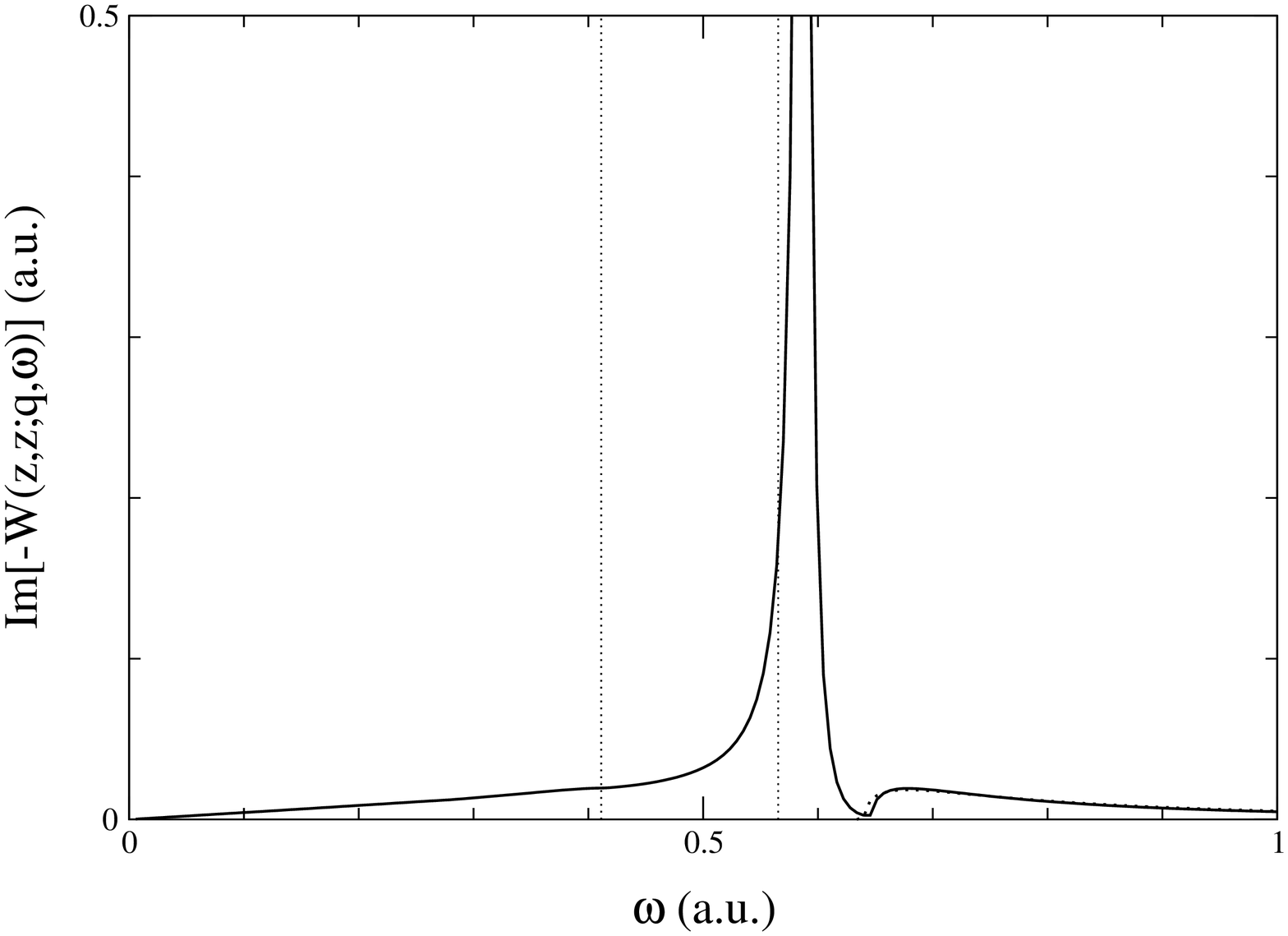} \caption{The solid line represents the energy-loss
function, ${\rm Im}[-W(z,z';q,\omega)]$, versus $\omega$, as
obtained at $z=z'=\lambda_F$, $q=0.4q_F$, and $r_s=2.07$ from
Eq.~(\ref{out}) by using the full RPA dielectric function
$\epsilon^{RPA}(Q,\omega)$. The thick dotted line, which is nearly
indistinguishable from the solid line covering the same part of the
spectrum, represents the hydrodynamic prediction from
Eq.~(\ref{hdp}). The thin dotted vertical lines represent the
energies $\omega_s=11.2\,{\rm eV}$ and $\omega_Q^S=15.4\,{\rm eV}$
of Eq.~(\ref{wqshd}) at which long-lived surface plasmons would
occur in a semi-infinite metal described by a Drude and a
hydrodynamic model, respectively. Introduction of the full RPA
dielectric function $\epsilon^{RPA}(Q,\omega)$ into Eq.~(\ref{gsrm})
yields a maximum of the surface-loss function ${\rm Im}g(q,\omega)$
(and, therefore, ${\rm Im}[-W(z,z';q,\omega)]$) at the
surface-plasmon energy $\omega_Q^S=16.0\,{\rm eV}$, which is
slightly larger than its hydrodynamic counterpart.}\label{fig7}
\end{figure}

Figures~\ref{fig6} and \ref{fig7} show the energy-loss function
${\rm Im}[-W(z,z';q,\omega)]$ that we have obtained at $z=z'$,
$q=0.4q_F$, and $r_s=2.07$ from Eqs.~(\ref{in}) and (\ref{out}),
respectively, by using the full RPA dielectric function
$\epsilon^{RPA}(Q,\omega)$. For $z$ coordinates well inside the
solid (Fig.~\ref{fig6}), instead of the single {\it classical}
collective excitation at $\omega_p$ (dotted vertical line)
predicted by Eq.~(\ref{inclas}) the RPA energy-loss spectrum
(solid line) is composed of (i) a continuum of bulk collective
excitations (dashed line) occurring at energies
$\omega_{Q=q}^B<\omega<\omega_{Q=Q_c}^B$~\cite{inside}, and  (ii)
the excitation of electron-hole (e-h) pairs represented by a thick
dotted line.

For $z$ coordinates that are outside the surface, it had been
generally believed that only surface plasmons and e-h pairs can be
excited. However, it was shown explicitly in Refs.~\cite{aitor} and
\cite{aran1} that the continuum of bulk-plasmon excitations
dominating the energy-loss spectrum inside the solid [see
Fig.~\ref{fig6}] is still present for $z$ coordinates outside, as
shown in Fig.~\ref{fig7} for $z=z'=\lambda_F$~\cite{lambdaf}. This
continuum, which covers the excitation spectrum at energies
$\omega\ge\omega_{Q=q}^B$ and is well separated from the
lower-energy spectrum arising from the excitation of surface
plasmons and e-h pairs, is accurately described by using the quantal
hydrodynamic surface energy-loss function of Eq.~(\ref{hdp}), which
has been represented in Fig.~\ref{fig7} by a thick dotted line.

Nonetheless, the main contribution to the energy-loss spectrum outside
the solid comes from the excitation of surface plasmons, which are
damped by the presence of e-h pairs. These
e-h pair excitations are not present in the classical and
hydrodynamic schemes described above, which predict
the existence of long-lived surface plasmons at the energies
represented
in Fig.~\ref{fig7} by thin dotted vertical lines:
$\omega_s=\omega_p/\sqrt{2}$ [see Eq.~(\ref{outclas})] and $\omega_q^S$
of Eq.~(\ref{wqshd}) [see Eqs.~(\ref{hds})-(\ref{hydrot2})],
respectively.

\subsubsection{Self-consistent scheme}\label{self}

For an accurate quantal description of the electronic excitations that
can occur in a semi-infinite metal, we need to consider the true
density-response function $\chi({\bf r},{\bf r}';\omega)$ entering
Eq.~(\ref{screened}), which is known to fulfill the $f$-sum rule of
Eq.~(\ref{fsumrule}).

\paragraph{Jellium surface.} In the case of a free-electron gas
bounded by a semi-infinite positive background of density
\begin{equation}\label{density+}
n_+(z)=\cases{\bar n,&$z\leq 0$,\cr\cr
0,&$z>0$,}
\end{equation}
translationally invariance in the plane of the surface allows to
define the 2D Fourier transform $W(z,z';q,\omega)$, which according to
Eq.~(\ref{screened}) can be obtained as follows
\begin{eqnarray}\label{wsurf}
&&W(z,z';q,\omega)=v(z,z';q)+\int dz_1\int dz_2\,v(z,z_1;q)\cr\cr
&&\times\chi(z_1,z_2;q,\omega)\,v(z_2,z';q),
\end{eqnarray}
where $v(z,z';q)$ is the 2D Fourier transform of the bare Coulomb
interaction
$v({\bf r},{\bf r}')$:
\begin{equation}
v(z,z';q)={2\pi\over q}\,{\rm e}^{-q|z-z'|},
\end{equation}
and $\chi(z,z';q,\omega)$ denotes the 2D Fourier transform of the
interacting density-response function $\chi({\bf r},{\bf r}';
\omega)$. In the framework of TDDFT, one uses Eq.~(\ref{eq:Xalda2})
to find:
\begin{eqnarray}\label{eq:Xalda3}
&&\chi(z,z';q,\omega)=\chi^0(z,z';q,\omega)+
\int dz_1\int dz_2\,\chi^0(z,z_1;q,\omega)\cr\cr
&&\times\left\{v(z_1,z_2;q)+f_{xc}[n_0](z_1,z_2;q,\omega)\right\}
\chi(z_2,z';q,\omega),
\end{eqnarray}
where $\chi^0(z,z';q,\omega)$ and $f_{xc}[n_0](z,z';q,\omega)$ denote
the 2D Fourier transforms of the noninteracting density-response
function
$\chi^0({\bf r},{\bf r}';\omega)$ and the XC kernel
$f_{xc}[n_0]({\bf r},{\bf r}';\omega)$, respectively. Using
Eq.~(\ref{chi0}), and noting that the single-particle orbitals
$\psi_i({\bf r})$ now take the form
\begin{equation}\label{psi0}
\psi_{{\bf k},i}({\bf r})={\rm e}^{i{\bf k}\cdot{\bf r}_\parallel}\,\psi_i(z),
\end{equation}
one finds:
\begin{eqnarray}
&&\chi^{0}(z,z';q,\omega)={2\over A}\mathrel
{\mathop{\sum}\limits_{i,j}}\psi_{i}(z)\psi_{j}^*(z)\psi_{j}(z')
\psi_{i}^*(z')\cr\cr &&\times \mathop{\sum}\limits_{{\bf k}}
\frac{f_{{\bf k},i}-f_{{\bf k}+{\bf q},j}}{E_{{\bf k},i}- E_{{\bf
k}+{\bf q},j}+\omega+i\eta}, \label{chi0zz1}
\end{eqnarray}
where
\begin{equation}
E_{{\bf k},i}=\varepsilon_i+{k^2\over 2},
\end{equation}
the single-particle orbitals $\psi_i(z)$ and energies $\varepsilon_i$
now being the solutions of the one-dimensional Kohn-Sham equation
\begin{equation}\label{ks1}
\left[-{1\over 2}{d^2\over dz^2}+
v_{KS}[n_0](z)\right]\psi_i(z)=\varepsilon_i\,\psi_i(z),
\end{equation}
with
\begin{equation}\label{ks0}
v_{KS}[n_0](z)=v_H[n_0](z)+v_{xc}[n_0](z),
\end{equation}
\begin{equation}\label{kssurf}
v_H(z)=-2\pi\int_{-\infty}^\infty dz'\,|z-z'|\,
\left[n_0(z')-n_+(z')\right],
\end{equation}
\begin{equation}
v_{xc}[n_0](z)=\left.{\delta E_{xc}[n]\over\delta n(z)}\right|_{n=n_0},
\end{equation}
and
\begin{equation}
n_0(z)={1\over\pi}\sum_i
(\varepsilon_F-\varepsilon_i)\,\psi_i^2(z)\
\theta(\varepsilon_F-\varepsilon_i).
\end{equation}
From Eq.~(\ref{fsumrule}), the imaginary part of the density-response
function
$\chi(z,z';q,\omega)$ is easily found to
fulfill the following sum rule:
\begin{eqnarray}\label{fsumrule2}
\int_{-\infty}^\infty d\omega\,\omega\,{\rm Im}\chi(z,z';q,\omega)&=&
-\pi\left[q^2+{d^2\over dzdz'}\right]\cr\cr
&\times&n_0(z)\,\delta(z-z').
\end{eqnarray}

Within this scheme, the simplest possible approximation is to neglect
XC effects altogether and set the XC potential $v_{xc}[n_0](z)$ and
kernel
$f_{xc}[n_0](z,z';q,\omega)$ equal to zero. In this case, the
one-dimensional single-particle wave functions $\psi_i(z)$ and
energies $\varepsilon_i$ are the self-consistent eigenfunctions and
eigenvalues of a one-dimensional Hartree Hamiltonian. The calculation
of the density-response function is further simplified if the
Hartree potential $v_H[n_0](z)$ of Eq.~(\ref{kssurf}) is replaced by
\begin{equation}\label{ibm}
v_{IBM}(z)=\cases{v_0,&$z\leq z_0$,\cr\cr
\infty,&$z>z_0$,}
\end{equation}
where the value $z_0=(3/16)\lambda_F$ is chosen so as to ensure charge
neutrality. This is the so-called inifinite-barrier model
(IBM)~\cite{lang}, in which the single-particle orbitals $\psi_i(z)$
are simply sines. If one further neglects interference between
incident and scattered electrons at the surface, this model yields
the classical IBM (CIBM)~\cite{griffin} which can be shown to be
equivalent to the SRM described in Section~\ref{srm}.

Alternatively, and with the aim of incorporating band-structure
effects (such as the presence of energy gaps and surface states)
approximately, the self-consistent jellium-like Kohn-Sham potential
of Eq.~(\ref{ks0}) can be replaced by a physically motivated model
potential $v_{MP}(z)$. Examples are the parameterized model
potential reported by Chulkov {\it el al.}~\cite{chulkov0}, which
was successful in the description of the lifetimes of image and
Shockley states in a variety of metal
surfaces~\cite{chulkov1,chemphys,science,others,asier02,Echen04,crampin1},
and the $q$-dependent model potential that has been reported
recently to investigate the momentum-resolved lifetimes of Shockley
states at the Cu(111) surface~\cite{crampin2}.

At this point, we note that for $z$ and $z'$ coordinates that are far
from the surface into the vacuum, where the electron density
vanishes,
Eq.~(\ref{wsurf}) takes the form of Eq.~(\ref{out}) (which within the
SRM is true for all $z,z'>0$), i.e.:
\begin{equation}
W(z,z';q,\omega)=v(z,z';q)-{2\pi\over q}\,{\rm e}^{-q(z+z')}\,
g(q,\omega),
\end{equation}
but with the surface-response function $g(q,\omega)$ now being given
by the general expression~\cite{pz}:
\begin{equation}\label{g}
g(q,\omega)=-{2\pi\over q}\,\int dz_1\int dz_2\,{\rm e}^{q(z_1+z_2)}
\,\chi(z_1,z_2;q,\omega),
\end{equation}
which according to Eq.~(\ref{eqa1}) can be expressed as follows
\begin{equation}\label{g2}
g(q,\omega)=\int dz\,{\rm e}^{qz}\,\delta n(z;q,\omega),
\end{equation}
with $\delta n(z;q,\omega)$ being the electron density induced by an
external potential of the form
\begin{equation}\label{ext2}
\phi^{ext}(z;q,\omega)=-{2\pi\over q}\,{\rm e}^{qz}.
\end{equation}
In the framework of TDDFT, the induced electron density is obtained as in the
RPA [see Eq.~(\ref{eqa1new})], but with the XC kernel
$f_{xc}[n_0]({\bf r},{\bf r}';q,\omega)$ added to the bare Coulomb interaction
$v({\bf r},{\bf r}')$. Hence, after Fourier transforming one writes
\begin{eqnarray}\label{eqa1new2}
&&\delta n(z;q,\omega)=\int{\rm d}z'\,
\chi^0(z,z';q,\omega)\left\{\phi^{ext}(z';q,\omega)+\int dz''\right.\cr\cr
&&\left.\times\left[v(z',z'';q)+f_{xc}[n_0](z',z'';q,\omega)\right]\delta
n(z'';q,\omega)\right\}.
\end{eqnarray}

Using Eqs.~(\ref{fsumrule2}) and (\ref{g}), the surface loss
function
${\rm Im}g(q,\omega)$ is easily found to fulfill the following sum
rule:
\begin{equation}\label{sumrule2}
\int_0^\infty d\omega\,\omega\,{\rm Im}g(q,\omega)=2\pi^2\,q\int dz\
{\rm
e}^{2qz}\,n_0(z),
\end{equation}
which for a step-like electron density $n_0(z)$ of the form of
Eq.~(\ref{density0}) reduces to Eq.~(\ref{sumrule}), as expected.

\paragraph{Periodic surface.} For a periodic surface, single-particle
wave functions are of the form
\begin{equation}
\psi_{{\bf k},n;i}({\bf r})=\psi_{{\bf k},n}({\bf r}_\parallel)\
\psi_i(z),
\end{equation}
where $\psi_{{\bf k},n}({\bf r}_\parallel)$ are Bloch states:
\begin{equation}
\psi_{{\bf k},n}({\bf r}_\parallel)={1\over\sqrt{A}}\,{\rm e}^{{\bf
k}\cdot{\bf r}_\parallel}\,u_{{\bf k},n}({\bf r}_\parallel),
\end{equation}
with ${\bf r}_\parallel$ and ${\bf k}$ being 2D vectors in the plane
of the surface. Hence, one may introduce the following Fourier
expansion of the screened interaction:
\begin{eqnarray}
W({\bf r},{\bf r}';\omega)&=&{1\over A}\sum_{\bf q}^{SBZ}\sum_{{\bf
g},{\bf g}'}
{\rm e}^{i({\bf q}+{\bf g})\cdot{\bf r}_\parallel}
{\rm e}^{-i({\bf q}+{\bf g}')\cdot{\bf r}_\parallel'}\cr\cr
&\times&W_{{\bf g},{\bf g}'}(z,z';{\bf q},\omega),
\end{eqnarray}
where ${\bf q}$ is a 2D wave vector in the surface Brillouin zone
(SBZ),
and ${\bf g}$ and ${\bf g}'$ denote 2D reciprocal-lattice vectors.
According to Eq.~(\ref{screened}), the 2D Fourier coefficients
$W_{{\bf g},{\bf g}'}(z,z';{\bf q},\omega)$ are given by the following
expression:
\begin{eqnarray}\label{screenedg}
&&W_{{\bf g},{\bf g}'}(z,z';{\bf q},\omega)=
v_{\bf g}(z,z';{\bf q})\,\delta_{{\bf g},{\bf g}'}+
\int dz_1\int dz_2\cr\cr
&&\times v_{\bf g}(z,z_1;{\bf q})\,
\chi_{{\bf g},{\bf g}'}(z_1,z_2;{\bf q},\omega)\,
v_{{\bf g}'}(z_2,z';{\bf q}),
\end{eqnarray}
where $v_{\bf g}(z,z';{\bf q})$ denote
the 2D Fourier coefficients of the bare Coulomb interaction
$v({\bf r},{\bf r}')$:
\begin{equation}
v_{\bf g}(z,z';{\bf q})={2\pi\over|{\bf q}+{\bf g}|}\,
{\rm e}^{-|{\bf q}+{\bf g}|\,|z-z'|},
\end{equation}
and $\chi_{{\bf g},{\bf g}'}(z,z';{\bf q},\omega)$ are the Fourier
coefficients of the interacting density-response function
$\chi({\bf r},{\bf r}';\omega)$. In the framework of TDDFT, one uses
Eq.~(\ref{eq:Xalda2}) to find:
\begin{eqnarray}\label{eq:Xalda4}
&&\chi_{{\bf g},{\bf g}'}(z,z';{\bf q},\omega)=
\chi_{{\bf g},{\bf g}'}^0(z,z';{\bf q},\omega)+
\int dz_1\int dz_2\cr\cr
&&\times\chi_{{\bf g},{\bf g}'}^0(z,z_1;{\bf q},\omega)
\,\times\left[v_{{\bf g}_1}(z_1,z_2;{\bf q})\,\delta_{{\bf g}_1,{\bf
g}_2}\right.\cr\cr
&&\left.+f_{{\bf g}_1,{\bf g}_2}^{xc}[n_0]
(z_1,z_2;{\bf q},\omega)\right]
\chi_{{\bf g}_2,{\bf g}'}(z_2,z';{\bf q},\omega),
\end{eqnarray}
where $\chi_{{\bf g},{\bf g}'}^0(z,z';{\bf q},\omega)$ and
$f_{{\bf g},{\bf g}'}^{xc}[n_0](z,z';{\bf q},\omega)$ denote the
Fourier coefficients of the noninteracting density-response function
$\chi^0({\bf r},{\bf r}';\omega)$ and the XC kernel
$f_{xc}[n_0]({\bf r},{\bf r}';\omega)$, respectively. Using
Eq.~(\ref{chi0}), one finds:
\begin{eqnarray}\label{eq9}
&&\chi_{{\bf g},{\bf g}'}^0(z,z';{\bf q},\omega)=\frac{2}{A}
\sum_{i,j}\psi_i(z)\psi_j^*(z)\psi_j(z')\psi_i^*(z')\cr\cr
&&\times\sum_{\bf k}^{\rm SBZ}\sum_{n,n'}\frac{f_{{\bf k},n;i}-f_{{\bf
k}+{\bf q},n';j}}
{\varepsilon_{{\bf
k},n;i}-\varepsilon_{{\bf k}+{\bf q},n';j} +\hbar(\omega +
{ i}\eta)}\cr\cr
&&\times\langle\psi_{{\bf k},n}|e^{-{ i}({\bf q}+{\bf g})\cdot{\bf
r}_\parallel}|\psi_{{\bf k}+{\bf q},n'}\rangle\cr\cr
&&\times\langle\psi_{{\bf k}+{\bf q},n'}|e^{{ i}({\bf q}+{\bf
g}')\cdot{\bf
r}_\parallel}|\psi_{{\bf k},n}\rangle,
\end{eqnarray}
the single-particle orbitals
$\psi_{{\bf k},n;i}({\bf r})=\psi_{{\bf k},n}({\bf r}_\parallel)\
\psi_i(z)$ and energies
$\varepsilon_{{\bf k},n;i}$ being the eigenfunctions and eigenvalues
of a 3D Kohn-Sham Hamiltonian with an effective potential that is
periodic in the plane of the surface.

As in the case of the jellium surface, we can focus on the special
situation where both $z$ and $z'$ coordinates are located far from
the surface into the vacuum. Eq.~(\ref{screened}) shows that under
such conditions the Fourier coefficients
$W_{{\bf g},{\bf g}'}(z,z';{\bf q},\omega)$ take the following
form:
\begin{eqnarray}
&&W_{{\bf g},{\bf g}'}(z,z';{\bf q},\omega)=
v_{{\bf g}}(z,z';{\bf q})\,\delta_{{\bf g},{\bf g}'}-
{2\pi q\over|{\bf q}+{\bf g}|\,|{\bf q}+{\bf g}'|}\cr\cr
&&\,\,\,\,\,\,\times\,g_{{\bf g},{\bf g}'}({\bf q},\omega)\,
{\rm e}^{-|{\bf q}+{\bf g}|z}\,{\rm e}^{-|{\bf q}+{\bf g}'|z'},
\end{eqnarray}
where
\begin{equation}\label{ggg}
g_{{\bf g},{\bf g}'}({\bf q},\omega)=
-{2\pi\over q}\,\int dz_1\int dz_2\,{\rm e}^{q(z_1+z_2)}
\chi_{{\bf g},{\bf g}'}(z_1,z_2;{\bf q},\omega).
\end{equation}
In particular,
\begin{equation}\label{g00}
g_{{\bf g}=0,{\bf g}=0}({\bf q},\omega)=\int dz\,{\rm e}^{qz}\,
\delta n_{{\bf g}=0}(z;{\bf q},\omega),
\end{equation}
with $\delta n_{\bf g}(z;{\bf q},\omega)$ being the Fourier coefficients of
the electron density induced by an external potential of the form
\begin{equation}\label{extg}
\phi_{\bf g}^{ext}(z;{\bf q},\omega)=-{2\pi\over q}\,{\rm e}^{qz}\
\delta_{{\bf g},{\bf 0}}.
\end{equation}

Finally, one finds from Eq.~(\ref{fsumrule}) that the Fourier
coefficients
$\chi_{{\bf g},{\bf g}'}(z,z';{\bf q},\omega)$ fulfill the following
sum rule:
\begin{eqnarray}\label{fsumrule3}
\int_{-\infty}^\infty d\omega&\omega&{\rm Im}
\chi_{{\bf g},{\bf g}'}(z,z';q,\omega)=
-\pi\left[q^2+{d^2\over dzdz'}\right]\cr\cr
&\times&n_{{\bf g}-{\bf g}'}^0(z;{\bf q})\,\delta(z-z'),
\end{eqnarray}
where the coefficients $n_{\bf g}^0(z;{\bf q})$ denote the 2D Fourier
components of the ground-state electron density $n_0({\bf r})$.
Furthermore, combining Eqs.~(\ref{ggg}) and (\ref{fsumrule3}), one
writes
\begin{equation}\label{sumrulep}
\int_0^\infty d\omega\,\omega\,{\rm Im}g_{{\bf g},{\bf g}'}(q,\omega)=
2\,\pi^2\,q\int dz\,{\rm e}^{2q_z}\,n_{{\bf g}-{\bf g}'}(z;{\bf q}),
\end{equation}
which is a generalization of the sum rule of Eq.~(\ref{sumrule}) to
the more general case of a real solid in which the crystal structure
parallel to the surface is taken into account.

\section{Surface-response function}

The central quantity that is involved in a description of surface
collective
excitations is the surface-response function introduced in the
preceding
section: $g(q,\omega)$ for a jellium surface [see Eq.~(\ref{g})] and
$g_{{\bf g},{\bf g}'}({\bf q},\omega)$ for a periodic surface [see
Eq.~(\ref{ggg})].

\subsection{Generation-rate of electronic excitations}

In the framework of TDDFT, an interacting many-electron system exposed
to a frequency-dependent external potential
$\phi^{ext}({\bf r},\omega)$ is replaced by a fictitious system of
{\it noninteracting}
electrons exposed to an effective self-consistent potential
$\phi^{sc}({\bf r},\omega)$ of the form
\begin{equation}
\phi^{sc}({\bf r},\omega)=\phi^{ext}({\bf r},\omega)+\delta\phi({\bf
r},\omega),
\end{equation}
where
\begin{equation}
\delta\phi({\bf r},\omega)=\int d{\bf r}'\,\left[v({\bf r},{\bf r}')+
f_{xc}[n_0]({\bf r},{\bf r}';\omega)\right]\,
\delta n({\bf r}';\omega).
\end{equation}

Hence, the rate at which a frequency-dependent external potential
$\phi^{ext}({\bf r},\omega)$ generates electronic excitations in the
many-electron system can be obtained, within lowest-order perturbation
theory, as follows
\begin{equation}\label{wij}
w(\omega)=2\pi\sum_{i,j}f_i(1-f_i)\left|\langle\psi_j({\bf r})|
\phi^{sc}({\bf r},\omega)\psi_i({\bf r})\rangle\right|^2\
\delta(\epsilon_i-\epsilon_f),
\end{equation}
with $\psi_i({\bf r})$ and $\epsilon_i$ being the eigenfunctions and
eigenvalues of a 3D Kohn-Sham Hamiltonian. In terms of the induced
electron density $\delta n({\bf r},\omega)$ of Eq.~(\ref{eqa1}), one
finds~\cite{liebsch0}
\begin{equation}
w(\omega)=-2\,{\rm Im}\int d{\bf r}\,\phi_{ext}^*({\bf r},\omega)\,
\delta n({\bf r},\omega),
\end{equation}
which in the case of a periodic surface takes the following form:
\begin{equation}
w(\omega)=\sum_{\bf g}\sum_{\bf q}^{SBZ}\,w_{\bf g}({\bf q},\omega),
\end{equation}
where $w_{\bf g}({\bf q},\omega)$ denotes the rate at which the
external potential
generates electronic excitations of frequency $\omega$ and parallel
wave vector
${\bf q}+{\bf g}$:
\begin{equation}
w_{\bf g}({\bf q},\omega)=-{2\over A}\,{\rm Im}\,\int dz\,
\phi_{\bf g}^{ext}(z,{\bf q})\,\delta n_{\bf g}(z,{\bf q}),
\end{equation}
$\phi_{\bf g}^{ext}(z,{\bf q})$ and $\delta n_{\bf g}(z,{\bf q})$
being the Fourier coefficients of the external potential and the
induced electron density, respectively.

In particular, if the external potential is of the form of
Eq.~(\ref{extg}), the rate $w_{\bf g}({\bf q},\omega)$ can be
expressed in
terms of the surface-response function
$g_{{\bf g},{\bf g}'}({\bf q},\omega)$ of Eq.~(\ref{ggg}), as follows
\begin{equation}\label{wg}
w_{\bf g}({\bf q},\omega)={4\pi\over qA}\,{\rm Im}
g_{{\bf g},{\bf g}}({\bf q},\omega)\,\delta_{{\bf g},0},
\end{equation}
which for a jellium surface reduces to
\begin{equation}\label{wg2}
w(q,\omega)={4\pi\over qA}\,{\rm Im}g(q,\omega),
\end{equation}
with $g(q,\omega)$ being the surface-response function of
Eq.~(\ref{g}). We note that although only the coefficient
$\chi_{{\bf g}=0,{\bf g}'=0}(z,z';{\bf q},\omega)$ enters into the
evaluation of the more realistic Eq.~(\ref{wg}), the full
$\chi_{{\bf g},{\bf g}'}^0(z,z';{\bf q},\omega)$ matrix is implicitly
included through Eq.~(\ref{eq:Xalda4}).

\subsection{Inelastic electron scattering}

\begin{figure}
\includegraphics[width=0.5\linewidth]
{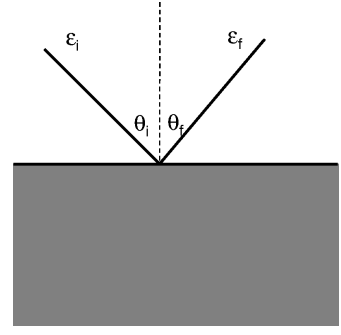}
\caption{A schematic drawing of the scattering geometry in
angle-resolved inelastic electron scattering experiments. On exciting
a surface mode of frequency $\omega(q)$, the energy of detected
electrons becomes $\varepsilon_f=\varepsilon_i-\omega(q)$, with the
momentum $q$ being determined by Eq.~(\ref{momq}).}\label{fig8}
\end{figure}

The most commonly used experimental arrangement for the detection of
surface collective excitations by the fields of moving charged
particles is based on angle-resolved inelastic electron
scattering~\cite{ibach}. Fig.~\ref{fig8} shows a schematic drawing
of the scattering geometry. A monochromatic beam of electrons of
energy $\varepsilon_i$, incident on a flat surface at an angle
$\theta_i$, is back scattered and detected by an angle-resolved
energy analyzer positioned at an angle $\theta_f$ and energy
$\varepsilon_f$. Inelastic events can occur, either before or after
the elastic event, on exciting a surface mode of frequency
$\omega(q)=\varepsilon_i-\varepsilon_f$. The energy and lifetime of
this mode are determined by the corresponding energy-loss peak in
the spectra, and the momentum $q$ parallel to the surface is
obtained from the measured angles $\theta_i$ and $\theta_f$, as
follows:
\begin{equation}\label{momq}
q=\sqrt{2}\left[\sqrt{\varepsilon_i}\
\sin\theta_i-\sqrt{\varepsilon_f}\,
\sin\theta_f\right].
\end{equation}

The inelastic scattering cross section corresponding to a process in
which an electronic excitation of energy $\omega$ and parallel wave
vector ${\bf q}$ is created at a semi-infinite solid surface can be
found to be
proportional to the rate given by Eq.~(\ref{wg}) [or Eq.~(\ref{wg2})
if the
medium
is represented by a jellium surface] and is, therefore, proportional
to the
imaginary part of the surface-response function, i.e., the so-called
surface-loss function. Hence, apart from kinematic factors (which can
indeed vary with energy and momentum~\cite{mills,gaspar}), the
inelastic scattering cross section is dictated by the rate
$w_{\bf g}({\bf q},\omega)$ [or $w(q,\omega)$ if the medium is
represented by a
jellium surface] at which an external potential of the form of
Eq.~(\ref{extg}) generates electronic excitations of frequency
$\omega$ and parallel wave vector ${\bf q}$.

In the following sections, we focus on the behavior of the
energy-loss function ${\rm Im}g_{{\bf g}=0,{\bf g}=0}({\bf
q},\omega)$ [or ${\rm Im}g(q,\omega)$] and its maxima, which account
for the presence and momentum-dispersion of surface collective
excitations.

\subsection{Surface plasmons: jellium surface}

\subsubsection{Simple models}

In the simplest possible model of a jellium surface in vacuum, in
which a semi-infinite medium with local dielectric function
$\epsilon(\omega)$ at $z\leq 0$ is terminated at $z=0$, the
surface-response function $g(q,\omega)$ is obtained from Eq.~(\ref{g1})
with $\epsilon_2=1$, or, equivalently, from Eq.~(\ref{gsrm}) with $q=0$, i.e.,
\begin{equation}\label{long2}
g(q,\omega)={\epsilon(\omega)-1\over\epsilon(\omega)+1},
\end{equation}
which for a Drude dielectric function [see Eq.~(\ref{drude})] leads to
the
surface-loss function
\begin{equation}\label{drudeg}
{\rm Im}g(q,\omega)={\pi\over 2}\,\omega_s\,\delta(\omega-\omega_s)
\end{equation}
peaked at the surface-plasmon energy $\omega_s=\omega_p/\sqrt{2}$.

The {\it classical} energy-loss function of Eq.~(\ref{long2})
represents indeed the true long-wavelength ($q\to 0$) limit of the
actual self-consistent
surface-loss function of a jellium surface. Nevertheless, the
classical picture leading to Eq.~(\ref{long2}) ignores both the
nonlocality of the electronic response of the system and the
microscopic spatial distribution of the electron density near the
surface. Nonlocal effects can be incorporated within the hydrodynamic
and specular-reflection models described in Sections~\ref{hydrog} and
\ref{srm}.

Within a one-step hydrodynamic approach, the surface-loss
function is also dominated by a delta function [see Eq.~(\ref{hds})]
but peaked at the momentum-dependent surface-plasmon energy of
Eq.~(\ref{wqshd}) [see also Eq.~(\ref{plasmons})]:
\begin{equation}\label{wqshd2}
\omega^2={1\over 2}\left[\omega_p^2+
\beta^2\,q^2+\beta\,q\,
\sqrt{2\omega_p^2+\beta^2\,q^2}\right],
\end{equation}
 which at long wavelengths yields
\begin{equation}\label{hdr2}
\omega=\omega_p/\sqrt{2}+\beta\,q/2,
\end{equation}
$\beta$ representing the speed of propagation of hydrodynamic
disturbances in
the electron system~\cite{noteh}.

In the SRM (with the bulk dielectric function
being described within the RPA), surface plasmons, which occur at a
momentum-dependent energy slightly different from its
hydrodynamic counterpart, are damped by the presence of e-h pair
excitations, as shown in Fig.~\ref{fig7}~\cite{noteg}.

The one-step hydrodynamic Eqs.~(\ref{wqshd2}) and (\ref{hdr2}) and a numerical
evaluation of the
imaginary part of the SRM surface-response function of
Eq.~(\ref{gsrm}) (see Fig.~\ref{fig7}) both yield a {\it positive}
surface-plasmon energy dispersion at all wave vectors. Nonetheless,
Bennett used a hydrodynamic model with a continuum decrease of the
electron
density at the metal surface, and found that a continuous
electron-density
variation yields a {\it monopole} surface plasmon with a {\it
negative} dispersion at low wave vectors~\cite{bennett}.

\subsubsection{Self-consistent calculations: long wavelengths}

Within a self-consistent long-wavelength description of the
jellium-surface electronic response, Feibelman showed that up to
first order in an expansion in powers of the magnitude $q$ of the
wave vector, the surface-response function of Eq.~(\ref{g}) can be
written
as~\cite{feibelman0}
\begin{equation}\label{expansion}
g(q,\omega)={
\left[\epsilon(\omega)-1\right]\,\left[1+qd_\perp(\omega)\right]
\over
\epsilon(\omega)+1-
\left[\epsilon(\omega)-1\right]\,qd_\perp(\omega)},
\end{equation}
where $\epsilon(\omega)$ represents the long-wavelength limit of the
dielectric function of the bulk material~\cite{drudenote} and
$d_\perp(\omega)$ denotes the centroid of the induced electron
density [see Eq.~(\ref{dperp})] with respect to the jellium
edge~\cite{dper}.
Eq.~(\ref{expansion}) shows that at long wavelengths the poles of the
surface-response function $g(q,\omega)$ are determined by the
non-retarded
surface-plasmon condition of Eq.~(\ref{ds}) with $\epsilon_2=1$, which
for a
semi-infinite free-electron metal in vacuum yields the surface-plasmon
dispersion
relation of Eq.~(\ref{sdr}), i.e.:
\begin{equation}\label{sdr2}
\omega=\omega_s\left(1+\alpha\,q\right),
\end{equation}
with
\begin{equation}\label{alpha}
\alpha=-{\rm Re}\left[d_\perp(\omega_s)\right]/2.
\end{equation}

Equations~(\ref{sdr2})-(\ref{alpha}) show that the long-wavelength
surface-plasmon energy dispersion is dictated by the position of the
centroid of the induced electron density with respect to the jellium edge.
This can be understood by noting that the potential associated with the
surface-plasmon charge attenuates on either side of
${\rm Re}\left[d_\perp(\omega_s)\right]$ with the attenuation constant $q$. If
the fluctuating charge lies inside the jellium edge
(${\rm Re}\left[d_\perp(\omega_s)\right]<0$), as $q$ increases
(thus the surface-plasmon potential attenuating faster) more of the
field overlaps the metal giving rise to more interchange of energy between
the electric field and the metal and resulting in a positive energy dispersion.
However, if the
fluctuating charge lies outside the jellium edge (${\rm
Re}\left[d_\perp(\omega_s)\right]>0$), as $q$ increases less of the
metal is
subject to the plasmon's electric field (i.e., there is a decreasing
overlap
of the fluctuating potential and the unperturbed electron density) which
results in less interchange of energy between the electric field and
the metal and a negative dispersion coefficient~\cite{forst,feibi}.

Quantitative RPA calculations of the surface-plasmon linear-dispersion
coefficient $\alpha$ entering Eq.~(\ref{sdr2}) were carried out by
several
authors by using the specular-reflection and
infinite-barrier models of the surface~\cite{ritchie0,kanazawa,beck1},
a step
potential~\cite{beck2,inglessp}, and the more
realistic Lang-Kohn self-consistent
surface potential~\cite{feibelman1,liebsch1,ks}. Both Feibelman's RPA
self-consistent
calculations~\cite{feibelman1} and the ALDA calculations carried out later by
Liebsch~\cite{liebsch1} and by Kempa and
Schaich~\cite{ks} demonstrated that in the range of typical bulk
densities
($r_s=2-6$) the centroid $d_\perp$ of the induced electron density at
$\omega_s$ lies outside the jellium edge, which leads to a {\it
negative}
long-wavelength dispersion of the surface plasmon. These calculations,
which corroborated Bennett's prediction~\cite{bennett}, also
demonstrated that
the long-wavelength surface-plasmon dispersion is markedly sensitive
to the
shape of the barrier and to the presence of short-range XC effects.

\begin{figure}
\includegraphics[width=0.95\linewidth]
{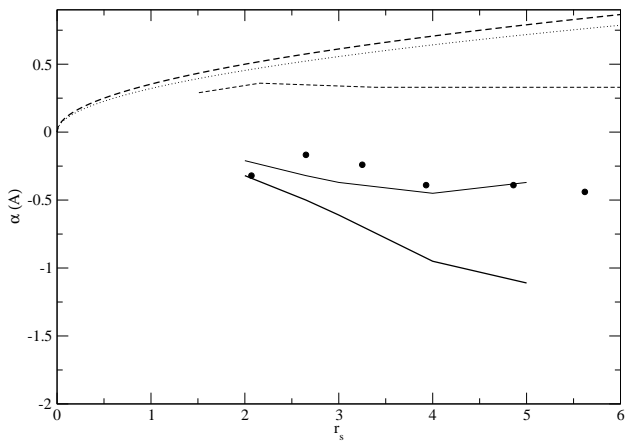}
\caption{Long-wavelength surface-plasmon dispersion coefficient
$\alpha$
entering Eq.~(\ref{sdr2}), versus the electron-density parameter
$r_s$, as
obtained from Eq.~(\ref{alphahd}) (dotted line) and from SRM (thick
dashed
line), IBM (thin dashed line), and self-consistent RPA and ALDA (thin
and
thick solid lines, respectively) calculations of the centroid of the
induced
electron density at $\omega=\omega_s$. The solid circles represent the
angle-resolved low-energy inelastic electron scattering measurements
reported
in Refs.~\cite{plummer1} for Na and K, in Ref.~\cite{plummer3} for Cs,
in
Ref.~\cite{plummer5} for Li and Mg, and in Ref.~\cite{alnew} for Al.
The IBM
and self-consistent (RPA and ALDA) calculations have been taken from
Refs.~\cite{beck1} and \cite{liebsch1}, respectively.}\label{fig9}
\end{figure}

Existing calculations of the long-wavelength dispersion coefficient
$\alpha$
entering Eq.~(\ref{sdr}) are shown in Fig.~\ref{fig9} and summarized
in
Table~\ref{table1}. At one
extreme, the dotted line of Fig.~\ref{fig9} gives the single-step
hydrodynamic coefficient $\alpha^{HD}$ obtained from Eq.~(\ref{hdr}),
i.e.,
\begin{equation}\label{alphahd}
\alpha^{HD}={\beta\over 2\omega_s},
\end{equation}
with $\beta=\sqrt{3/5}(3\pi^2n_0)^{1/3}$; within this model, the
long-wavelength surface-plasmon dispersion is always positive. As in
the single-step hydrodynamic approach, the SRM equilibrium density
profile is of the form of Eq.~(\ref{density0}), and in the IBM the
electron density still varies too rapidly; as a result, the
corresponding SRM and IBM $\alpha$ coefficients (represented by
thick and thin dashed lines, respectively) are both positive. At the
other extreme are the more realistic self-consistent RPA and ALDA
calculations, represented by thin and thick solid lines,
respectively, which yield a linear-dispersion coefficient $\alpha$
that is negative in the whole metallic density range.

\begin{table}
\caption{Long-wavelength surface-plasmon dispersion coefficient
$\alpha$ in
$\rm\AA$ for various simple metal surfaces, as obtained from
Eq.~(\ref{alphahd})
(HD) and from SRM, IBM, and self-consistent RPA and ALDA calculations
of the
centroid of the induced electron density at $\omega=\omega_s$ [see
Eq.~(\ref{alpha})], and from angle-resolved low-energy inelastic
electron
scattering measurements reported in Refs.~\cite{plummer1} for Na and
K, in
Ref.~\cite{plummer3} for Cs, in Ref.~\cite{plummer5} for Li and Mg,
and in
Ref.~\cite{alnew} for Al. As in Fig.~\ref{fig9}, the IBM and
self-consistent
(RPA and ALDA) calculations have been taken from Refs.~\cite{beck1}
and
\cite{liebsch1}, respectively. Also shown in this table are the
measured
values of the surface-plasmon energy $\omega_s$ at $q=0$, which are
all
slightly below the jellium prediction:
$\omega_p/\sqrt{2}=\sqrt{3/2r_s^3}e^2/a_0$ due to band-structure effects.}
\begin{ruledtabular} \begin{tabular}{lcccccccc}
&$r_s$&$\omega_s$&HD&SRM&IBM&RPA&ALDA&Exp.\\ \hline
Al&2.07&10.86&0.46&0.50&0.36&-0.21&-0.32&-0.32\\
Mg&2.66&7.38&0.52&0.57&&-0.30&-0.50&-0.41\\
Li&3.25&4.28&0.58&0.63&0.33&-0.40&-0.70&-0.24\\
Na&3.93&3.99&0.64&0.70&&-0.45&-0.85&-0.39\\
K&4.86&2.74&0.71&0.78&&-0.35&-1.10&-0.39\\
Cs&5.62&1.99&0.76&0.84&0.33&-0.26&-1.14&-0.44\\
\end{tabular} \end{ruledtabular} \label{table1}
\end{table}

Conclusive experimental confirmation that the original Bennett's
prediction~\cite{bennett} was correct came with a series of
measurements based on angle-resolved low-energy inelastic electron
scattering~\cite{plummer1,plummer2,plummer3,plummer5,alnew}.
It has
been demonstrated that the surface-plasmon energy of simple metals
disperses
downward in energy at small momentum $q$ parallel to the surface, the
dispersion coefficients (represented in Fig.~\ref{fig9} by solid
circles)
being in reasonable agreement with self-consistent jellium
calculations, as
shown in Fig.~\ref{fig9} and Table~\ref{table1}.

\subsubsection{Self-consistent calculations: arbitrary wavelengths}

\begin{figure}
\includegraphics[width=0.95\linewidth]
{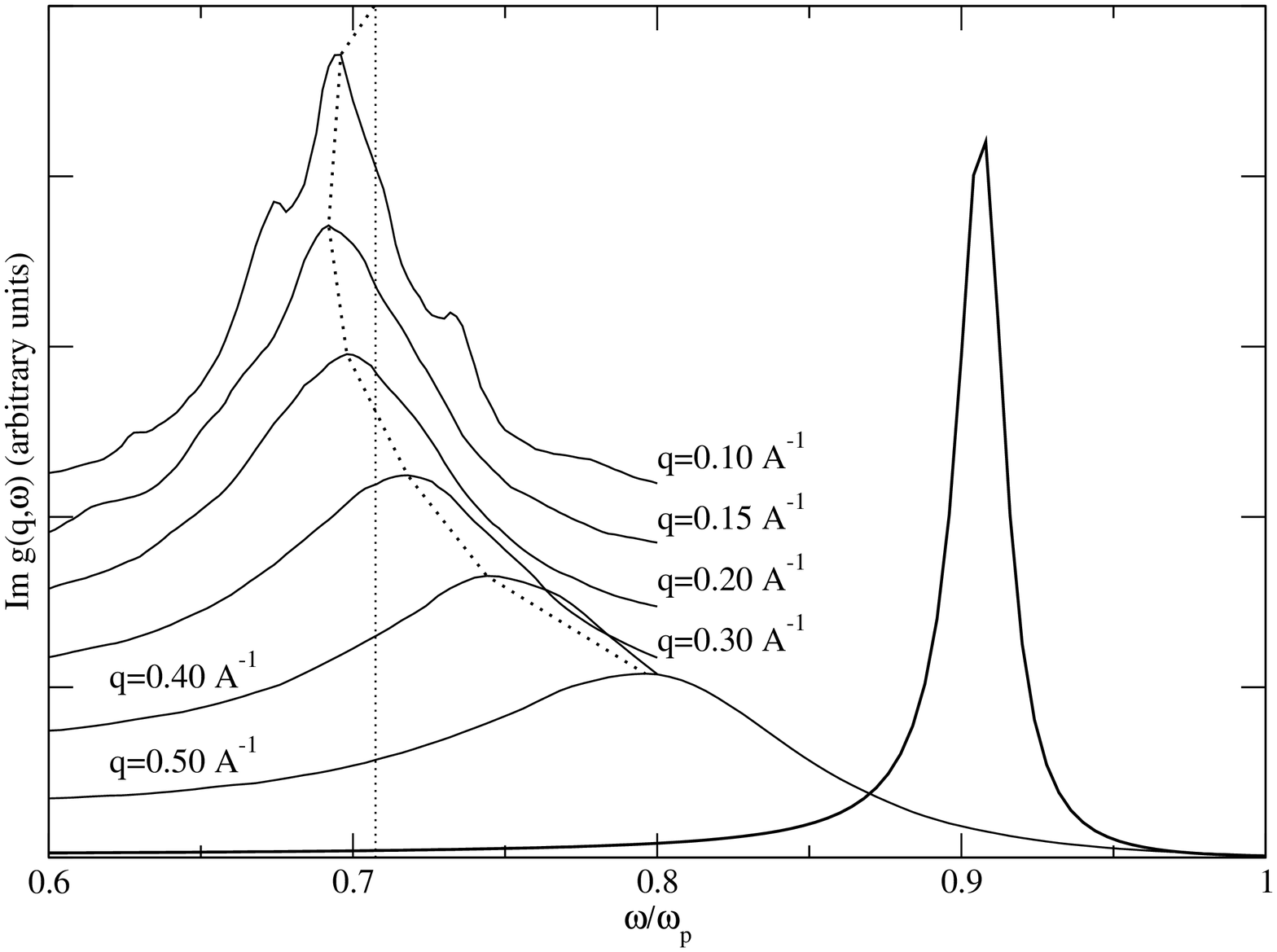} \caption{Self-consistent RPA calculations (thin solid
lines) of the surface loss function ${\rm Im}g(q,\omega)$, as
obtained from Eq.~(\ref{g}), versus the energy $\omega$ for a
semi-infinite free-electron gas with the electron density equal to
that of valence electrons in Al ($r_s=2.07$) and for various
magnitudes of the 2D wave vector ${\bf q}$. The thick solid line
represents the SRM surface loss function for $r_s=2.07$ and
$q=0.5\,{\rm \AA^{-1}}$.}\label{fig10}
\end{figure}

At arbitrary wavelengths, surface-plasmon energies can be derived from
the
maxima of the surface loss function ${\rm Im}g(q,\omega)$ and compared
to the
peak positions observed in experimental electron energy-loss spectra.
Fig.~\ref{fig10} shows the self-consistent calculations of ${\rm Im}
g(q,\omega)$ that we have obtained
in the RPA for a semi-infinite free-electron gas (jellium surface)
with the
electron density equal to that of valence electrons in Al
($r_s=2.07$). For 2D
wave vectors of magnitude in the range $q=0-0.5\,{\rm\AA}^{-1}$, all
spectra are
clearly dominated by a surface-plasmon excitation, which is seen to
first
shift to lower frequencies as $q$ increases [as dictated by
Eqs.~(\ref{sdr2})-(\ref{alpha})] and then, from about $q=0.15\,{\rm\AA}^{-1}$
on, towards
higher frequencies.

For the numerical evaluation of the spectra shown in Fig.~\ref{fig10} we have
first computed the interacting density-response function $\chi(z,z';q,\omega)$
of a sufficiently thick jellium slab by following the method described in
Ref.~\cite{eguiluz0}, and we have then derived the surface-loss function from
Eq.~(\ref{g}). Alternatively, the self-consistent energy-loss spectra reported
in Ref.~\cite{plummer3} (see also Ref.~\cite{liebsch0}) were obtained by first
computing the noninteracting density-response function $\chi^0(z,z';q,\omega)$
of a semi-infinite electron system in terms of Green's functions, then
performing a matrix inversion of Eq.~(\ref{eqa1new2}) with the external
potential $\phi^{ext}(z;q,\omega)$ of Eq.~(\ref{ext2}), and finally deriving
the surface-loss function from Eq.~(\ref{g2}). As expected, both approaches
yield the same results.

\begin{figure}
\begin{tabular}{ll}
\includegraphics*[width=0.5\linewidth] {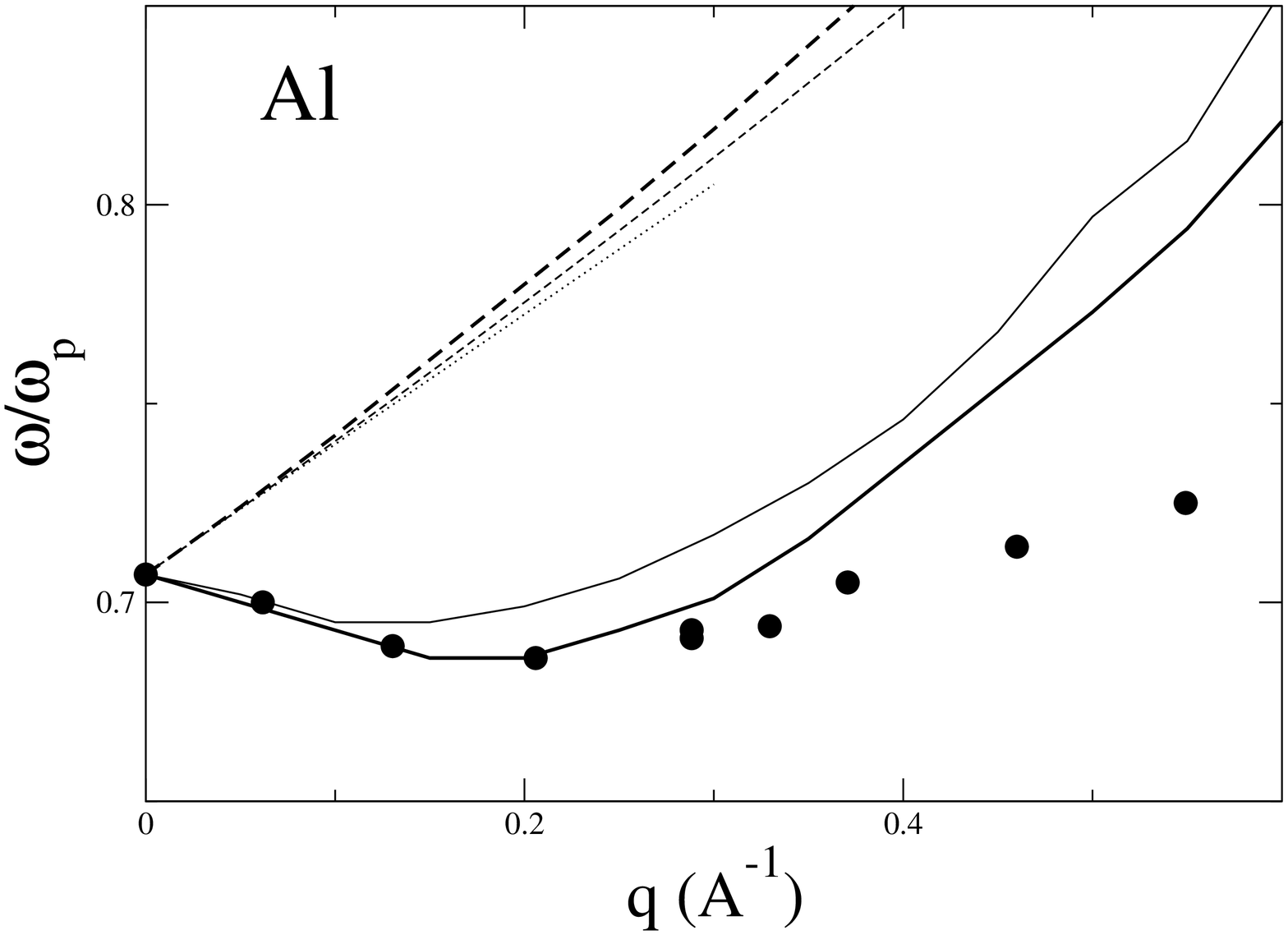}
&
\includegraphics*[width=0.5\linewidth] {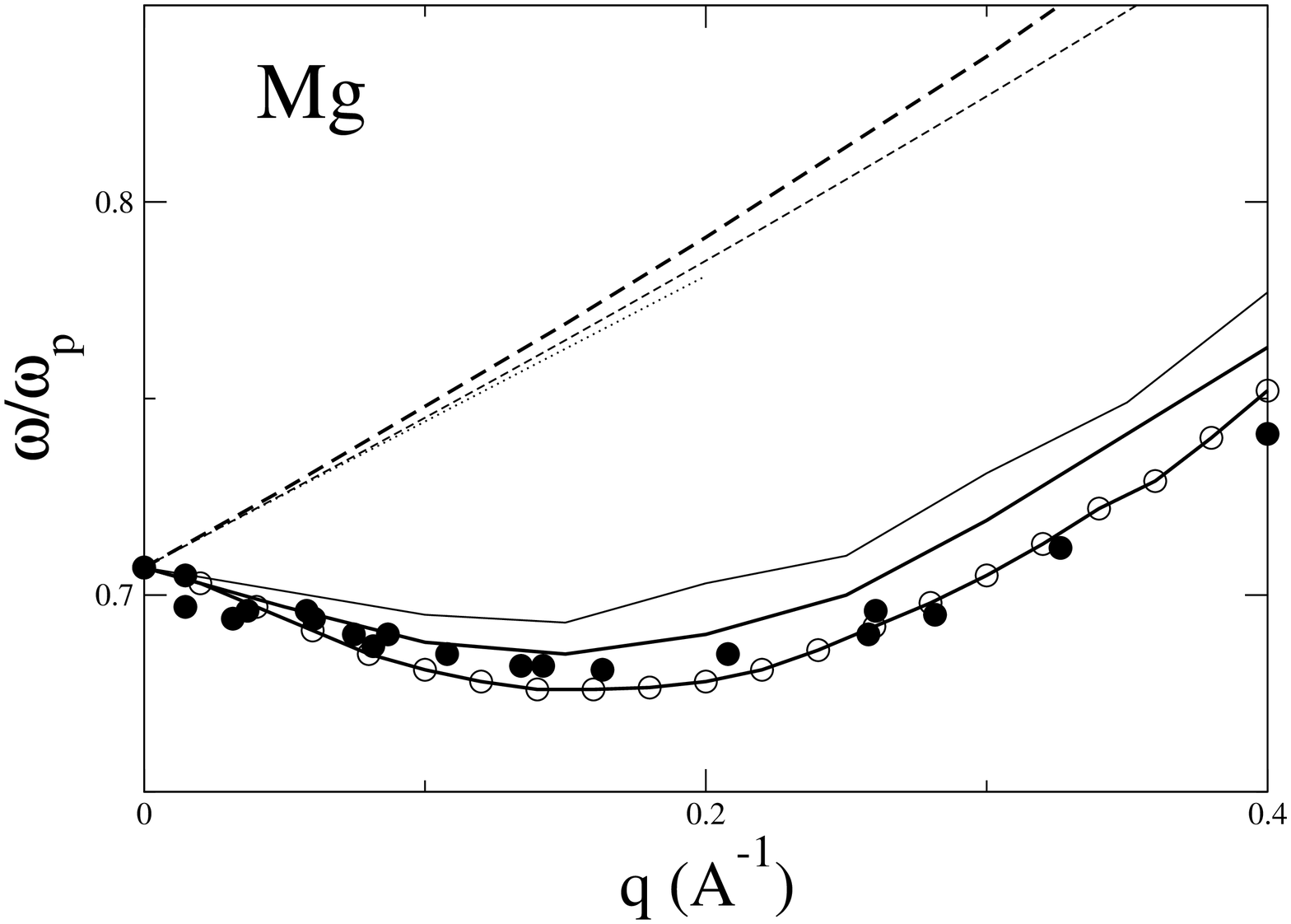}
\\
\includegraphics*[width=0.5\linewidth] {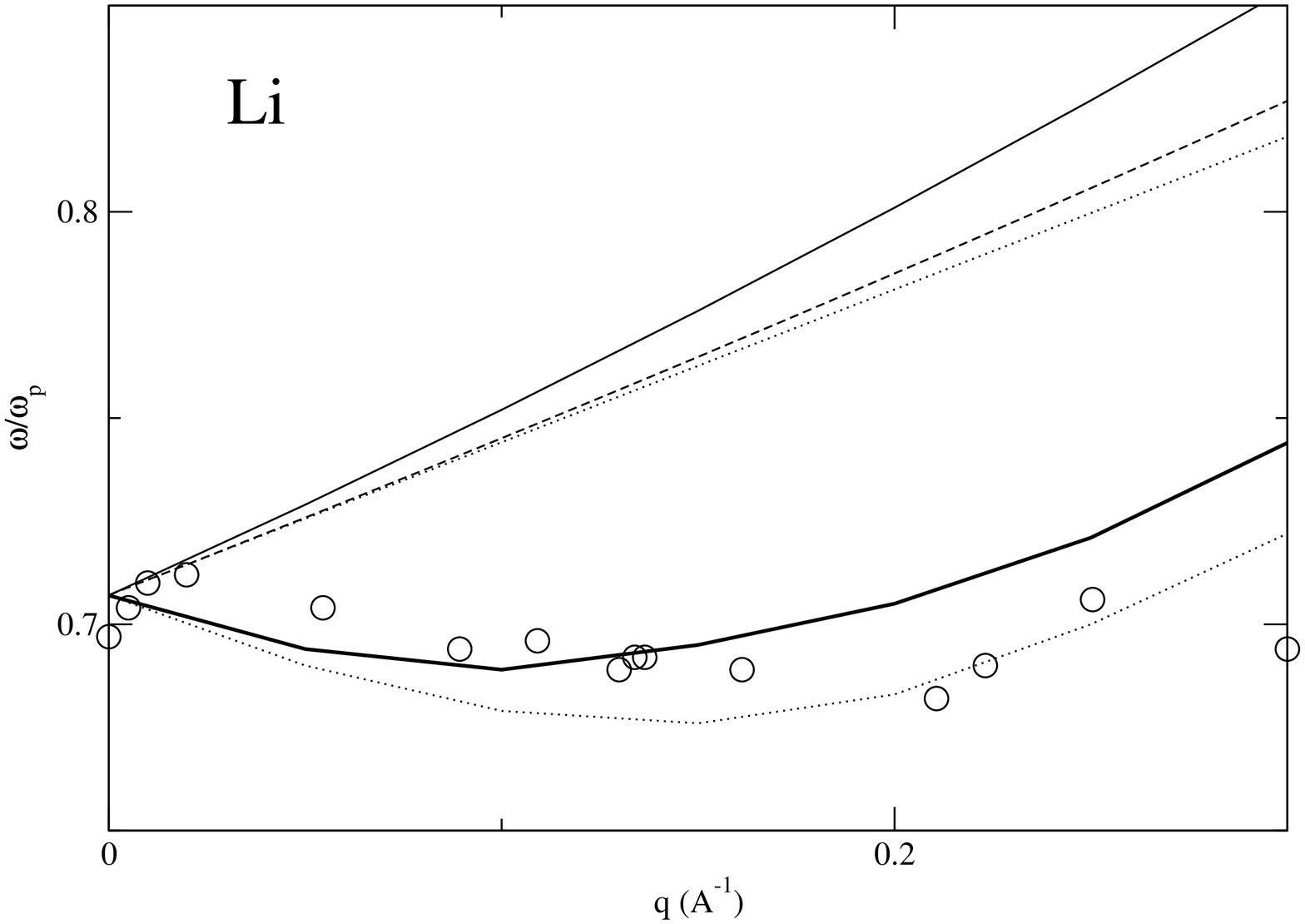}
&
\includegraphics*[width=0.5\linewidth] {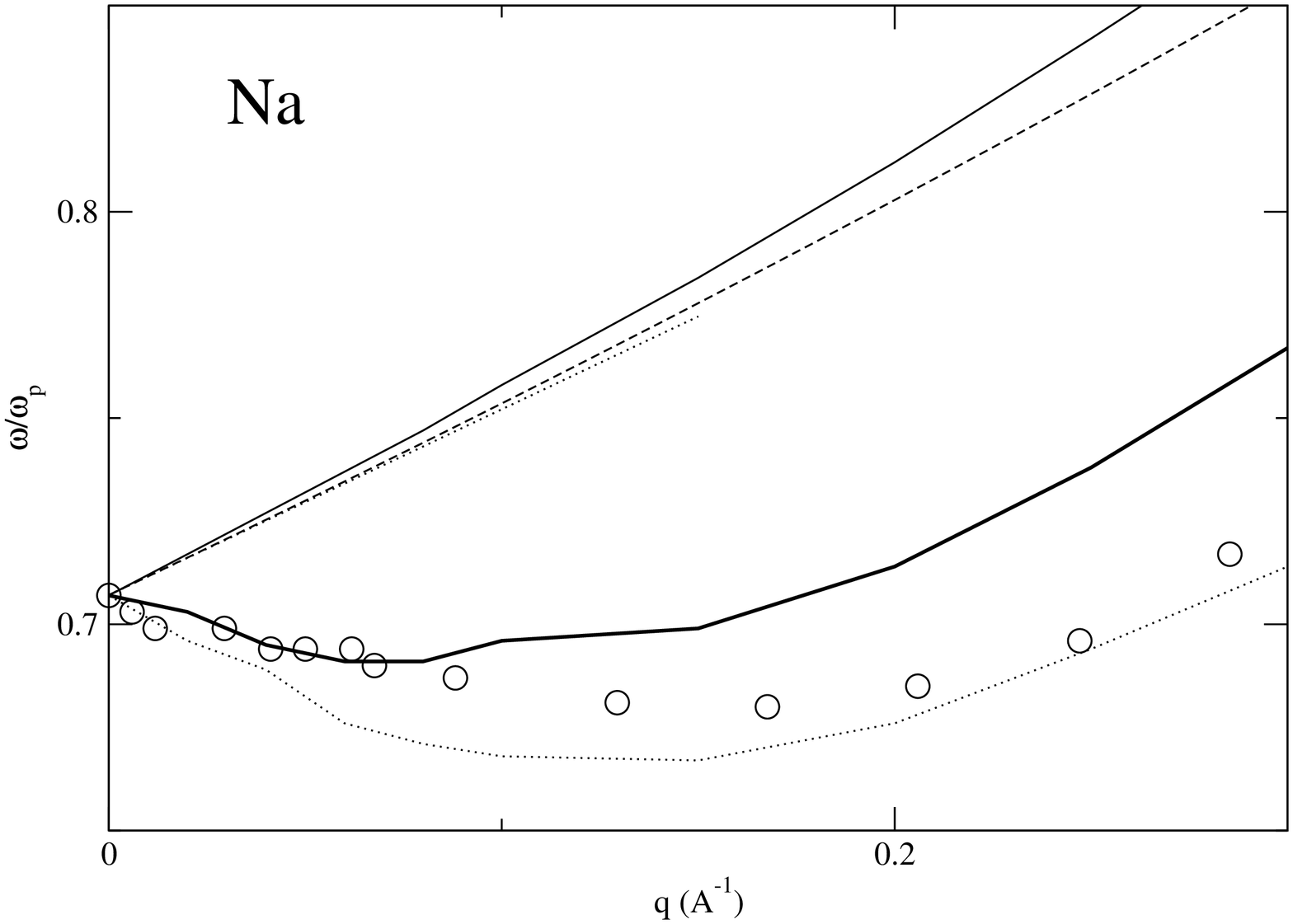}
\\
\includegraphics*[width=0.5\linewidth] {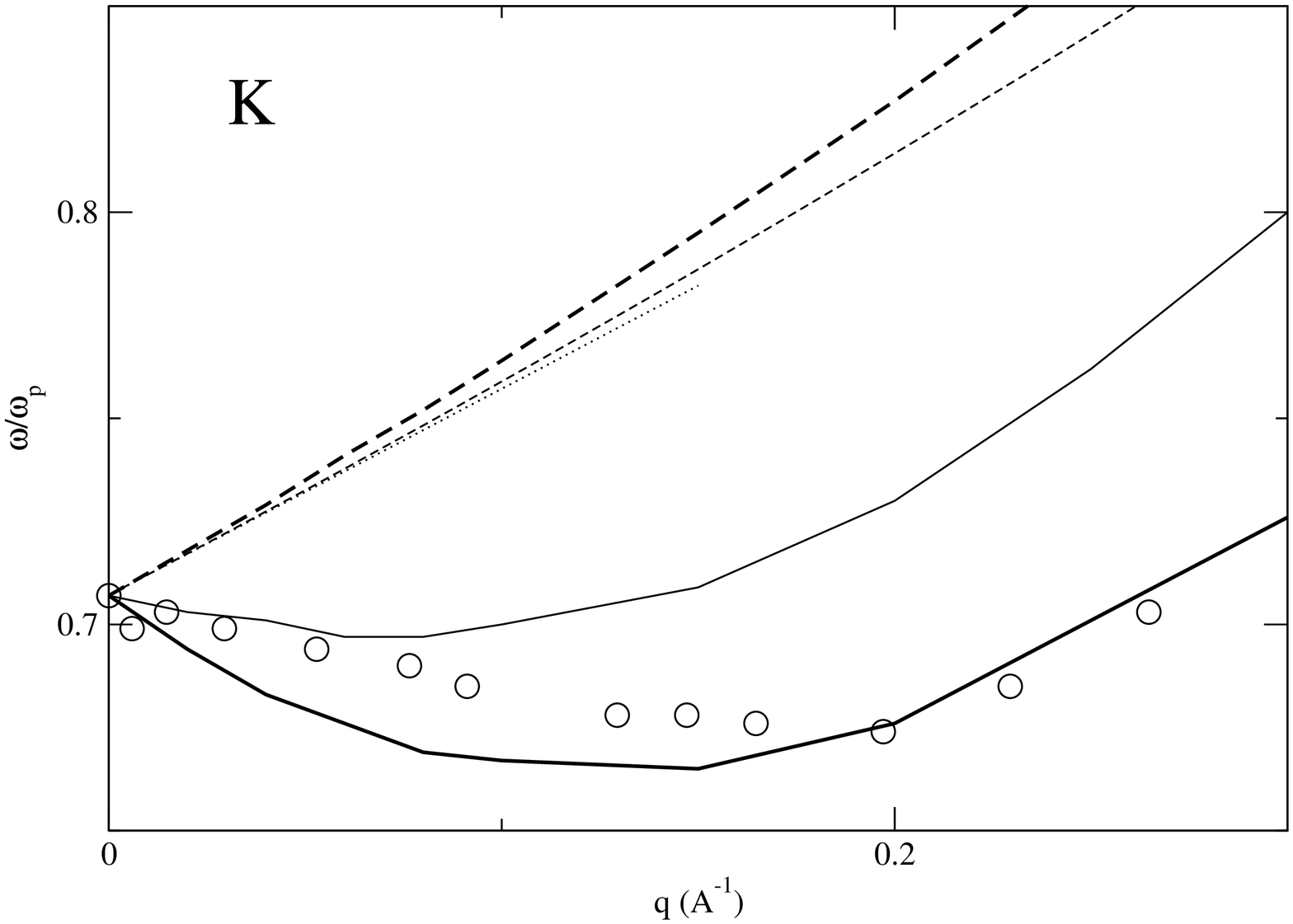}
&
\includegraphics*[width=0.5\linewidth] {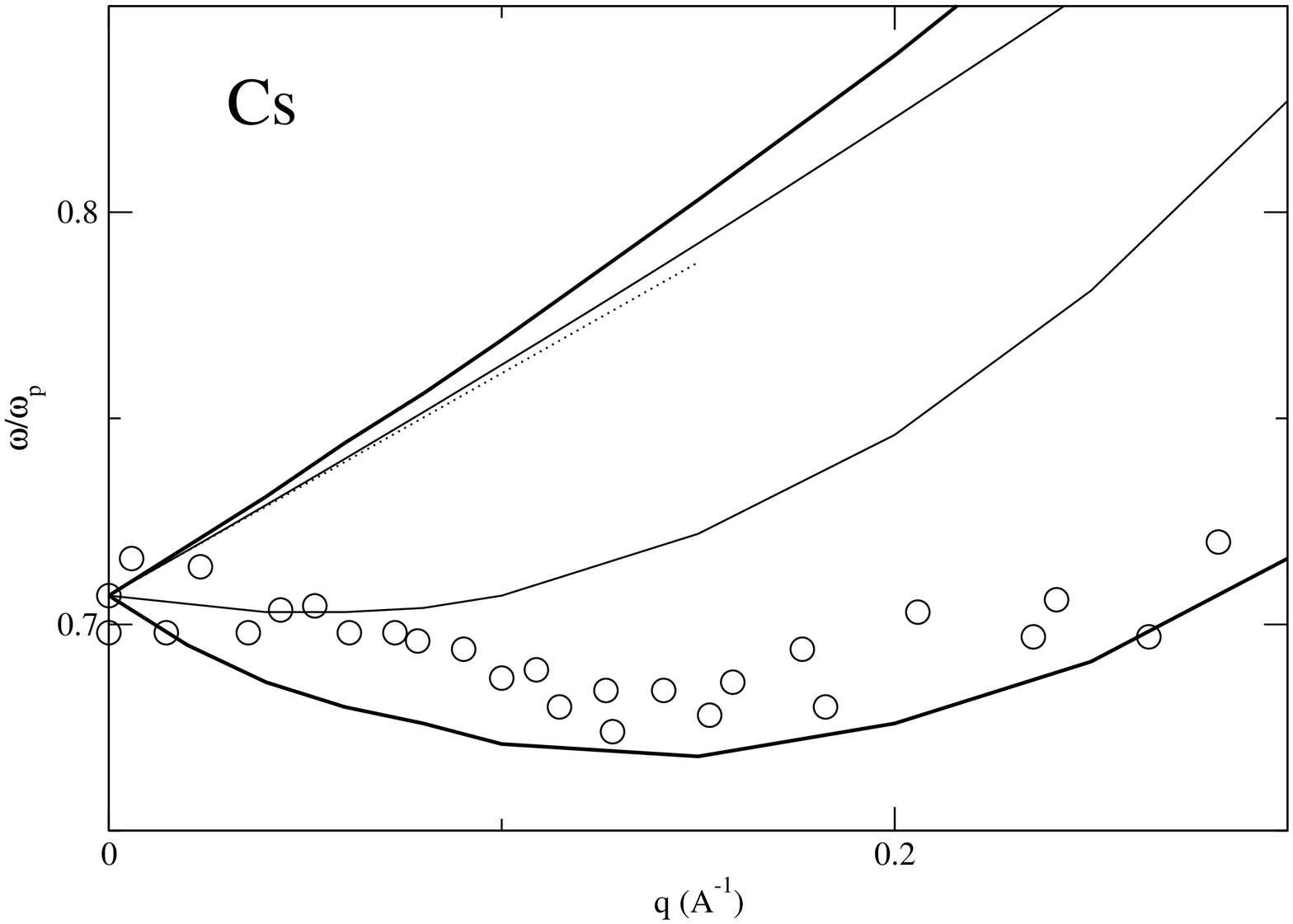}
\end{tabular}
\caption{Surface-plasmon energy dispersion for the simple metals Al,
Mg, Li,
Na, K, and Cs, as obtained from Eq.~(\ref{wqshd2}) (thin dashed lines)
and
from SRM (thick dashed lines) and self-consistent RPA (thin solid
lines) and
ALDA (thick solid lines) calculations of the surface loss function
${\rm Im}g(q,\omega)$, and from the peak positions observed in
experimental
electron energy-loss spectra (solid circles). The dotted lines
represent the
initial slope of the HD surface-plasmon energy dispersion, as obtained
from
Eq.~(\ref{hdr}. In the case of Mg, the thick solid line with open
circles
represents the jellium TDDFT calculations reported in Ref.~\cite{mg}
and
obtained with the use of the nonlocal (momentum-dependent) static XC
local-field factor of Eq.~(\ref{corradini}). All frequencies have been
normalized to the measured value $\omega_s$ of the $q=0$
surface-plasmon
energy.}\label{fig11}
\end{figure}

Figure~\ref{fig11} shows the calculated and measured dispersion of
surface plasmons in Al, Mg, Li, Na, K, and Cs. The jellium
calculations presented here do not include effects due to
band-structure effects; thus, all frequencies have been normalized
to the measured value $\omega_s$ of the $q=0$ surface-plasmon energy
(see Table~\ref{table1}). This figure shows that the single-step HD
and SRM surface-plasmon dispersions (thin and thick dashed lines,
respectively) are always upward and nearly linear. However,
self-consistent RPA and ALDA calculations (thin and thick solid
lines, respectively), which are based on a self-consistent treatment
of the surface density profile, show that the surface-plasmon
dispersion is initially downward (as also shown in Fig.~\ref{fig9}
and Table~\ref{table1}), then flattens out, and rises thereafter, in
agreement with experiment (solid circles). A comparison between
self-consistent RPA and ALDA calculations show that although there
is no qualitative difference between them XC effects tend to reduce
the surface-plasmon energy, thereby improving the agreement with the
measured plasmon frequencies. This lowering of the surface-plasmon
energies shows that dynamic XC effects combine to lower the energy
of the electron system, which is due to the weakening of the Coulomb
e-e interaction by these effects.

Recently, XC effects on the surface-plasmon dispersion of Mg and Al were
introduced
still in the framework of TDDFT (but beyond the ALDA) by using the
nonlocal
(momentum-dependent) static XC local-field factor of
Eq.~(\ref{corradini})~\cite{mg,vac}. At low wave vectors, this calculation
(thick
solid line with open circles) nearly coincides with the ALDA calculation
(thick
solid line), as expected. At larger wave vectors, however, the {\it
nonlocal}
calculation begins to deviate from the ALDA, bringing the
surface-plasmon
dispersion to nearly perfect agreement with the data for all values of
the 2D
wave vector. {\it Ab initio} calculations of the surface-plasmon
dispersion of real Al and Mg were also reported in Refs.~\cite{mg}
and~\cite{vac}. For plasmon frequencies that are
normalized to the measured value $\omega_s$ at $q=0$ (as in
Fig.~\ref{fig11}),
{\it ab initio} and jellium calculations are found to be nearly
indistinguishable; however, only the {\it ab initio} calculations account
for an overall lowering of the surface-plasmon dispersion that is due to core
polarization.

In the case of the alkali metals Li, Na, K, and Cs, there is a
mismatch in the
linear (low $q$) region of the surface-plasmon dispersion curve between jellium
ALDA calculations and the experiment. The
theoretical challenges for the future are therefore to understand the
impact
of band-structure and many-body effects on the energy of surface
plasmons in
these materials, which will require to pursue complete
first-principles
calculations of the surface electronic response at the level of those
reported
in Refs.~\cite{mg} and ~\cite{vac} for Mg and Al, respectively.

Thin films of jellium have been considered recently, in order to investigate
the influence of the slab thickness on the excitation spectra and the
surface-plasmon energy dispersion~\cite{marusik}. Oscillatory structures were
found, corresponding to electronic interband transitions, and it was concluded
that in the case of a slab thickness larger than $\sim 100\,{\rm a.u.}$
surface plasmons behave like the surface plasmon of a semi-infinite system.

\subsubsection{Surface-plasmon linewidth}

\begin{figure}
\begin{tabular}{ll}
\includegraphics*[width=0.5\linewidth] {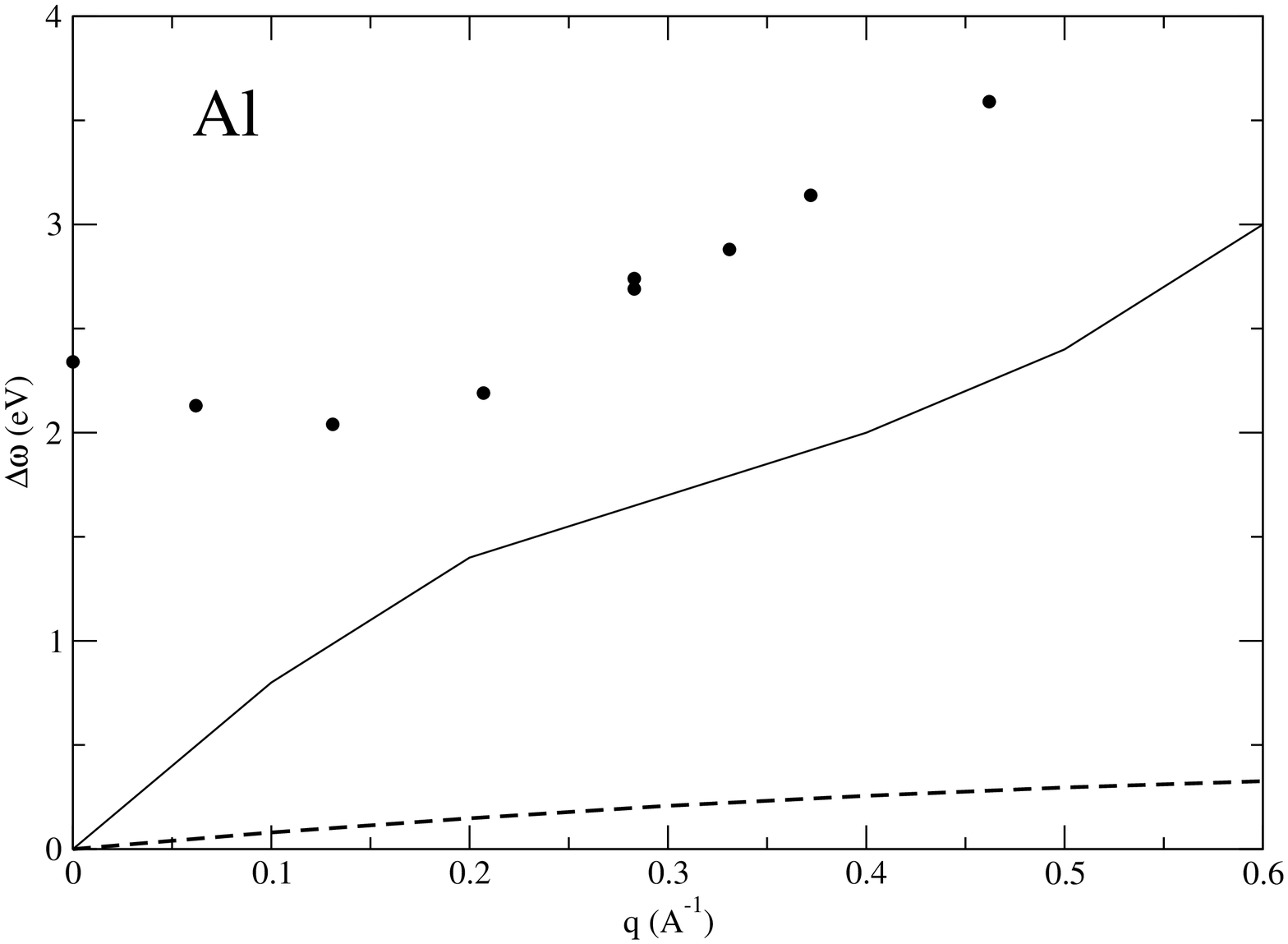}
&
\includegraphics*[width=0.5\linewidth] {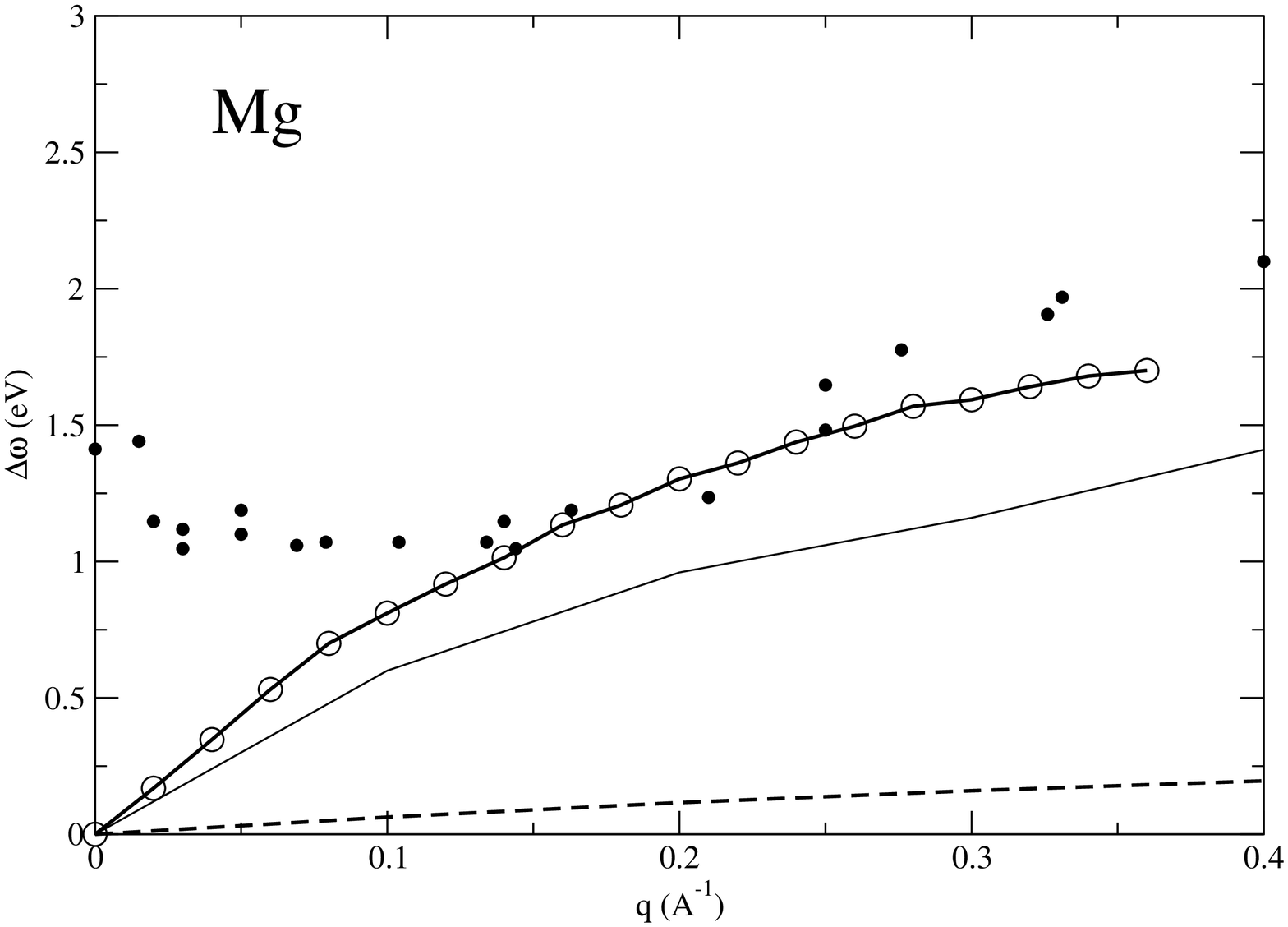}
\\
\includegraphics*[width=0.5\linewidth] {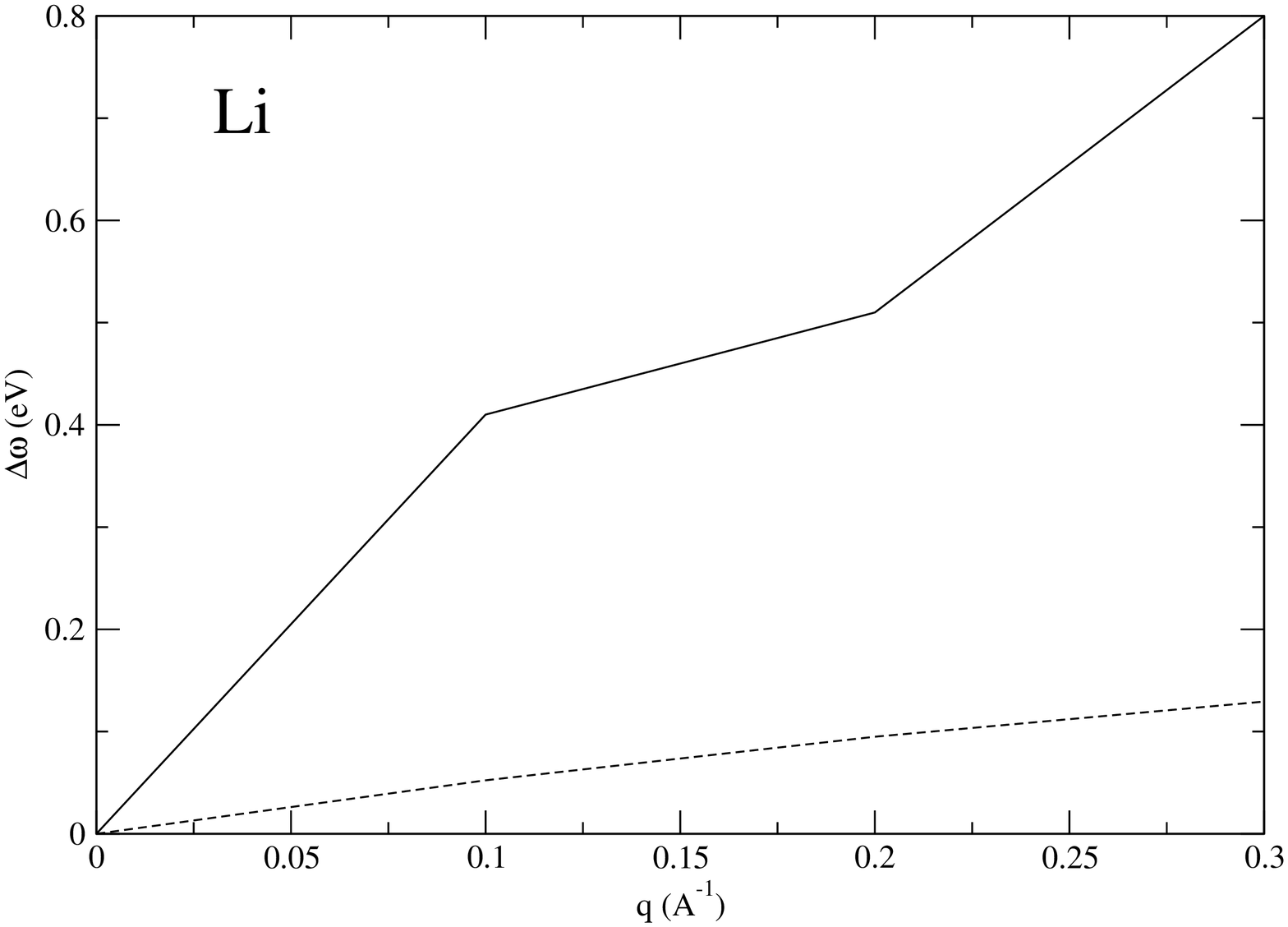}
&
\includegraphics*[width=0.5\linewidth] {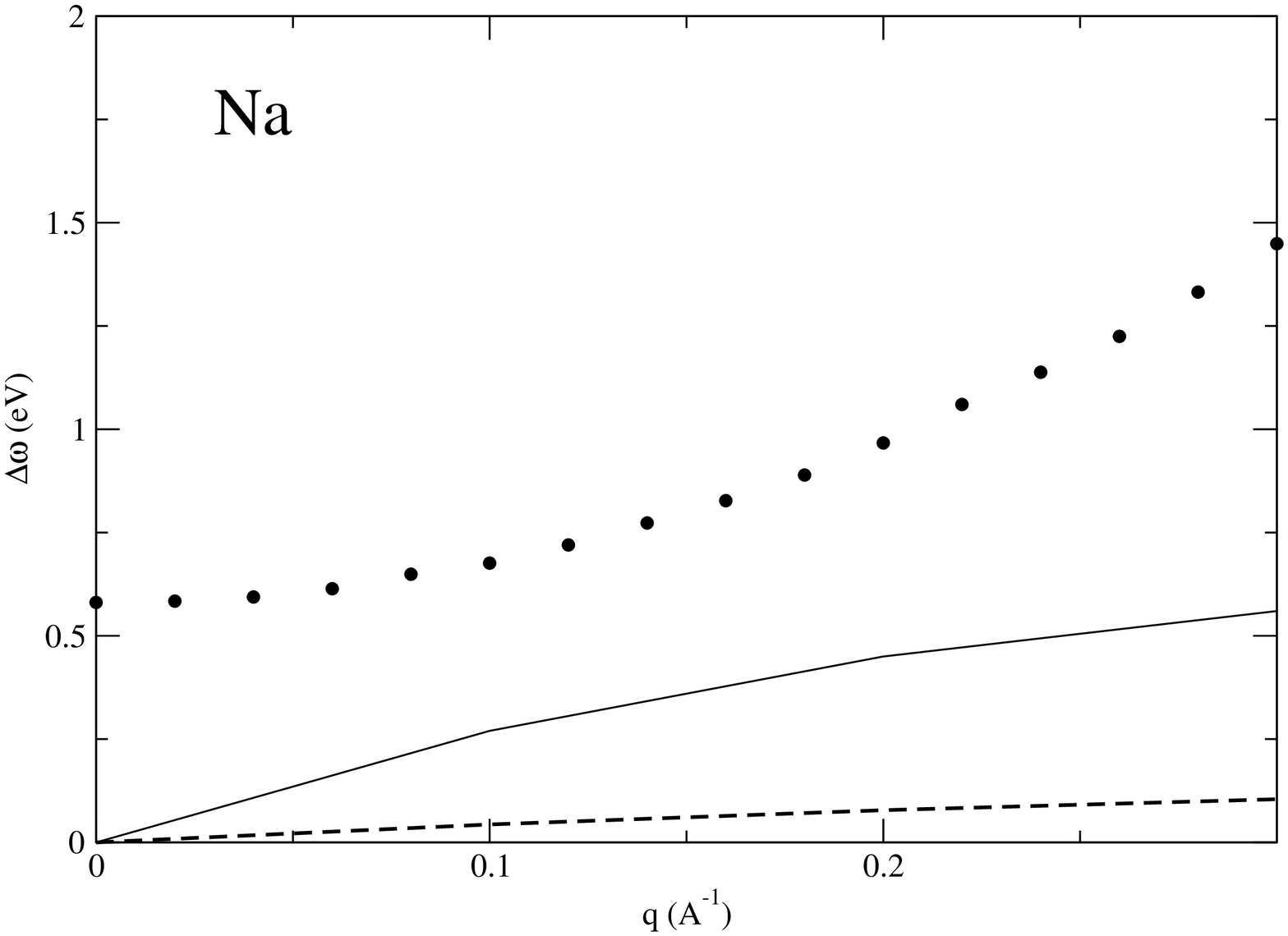}
\\
\includegraphics*[width=0.5\linewidth] {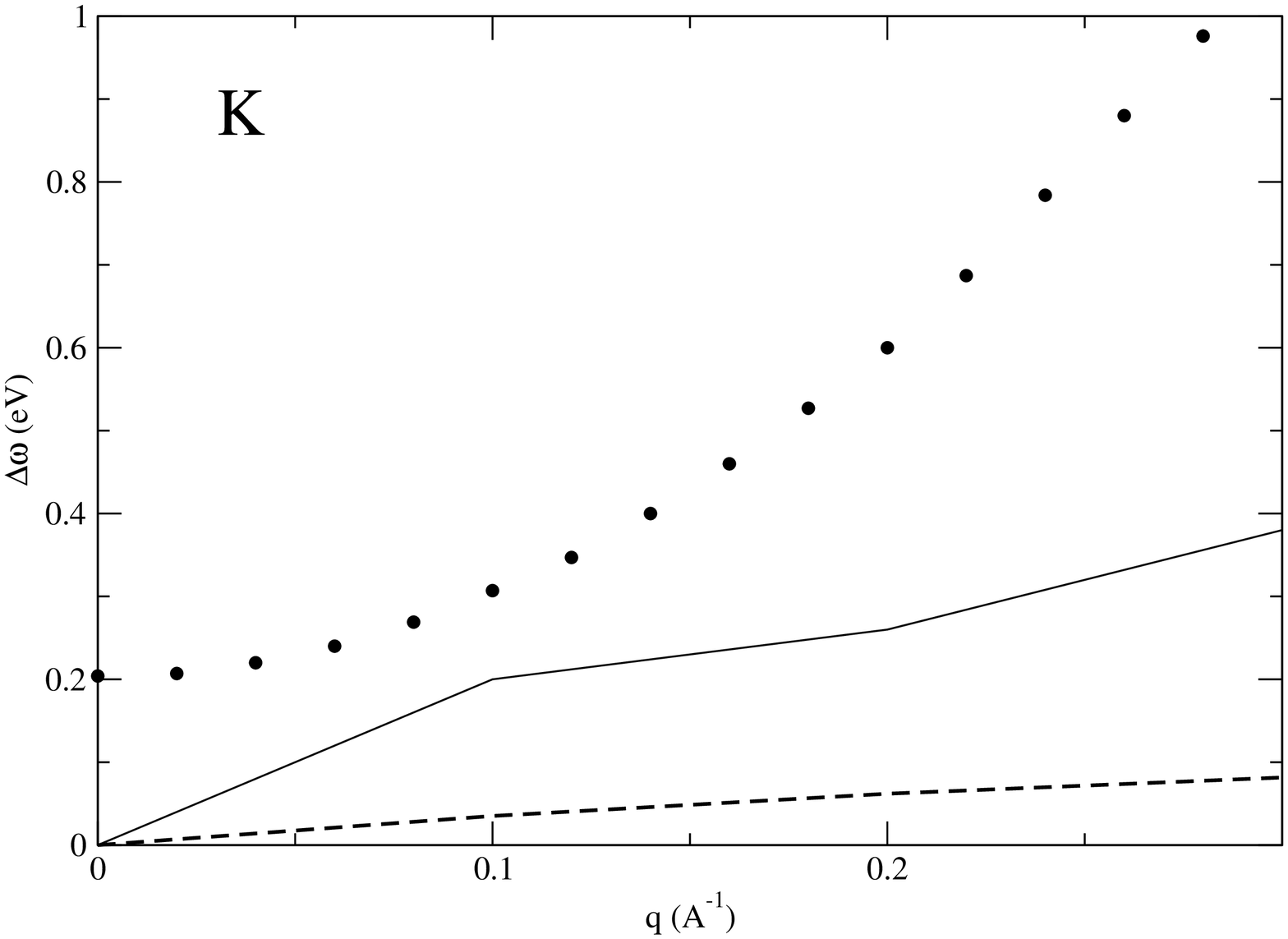}
&
\includegraphics*[width=0.5\linewidth] {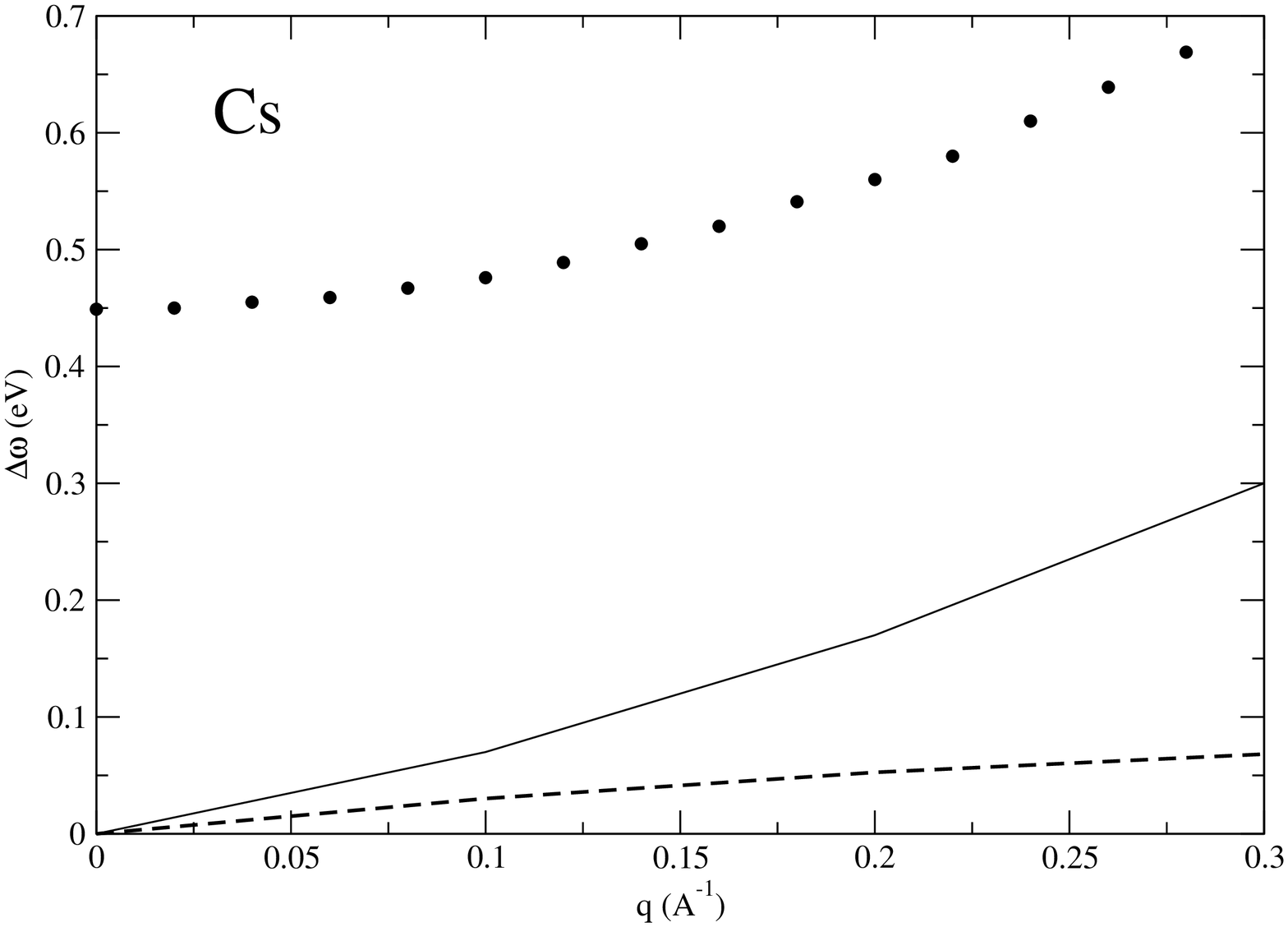}
\end{tabular}
\caption{Surface-plasmon full width at half maximum (FWHM),
$\Delta\omega$, for the simple metals Al,
Mg, Li,
Na, K, and Cs, as obtained from SRM (thick dashed lines) and
self-consistent RPA (thin solid lines) calculations of the surface
loss function ${\rm Im}g(q,\omega)$, and from the experimental
electron energy-loss spectra at different scattering angles (solid
circles) reported in Refs.~\cite{plummer1} for Na and
K, in
Ref.~\cite{plummer3} for Cs, in Ref.~\cite{plummer5} for Li and Mg,
and in
Ref.~\cite{alnew} for Al. In the case of Mg, the thick solid line with
open
circles
represents the jellium TDDFT calculations reported in Ref.~\cite{mg}
and obtained with the use of the nonlocal (momentum-dependent) static
XC local-field factor of Eq.~(\ref{corradini}).}\label{fig12}
\end{figure}

At jellium surfaces, the actual self-consistent surface response
function ${\rm Im}g(q,\omega)$ reduces in the long-wavelength
($q\to 0$) limit to Eq.~(\ref{drudeg}), so that long-wavelength
surface-plasmons are expected to be infinitely long-lived
excitations. At finite wave vectors, however, surface plasmons are
damped (even at a jellium surface) by the presence of e-h pair
excitations~\cite{nazarov1}.

Figure~\ref{fig12} shows the results that we have obtained from SRM
(dashed lines) and self-consistent (solid lines) RPA calculations of
the full width at half maximum (FWHM), $\Delta\omega$, of the
surface loss function ${\rm Im}g(q,\omega)$. Also shown in this
figure are the corresponding linewidths that have been reported in
Refs.~\cite{plummer1,plummer3,plummer5,alnew} from experimental
electron energy-loss spectra at different scattering angles. This
figure clearly shows that at real surfaces the surface-plasmon peak
is considerably wider than predicted by self-consistent RPA jellium
calculations, especially at low wave vectors. This additional
broadening should be expected to be mainly caused by the presence of
short-range many-body XC effects and interband transitions, but also
by scattering from defects and phonons.

Many-body XC effects on the surface-plasmon linewidth of Mg and Al were
incorporated in Refs.~\cite{mg} and~\cite{vac} in the framework of TDDFT with
the use of the nonlocal (momentum-dependent) static XC local-field factor of
Eq.~(\ref{corradini}). These TDDFT calculations have been plotted in
Fig.~\ref{fig12} by solid lines with circles; a comparison of these results
with the corresponding RPA calculations (solid lines without circles) shows
that short-range XC correlation effects tend to increase the finite-$q$
surface-plasmon linewidth, bringing the jellium calculations into nice
agreement with experiment at the largest values of $q$. Nevertheless, jellium
calculations cannot possibly
account for the measured surface-plasmon linewidth at small $q$,
which deviates from zero even at $q=0$.

\begin{table}
\caption{Relative widths $\Delta\omega/\omega_s$ of surface plasmons, as
derived from the imaginary part of the surface-response function of
Eq.~(\ref{long2}) with measured values of the bulk dielectric function
$\epsilon(\omega)$ (theory)~\cite{liebsch0} and from the surface-loss
measurements at $q=0$ reported in Refs.~\cite{plummer1,plummer3,plummer5,alnew}
(experiment).}
\begin{ruledtabular} \begin{tabular}{lcccccccc}
&Al&Mg&Li&Na&K&Cs&Ag&Hg\\ \hline
Theory&0.035&0.16&0.33&&0.035&&0.027&0.18\\
Experiment&0.22&0.19&0.35&&0.07&&0.027&0.16\\
\end{tabular} \end{ruledtabular} \label{table2}
\vspace{0.5 cm}
\end{table}

In the long-wavelength ($q\to 0$) limit, surface plasmons are known to be
dictated by bulk properties through a long-wavelength bulk dielectric function
$\epsilon(\omega)$, as in Eq.~(\ref{long2}). Hence, the experimental
surface-plasmon widths $\Delta\omega$ at $q=0$ should be approximately
described by using in Eq.~(\ref{long2}) the measured bulk dielectric function
$\epsilon(\omega)$. Table~\ref{table2}
exhibits the relative widths $\Delta\omega/\omega_s$ derived in this
way~\cite{liebsch0}, together with available surface-loss measurements at
$q=0$. Since silver (Ag) and mercury (Hg) have partially occupied $d$-bands, a
jellium model, like the one leading to Eq.~(\ref{long2}), is not, in
principle, appropriate to describe these surfaces. However, Table~\ref{table2}
shows that the surface-plasmon width of these solid surfaces is very well
described by introducing the measured bulk dielectric function (which includes
band-structure effects due to the presence of $d$-electrons) into
Eq.~(\ref{long2}). Nevertheless, the surface-plasmon widths of simple metals
like K and Al, with no $d$-electrons, are considerably larger than predicted
in this simple way~\cite{beck3}. This shows that an understanding of
surface-plasmon broadening mechanisms requires a careful analysis of the
actual band structure of the solid.

Approximate treatments of the impact of the band structure on the
surface-plasmon energy dispersion have been developed by several
authors, but a first-principles description of the surface-plasmon
energy dispersion and linewidth has been reported only in the case
of the simple-metal prototype surfaces Mg(0001)~\cite{mg} and
Al(111)~\cite{vac}; these calculations will be discussed in
Section~\ref{abinitio}.

\subsubsection{Multipole surface plasmons}

In his attempt to incorporate the smoothly decreasing electron density profile
at the surface, Bennett~\cite{bennett} solved the equations of a simple
hydrodynamic model
with a density profile which decreases linearly through the surface region and
found that in addition to Ritchie's surface plasmon at $\omega\sim\omega_s$,
with a negative energy dispersion at low ${\bf q}$ wave vectors, there is an
upper surface plasmon at higher energies. This is the so-called
multipole surface plasmon, which shows a positive wave vector dispersion even
at small ${\bf q}$.

The possible existence and properties of multipole surface plasmons was later
investigated in the framework of hydrodynamical models for various choices of
the electron-density profile at the surface~\cite{eg1,eg2,ss1}. According to
these calculations, higher multipole excitations could indeed exist, for a
sufficiently diffuse surface, in addition to the usual surface plasmon at
$\omega\sim\omega_s$. However, approximate quantum-mechanical RPA calculations
gave no evidence for the existence of multipole surface
plasmons~\cite{beck2,inglesvw}, thereby leading to the speculation that
multipole surface plasmons might be an artifact of the hydrodynamic
approximation~\cite{ss1,ap0}.

The first experimental sign for the existence of multipole surface
modes was established by Schwartz and Schaich~\cite{ss2} in their
theoretical analysis of the photoemission yield spectra that had
been reported by Levinson {\it et al.}~\cite{levinson}. Later on,
Dobson and Harris~\cite{dh} used a DFT scheme to describe
realistically the electron-density response at a jellium surface, to
conclude that multipole surface plasmons should be expected to exist
even for a high-density metal such as Al (which presents a
considerably abrupt electron density profile at the surface). Two
years later, direct experimental evidence of the existence of
multipole surface plasmons was presented in inelastic reflection
electron scattering experiments on smooth films of the low-density
metals Na, K, and Cs~\cite{plummer2}, the intensity of these multipole surface
plasmons being in agreement with the DFT calculations reported later by
Nazarov~\cite{nazarov2}. The Al multipole surface
plasmon has been detected only recently by means of angle-resolved
high-resolution electron-energy-loss spectroscopy
(HREELS)~\cite{alnew}.

\begin{figure}
\includegraphics[width=0.95\linewidth]
{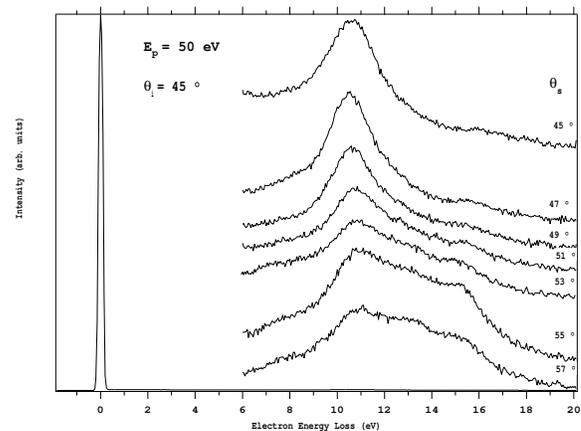}
\caption{Angle-resolved high-resolution electron energy-loss spectra of the
Al(111) surface at various scattering angles $\theta_s$. The primary beam
energy is $50\,{\rm eV}$ and the incident angle is $\theta_i=45^0$ (from
Ref.~\cite{alnew}, used with permission).}\label{fig13}
\end{figure}

\begin{figure}
\includegraphics[width=0.95\linewidth]
{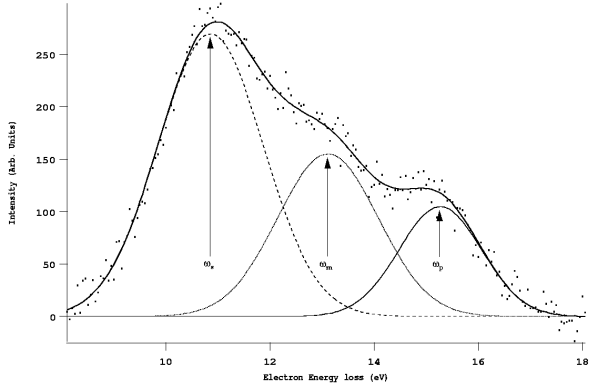}
\caption{Angle-resolved high-resolution electron energy-loss spectra of the
Al(111) surface at $\theta_s=53^0$, for $\theta_i=45^0$ and
$\varepsilon_i=50\,{\rm eV}$ (as in Fig.~\ref{fig13}) but now together with
the deconvolution into contributions corresponding to the excitation of the
conventional surface plasmon at $\omega_s=10.55\,{\rm eV}$, multipole surface
plasmon at $13.20\,{\rm eV}$, and bulk plasmon at $\omega_p=15.34\,{\rm eV}$
(from Ref.~\cite{alnew}, used with permission).}\label{fig14}
\end{figure}

Figure~\ref{fig13} shows the loss spectra of the Al(111) surface, as obtained
by Chiarello {\it al.}~\cite{alnew} with an incident electron energy of
$50\,{\rm eV}$ and an incident angle of $45^0$ with respect to the surface
normal. The loss spectrum obtained in the specular geometry ($\theta_s=45^0$)
is characterized mainly by a single peak at the conventional surface-plasmon
energy $\omega_s=10.55\,{\rm eV}$. For off-specular scattering angles
($\theta_s\neq 45^0$), the conventional surface plasmon exhibits a clear
energy dispersion and two other features arise in the loss spectra: the
multipole surface plasmon and the bulk plasmon. The loss spectrum obtained at
$\theta_s=53^0$, which corresponds to $q=0$ [see Eq.~(\ref{momq})] is
represented again in Fig.~\ref{fig14}, but now together with the background
substraction and Gaussian-fitting procedure reported in
Ref.~\cite{alnew}. The peak at $\omega_p=15.34$ corresponds to the excitation
of the Al bulk plasmon, and the multipole surface plasmon is located at
$13.20\,{\rm eV}$.

\begin{table}
\caption{Angle-resolved low-energy inelastic electron scattering measurements
of the energy $\omega_m$ and width $\Delta\omega$ of multipole surface
plasmons at $q=0$, as reported in Refs.~\cite{plummer2} for Na and
K, in Ref.~\cite{plummer3} for Cs, in Ref.~\cite{plummer5} for Li and Mg,
and in Ref.~\cite{alnew} for Al. Also shown is the ratio $\omega_m/\omega_p$
calculated in the RPA and ALDA and reported in Ref.~\cite{plummer3}.}
\begin{ruledtabular} \begin{tabular}{lcccccccc}
&$r_s$&$\omega_m$ (eV)&$\omega_m/\omega_p$&RPA&ALDA&
$\Delta\omega$ (eV)&$\Delta\omega/\omega_p$&\\ \hline
Al&2.07&13.20&0.86&0.821&0.782&2.1&0.14\\
Mg&2.66&&&0.825&0.784&&\\
Li&3.25&&&0.833&0.789&&\\
Na&3.93&4.67&0.81&0.849&0.798&1.23&0.21\\
K&4.86&3.20&0.84&0.883&0.814&0.68&0.18\\
Cs&5.62&2.40&0.83&0.914&0.837&0.64&0.22\\
\end{tabular} \end{ruledtabular} \label{table3}
\end{table}

The calculated and measured energies and linewidths of long-wavelength
($q\to 0$) multipole surface plasmons in simple metals are given in
Table~\ref{table3}. On the whole, the ratio $\omega_m/\omega_p$ agrees with
ALDA calculations. Good agreement between ALDA calculations and experiment is
also obtained for the entire dispersion of multipole surface plasmons, which
is found to be approximately linear and positive. This positive dispersion is
originated in the fact that the centroid of the induced electron density,
which at $\omega\sim\omega_s$ is located outside the jellium edge, is shifted
into the metal at the multipole resonance frequency $\omega_m$. We also note
that multipole surface plasmons have only been observed at wave vectors well
below the cutoff value for Landau damping, which has been argued to be due to
an interplay between Coulomb and kinetic energies~\cite{bar1}.

\begin{figure}
\includegraphics[width=0.95\linewidth]
{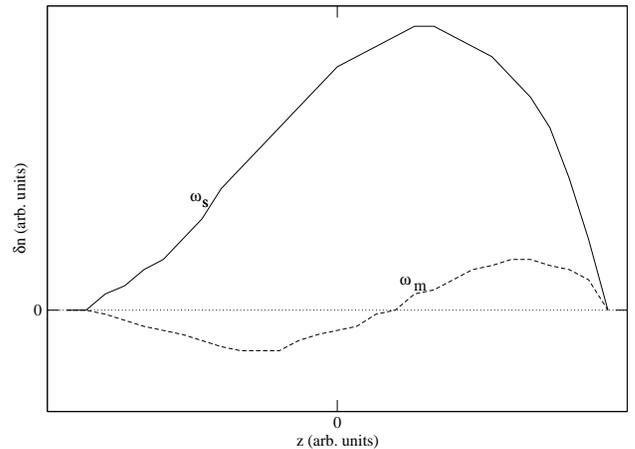}
\caption{Real part of the electron density induced at $\omega=\omega_s$ (solid
line) and $\omega=\omega_m$ (dashed line), as reported in Ref.~\cite{liebsch0}
for a model unperturbed electron density profile with linear decay over a
finite region. }\label{fig15}
\end{figure}

In a semi-infinite metal consisting of an abrupt step of the unperturbed
electron density at the surface (as in the hydrodynamic and
specular-reflection models described above), surface plasmons would be
localized at the surface and would propagate like plane waves, as
shown in Fig.~\ref{fig0}, with positive and negative surface charge regions
alternating periodically. In the real situation in which the electron density
decays smoothly at the surface, surface plasmons also propagate along the
surface as illustrated in Fig.~\ref{fig0}, but the finite width of the
electron-density profile yields fluctuating densities with finite widths, as
illustrated in Fig.~\ref{fig15} for a model surface electron density profile
with linear decay over a finite region.

Figure~\ref{fig15} qualitatively represents the real situation in which apart
from small Friedel oscillations the distribution of the conventional surface
plasmon at $\omega_s$ consists in the direction normal to the surface of a
single peak (i.e., it has a {\it monopole} character), while the charge
distribution of the upper mode at $\omega_m$ has decreasing oscillating
amplitude towards the interior of the metal, i.e., it has a {\it multipole}
character. Along the direction normal to the surface, the electronic density
associated with this multipole surface plasmon integrates to zero.

In the retarded region, where $q<\omega_s/c$, the surface-plasmon dispersion
curve deviates from the non-retarded limit (where $q>>\omega_s/c$) and
approaches the light line $\omega=cq$, as shown by the lower solid line of
Fig.~\ref{fig2}, thus going to zero at $q=0$; hence, in a light experiment the
external radiation dispersion line will never intersect the surface-plasmon
line, i.e., in an ideal flat surface the conventional {\it monopole} surface
plasmon cannot be excited in a photoyield experiment. However, the multipole
surface plasmon dispersion curve crosses the light line at $q\sim\omega_m/c$
and goes to $\omega_m$ at $q=0$. Consequently, angle- and energy-resolved
photoyield experiments are suitable to identify the multipole surface plasmon.
A large increase in the surface photoyield was observed from K and Rb at
$\omega_m=3.15$ and $2.84\,{\rm eV}$, respectively~\cite{monin}, from
Al(100) at $\omega_m=12.5\,{\rm eV}$~\cite{levinson}, and from Al(111) at
$\omega_m=13\,{\rm eV}$~\cite{barman1}, in nice agreement with the
multipole-plasmon energy observed with the HREELS technique by Chiarello {\it
et al.}~\cite{alnew}. HREELS measurements yield, however, a FWHM of $2.1\,{\rm
eV}$ for the $q=0$ multipole surface plasmon in Al(111)~\cite{alnew}, which is
considerably smaller than the FWHM of $3\,{\rm eV}$ measured  by photoyield
experiments~\cite{barman1}.

\subsection{Surface plasmons: real surfaces}

\subsubsection{Stabilized jellium model}

A simple way of including approximately the lattice potential that is absent in
the jellium model is the so-called stabilized jellium or structureless
pseudopotential model~\cite{stm1,stm2}, which yields energy stability against
changes in the background density.

In this model, a solid surface is assumed to be translationally invariant in
the plane of the surface, as in the jellium model. Hence, single-particle wave
functions can be separated as in Eq.~(\ref{psi0}) into a plane wave along the
surface and a component $\psi_i(z)$ describing motion normal to the surface.
In the framework of DFT and TDDFT, this component is obtained by solving
self-consistently a Kohn-Sham equation of the form of Eq.~(\ref{ks1}) but with
the effective Kohn-Sham potential of Eq.~(\ref{ks0}) being replaced by
\begin{equation}\label{ks00}
v_{KS}[n_0](z)=v_H[n_0](z)+v_{xc}[n_0](z)+<\delta v>_{WS},
\end{equation}
$<\delta v>_{WS}$ representing the difference between a local pseudopotential
and the jellium potential:
\begin{equation}
<\delta v>_{WS}={3r_c^2\over 2r_s^3}-{3Z^{2/3}\over 10r_s},
\end{equation}
where $Z$ is the chemical valence of the solid and $r_c$ is a core radius that
is chosen to stabilize the metal for given values of the parameters $r_s$ and
$Z$.

The stabilized jellium model was used by Ishida and Liebsch~\cite{ishida} to
carry out RPA and ALDA calculations of the dispersion of the energy and
linewidth of surface plasmons in Mg and Li. It is well known that the
stabilized jellium model gives considerably better work functions and surface
energies than the standard jellium model; however, the impact of a
structureless pseudopotential on the properties of surface plasmons is found
to be very small. In particular, this simple pseudopotential cannot explain
the presence of core polarization lowering the surface-plasmon frequency for
all wave vectors and does not account for the fact that the measured $q$
dependence of the Li surface-plasmon energy is very much flatter than
calculated within the jellium model (see Fig.~\ref{fig11}). This discrepancy
(not present in the case of Al and Mg) is attributed to the presence of an
interband transition in Li at $3.2\,{\rm eV}$, which is only slightly below
the measured surface-plasmon energy at $q=0$ ($\omega_s=4.28\,{\rm eV}$) and
allows, therefore, for interference between bulk single-particle and surface
collective modes, as discussed in Ref.~\cite{ishida}

\subsubsection{Occupied $d$-bands: simple models}

\begin{table}
\caption{Measured values of the surface-plasmon energy $\omega_s$ at
$q=0$ and the long-wavelength surface-plasmon dispersion coefficient
$\alpha$ of Eq.~(\ref{alpha}) for the noble metal Ag and the
transition metals Hg and Pd. Also shown in this table is the
coefficient $\alpha$  obtained from RPA and ALDA calculations of the
centroid of the electron density induced at $\omega=\omega_s$ [see
Eq.~(\ref{alpha})] in a homogeneous electron gas with the electron
density equal to that of valence $sp$ electrons in Ag and Hg. Note
that the measured values of the surface-plasmon energy $\omega_s$ at
$q=0$ are considerably below the jellium prediction:
$\omega_p/\sqrt{2}=\sqrt{3/2r_s^3}e^2/a_0$, mainly due to the
presence of $d$ electrons. $\omega_s$ and $\alpha$ are given in eV
and $\AA$, respectively.}
\begin{ruledtabular} \begin{tabular}{lccccccccc}
&$r_s$&$\omega_s$&Direction&$\alpha^{exp}$&$\alpha^{RPA}$&$\alpha^{ALDA}$\\
\hline
Ag(100)&3.02&3.7&                  & 0.377&-0.370&-0.609\\
Ag(111)&3.02&3.7&                  & 0.162&-0.370&-0.609\\
Ag(110)&3.02&3.7&$<100>$           & 0.305&-0.370&-0.609\\
       &    &   &$<1\overline{1}0>$& 0.114&-0.370&-0.609\\
Hg     &2.65&6.9&                  &-0.167&-0.32&-0.50\\
Pd     &    &7.4&                  &-1.02 &\\
\end{tabular} \end{ruledtabular} \label{table4}
\end{table}

Significant deviations from the dispersion of surface plasmons at jellium
surfaces occur on the noble metal Ag~\cite{rocca1,rocca2,rocca3,rocca4}, and
the transition metals Hg~\cite{plummer6} and Pd~\cite{pd}
(see Table~\ref{table4}). These deviations are mainly due to the presence in
these metals of filled $4d$ and $5d$ bands, which in the case of Ag yields an
anomalous positive and strongly crystal-face dependent dispersion.

\begin{figure}
\includegraphics[width=0.95\linewidth]
{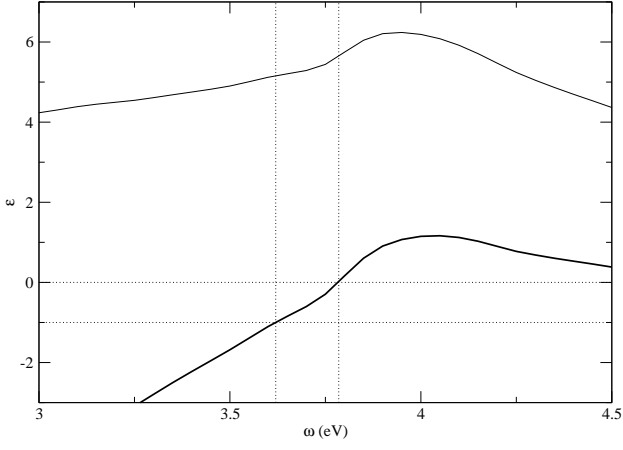}
\caption{The thin solid line represents the real part of the $d$ band
contribution $\epsilon_d(\omega)$ to the measured optical dielectric function
of Ag~\cite{jc}. The thick solid line represents $\epsilon(q
\omega)+\epsilon_d(\omega)-1$ at $q=0$, i.e.,
$\epsilon_d(\omega)-\omega_p^2/\omega^2$. The vertical dotted lines represent
the energies at which the plasmon conditions of Eqs.~(\ref{zerod}) and
(\ref{zeros}) are fulfilled: $\omega_p'=3.78\,{\rm eV}$ and
$\omega_s'=3.62\,{\rm eV}$, respectively. At these frequencies:
$\epsilon_d(\omega_p')=5.65$ and $\epsilon_d(\omega_p)=5.15$.}\label{fig16}
\end{figure}

\paragraph{Ag.} In order to describe the observed features of Ag surface
plasmons, Liebsch~\cite{ag3} considered a self-consistent jellium model for
valence $5s$ electrons and accounted for the presence of occupied $4d$ bands
via a polarizable background of $d$ electrons characterized by a local
dielectric function $\epsilon_d(\omega)$ that can be taken from bulk optical
data (see Fig.~\ref{fig16})~\cite{jc}.

For a homogeneous electron gas, one simply replaces in this model the
bare Coulomb interaction $v({\bf r},{\bf r}')$ by
\begin{equation}\label{vnew}
v'({\bf r},{\bf r}';\omega)=v({\bf r},{\bf r}')\,\epsilon_d^{-1}(\omega),
\end{equation}
which due to translational invariance yields the following expression for the
Fourier transform $W'(q,\omega)$ of the screened interaction
$W'({\bf r},{\bf r}';\omega)$ of the form of Eq.~(\ref{screened0}):
\begin{equation}
W'(q,\omega)={v(q)\over\epsilon(q,\omega)+\epsilon_d(\omega)-1},
\end{equation}
with $\epsilon(q,\omega)$ being the dielectric function of a homogeneous system
of $sp$ valence electrons ($5s$ electrons in the case of Ag). Hence, the
screened interaction $W'(q,\omega)$ has poles at the bulk-plasmon condition
\begin{equation}\label{zerod}
\epsilon(q,\omega)+\epsilon_d(\omega)-1=0,
\end{equation}
which in the case of Ag ($r_s=2.02$) and in the absence of $d$ electrons
($\epsilon_d=1$) yields the long-wavelength ($q\to 0$) bulk plasmon energy
$\omega_p=8.98\,{\rm eV}$. Instead, if $d$ electrons are characterized by the
frequency-dependent dielectric function $\epsilon_d(\omega)$ represented by a
thin solid line in Fig.~\ref{fig16}, Eq.~(\ref{zerod}) yields the observed
long-wavelength bulk plasmon at
\begin{equation}
\omega_p'={\omega_p\over\sqrt{\epsilon_d(\omega_p')}}=3.78\,{\rm eV}.
\end{equation}

In the case of a solid surface, one still uses a modified ($d$-screened)
Coulomb interaction $v'({\bf r},{\bf r}';\omega)$ of the form of
Eq.~(\ref{vnew}), but with $\epsilon_d(\omega)$ being replaced by
\begin{equation}\label{epsd}
\epsilon_d(z,\omega)=\cases{\epsilon_d(\omega),&$z\leq z_d$ \cr\cr
1, & $z > z_d$,}
\end{equation}
which represents a polarizable background of $d$ electrons that extends up to a
certain plane at $z=z_d$. Using Eq.~(\ref{epsd}), the 2D Fourier transform of
$v'({\bf r},{\bf r}';\omega)$ yields~\cite{catalina1}:
\begin{eqnarray}\label{ll}
v'(z,z';{\bf q}_\parallel,\omega)&=&\frac{2\pi}
{q_\parallel\,\epsilon_d(z',\omega)}\,
[{\rm e}^{-q_\parallel\,|z-z'|}+{\rm sgn}(z_d-z')\nonumber\\
&\times&\sigma_d(\omega)\,{\rm e}^{-q_\parallel|z-z_d|}{\rm
e}^{-q_\parallel|z_d-z'|}],
\end{eqnarray}
where
\begin{equation}\label{lll}
\sigma_d=[\epsilon_d(\omega)-1]/[\epsilon_d(\omega)+1].
\end{equation}
Introduction of Eqs.~(\ref{ll}) and (\ref{lll}) into and equation of the form
of Eq.~(\ref{wsurf}) yields the following expression for the modified RPA
surface-response function:
\begin{equation}\label{gm1}
g'(q,\omega)=\int dz\,{\rm e}^{qz}\,{\delta n(z;q,\omega)\over\epsilon_d(z
\omega)}+a(q,\omega),
\end{equation}
where $\delta n(z;q,\omega)$ represents the RPA induced density of $sp$ valence
electrons [which is given by Eq.~(\ref{eqa1new2}) with
$f_{xc}[n_0](z,z';q,\omega)=0$], and
\begin{eqnarray}\label{gm2}
&&a(q,\omega)=\sigma_d(\omega)\left[{\rm e}^{qz_d}\right.\cr\cr
&&\left.+\int dz{\rm e}^{-q|z_d-z|}{\rm sgn}(z_d-z)
{\delta n(z;q,\omega)\over\epsilon_d(z,\omega)}\right].
\end{eqnarray}

In the long-wavelength ($q\to 0$) limit, Eq.~(\ref{gm1}) takes the form of
Eq.~(\ref{long2}) but with $\epsilon(\omega)$ replaced by
$\epsilon(\omega)+\epsilon_d(\omega)-1$, which leads to the surface-plasmon
condition:
\begin{equation}\label{zeros}
\epsilon(\omega)+\epsilon_d(\omega)=0.
\end{equation}
For a Drude dielectric function $\epsilon(\omega)$ with
$\omega_p=8.98\,{\rm eV}$ ($r_s=3.02$) and in the absence of $d$ electrons
($\epsilon_d=1$), the surface-plasmon condition of Eq.~(\ref{zeros}) yields
the surface-plasmon energy $\omega_s=6.35\,{\rm eV}$. However, in the presence
of $d$ electrons characterized by the frequency-dependent dielectric function
$\epsilon_d(\omega)$, Eq.~(\ref{zeros}) leads to a modified ($d$-screened)
surface plasmon at
\begin{equation}\label{omegasp}
\omega_s'={\omega_p\over\sqrt{1+\epsilon_d(\omega_s')}}=3.62\,{\rm eV},
\end{equation}
which for $\omega_p=8.98\,{\rm eV}$ and the dielectric function
$\epsilon_d(\omega)$ represented by a thin solid line in Fig.~\ref{fig16}
yields $\omega_s'=3.62\,{\rm eV}$, only slightly below the energy
$3.7\,{\rm eV}$ of the measured Ag surface plasmon~\cite{rocca3}.

At finite wavelengths, the surface-plasmon dispersion was derived by Liebsch
from the peak positions of the imaginary part of the surface-response function
of Eq.~(\ref{gm1}), as obtained with a self-consistent RPA calculation of the
induced density $\delta n(z;q,\omega)$ of $5s^1$ valence electrons. The
surface-plasmon dispersion was found to be positive for $z_d\leq 0$ and to
best reproduce the observed linear dispersion of Ag surface plasmons for
$z_d=-0.8\,\rm{\AA}$. Hence, one finds that (i) the $s$-$d$ screened
interaction is responsible for the lowering of the surface-plasmon energy at
$q=0$ from the free-electron value of $6.35\,{\rm eV}$ to $3.62\,{\rm eV}$,
and (ii) the observed blueshift of the surface-plasmon frequency at
increasing $q$ can be interpreted as a reduction of the $s$-$d$ screened
interaction in the "selvedge" region that is due to a decreasing penetration
depth of the induced electric field.

With the aim of describing the strongly crystal-face dependence of the
surface-plasmon energy dispersion in Ag, Feibelman~\cite{ag2} thought of the
Ag surface plasmon as a collective mode that is split off the bottom of the
$4d$-to-$5s$ electron-hole excitation band, and considered the relation
between the surface-plasmon dispersion and the $4d$-to-$5s$ excitations
induced by the surface-plasmon's field. He calculated the $s$-$d$ matrix
elements using a one-dimensional surface perturbation of the jellium Lang-Kohn
potential, and he argued that for Ag(100) the dispersion coefficient is
increased relative to Ag(111) because in the case of Ag(100) the $4d$
electrons lie closer to the centroid of the oscillating free-electron
charge, and the probability of $4d$-to-$5s$ excitation is, therefore,
enhanced.

\begin{figure}
\includegraphics[width=0.95\linewidth]
{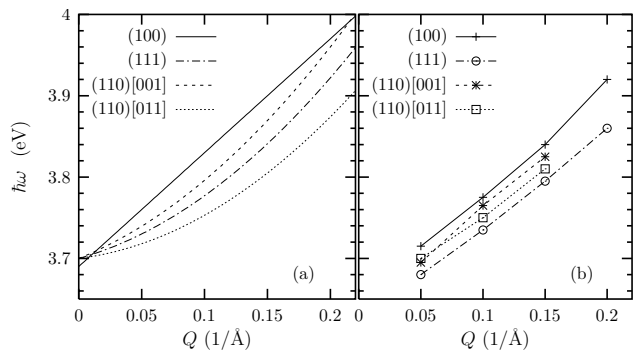}
\caption{Surface-plasmon energy dispersion for the low-index faces (100),
(111), and (110) of Ag. In the case of Ag(110), the exhibited surface-plasmon
energy dispersions correspond to wave vectors along the [001] and [110]
directions. (a) Experimental data from Ref.~\cite{rocca3}. (b) The results
obtained within the jellium-dipolium model (from Ref.~\cite{lopez}, used with
permission).}\label{fig17}
\end{figure}

Most recently, a self-consistent jellium model for valence $5s$ electrons in Ag
was combined with a so-called dipolium model in which the occupied $4d$ bands
are represented in terms of polarizable spheres located at the sites of a
semiinfinite fcc lattice~\cite{lopez} to calculate electron energy-loss
spectra for all three low-index faces of Ag. The surface-plasmon energy
dispersions obtained from these spectra are exhibited in Fig.~\ref{fig17}
together with the experimental measurements~\cite{rocca3}. This figure shows
that the trend obtained for the different crystal orientations is in
qualitative agreement with the data. On the one hand, the surface-plasmon
dispersion of Ag(100) lies above that of the (111) surface; on the other hand,
the slope for Ag(110) is larger for wave vectors along the [001] direction
than when the wave vector is taken to have the [110] direction, illustrating
the effect of the interplanar geometry on the effective local fields. The
observed overall variation of the measured positive slope with crystal
orientation is, however, considerably larger than predicted by the
jellium-dipolium model. This must be a signature of genuine band structure
effects not captured in the present model, which calls for a first-principles
{\it ab initio} description of the electronic response of real Ag surfaces.

The impact of the band structure on the surface-plasmon energy
dispersion in Ag was addressed by Moresco {\it et
al.}~\cite{moresco2} and by Savio {\it et al.}~\cite{savio}. These
authors performed test experiments with the K/Ag(110) and O/Ag(001)
systems and concluded that (i) the surface-plasmon dispersion can be
modified at will by manipulating the surface electronic structure
near $\omega_s$ and (ii) surface interband transitions between
Shockley states should be responsible for the anomalous linear
behavior of the surface-plasmon dispersion in Ag(100). Furthermore,
HREELS experiments on sputtered and nanostructured Ag(100) have
shown that the anomaly exhibited by surface plasmons on Ag(100) can
indeed be eliminated by modifying the surface
structure~\cite{savio2}.

\paragraph{Hg.} The $s$-$d$ polarization model devised and used by Liebsch to
describe the positive energy dispersion of surface plasmons in Ag~\cite{ag3}
was also employed for a description of surface plasmons in Hg~\cite{plummer6}.

At $q=0$, the measured surface-plasmon energy of this transition metal
($\sim 6.9\,{\rm eV}$) lies about $1\,{\rm eV}$ below the value expected
for a bounded electron gas with the density equal to the average density of
$6s^2$ valence electrons in Hg ($r_s=2.65$). This can be explained along the
lines described above for Ag [see Eq.~(\ref{omegasp})], with the use of a
polarizable background with $\epsilon_d(\omega_s')=1.6$.

\begin{figure}
\includegraphics[width=0.65\linewidth,angle=-90]
{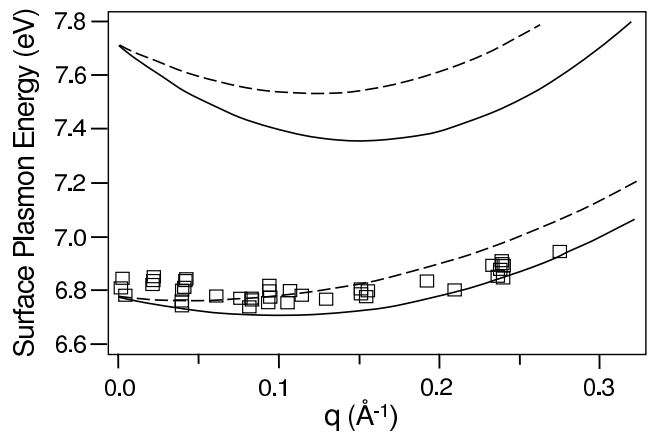}
\caption{Hg surface-plasmon energy dispersion, as obtained from standard
jellium calculations of the RPA and ALDA surface loss function $g(q,\omega)$
of Eq.~(\ref{g2}) (upper dashed and solid curves, respectively), from
stabilized jellium calculations of the RPA and ALDA surface loss function
$g'(q,\omega)$ of Eq.~(\ref{gm1}) (upper dashed and solid curves,
respectively), and from the peak positions observed in experimental electron
energy-loss spectra (from Ref.~\cite{plummer6}, used with
permission).}\label{fig18}
\end{figure}

At finite wave vectors, the $s$-$d$ screening in Hg is found to considerably
distort the surface-plasmon dispersion, as in the case of Ag, but now the
linear coefficient of the low-$q$ dispersion being still negative though much
smaller than in the absence of $d$ electrons. This is illustrated in
Fig.~\ref{fig18}, where the calculated (RPA and ALDA) and measured energy
dispersions of the Hg surface plasmon are compared to the corresponding
dispersions obtained within a standard jellium model (with $r_s=2.65$) in the
absence of a polarizable medium. This figure clearly shows that in addition to
the overall lowering of the surface-plasma frequencies by about $12\%$
relative to the jellium calculations (with no $d$ electrons) the $s$-$d$
screening leads to a significant flattening of the energy dispersion, thereby
bringing the jellium calculations into nice agreement with experiment.

Simple calculations have also allowed to describe correctly the measured
broadening of the Hg surface plasmon. As shown in Table~\ref{table2}, the Hg
surface-plasmon width at $q=0$ is very well described by simply introducing
the measured bulk dielectric function into Eq.~(\ref{long2}). For a
description of the surface-plasmon broadening at finite $q$,
Kim {\it at}~\cite{plummer6} introduced into Eq.~(\ref{long2}) a Drude
dielectric function of the form of Eq.~(\ref{drude}), but with a $q$-dependent
finite $\eta$ defined as
\begin{equation}
\eta(q)=\eta(0)\,{\rm e}^{-qa},
\end{equation}
with $a=3\,{\rm\AA}$ and $\eta(0)=\Delta\omega_s/\omega_s$ being the measured
relative width at $q=0$. This procedure gives a surface-plasmon linewidth of
about $1\,{\rm eV}$ for all values of $q$, in reasonable agreement with
experiment.

\paragraph{Pd.}

Collective excitations on transition metals with both $sp$ and $d$-bands
crossing the Fermi level are typically strongly damped by the presence of
interband transitions. Pd, however, is known to support collective excitations
that are relatively well defined.

Surface plasmons in Pd(110) were investigated with angle-resolved
electron-energy-loss spectroscopy by Rocca {\it et al.}~\cite{pd}.
These authors observed a prominent loss feature with a strongly
negative linear initial dispersion, which they attributed to a
surface-plasmon excitation. At $q=0$, they found
$\omega_s=7.37\,{\rm eV}$ and $\Delta\omega_s\sim 2\,{\rm eV}$, in
agreement with optical data~\cite{vehse}. At low wave vectors from
$q=0$ up to $q=0.2\,{\rm\AA}^{-1}$, they found a linear
surface-plasmon dispersion of the form of Eq.~(\ref{sdr2}) with
$\alpha=1\,{\rm\AA}$. This linear dispersion is considerably
stronger than in the case of simple metals (see Table~\ref{table1}),
and calls for a first-principles description of this material where
both occupied and unoccupied $sp$ and $d$ states be treated on an
equal footing.

\paragraph{Multipole surface plasmons.}

Among the noble and transition metals with occupied $d$ bands, multipole
surface plasmons have only been observed recently in the case of Ag. In an
improvement over previous HREELS experiments, Moresco {\it et
al.}~\cite{roccal} performed ELS-LEED experiments with both high-momentum and
high-energy resolution, and by substracting
the data for two different impact energies they found a peak at
$3.72\,{\rm eV}$, which was interpreted to be the Ag multipole plasmon.
However, Liebsch argued that the frequency of the Ag multipole surface plasmon
should be in the $6$-$8\,{\rm eV}$ range {\it above} rather than {\it below}
the bulk plasma frequency, and suggested that the observed peak at
$3.72\,{\rm eV}$ might not be associated with a multipole surface
plasmon~\cite{liebschl}.

Recently, the surface electronic structure and optical response of Ag have
been studied on the basis of angle- and energy-resolved photoyield
experiments~\cite{barman2}. In these experiments, the Ag multipole surface
plasmon is observed at $3.7\,{\rm eV}$, but no signature of the multipole
surface plasmon
is observed above the plasma frequency ($\omega_p=3.8\,{\rm eV}$) in
disagreement with the existing theoretical prediction~\cite{liebschl}.

\subsubsection{First-principles calculations}\label{abinitio}

First-principles calculations of the surface-plasmon energy and
linewidth dispersion of real solids have been carried out only in
the case of the simple-metal prototype surfaces Mg(0001)~\cite{mg}
and Al(111)~\cite{vac}. These calculations were performed by
employing a supercell geometry with slabs containing 16(27) atomic
layers of Mg(Al) separated by vacuum intervals. The matrix
$\chi_{{\bf g},{\bf g}'}^0(z,z';{\bf q},\omega)$ was calculated from
Eq.~(\ref{eq9}), with the sum over $n$ and $n'$ running over bands
up to energies of $30\,{\rm eV}$ above the Fermi level, the sum over
${\bf k}$ including 7812 points, and the single-particle orbitals
$\psi_{{\bf k},n;i}({\bf r})=\psi_{{\bf k},n}({\bf
r}_\parallel)\psi_i(z)$ being expanded in a plane-wave basis set
with a kinetic-energy cutoff of $12\,{\rm Ry}$. The surface-response
function $g_{{\bf g}=0,{\bf g}'=0}({\bf q},\omega)$ was then
calculated from Eq.~(\ref{ggg}) including $\sim 300$ 3D
reciprocal-lattice vectors in the evaluation of the RPA or TDDFT
Fourier coefficients $\chi_{{\bf g},{\bf g}'}(z,z';{\bf q},\omega)$
of Eq.~(\ref{eq:Xalda4}). Finally, the dispersion of the energy and
linewidth of surface plasmons was calculated from the maxima of the
imaginary part of $g_{{\bf g}=0,{\bf g}'=0}({\bf q},\omega)$ for
various values of the magnitude and the direction of the 2D wave
vector ${\bf q}$.

The {\it ab initio} calculations reported in Refs.~\cite{mg}
and~\cite{vac} for the surface-plasmon energy dispersion of Mg(0001)
and Al(111) with the 2D wave vector along various symmetry
directions show that (i) there is almost perfect isotropy of the
surface-plasmon energy dispersion, (ii) there is excellent agreement
with experiment (thereby accurately accounting for core polarization
not presented in jellium models), as long as the nonlocal
(momentum-dependent) static XC local-field factor of
Eq.~(\ref{corradini}) is employed in the evaluation of the
interacting density-response function, and (iii) if the
corresponding jellium calculations are normalized to the measured
value $\omega_s$ at $q=0$ (as shown in Fig.~\ref{fig11} by the thick
solid line with open circles) {\it ab initio} and jellium
calculations are found to be nearly indistinguishable.

\begin{figure}
\includegraphics[width=0.95\linewidth]
{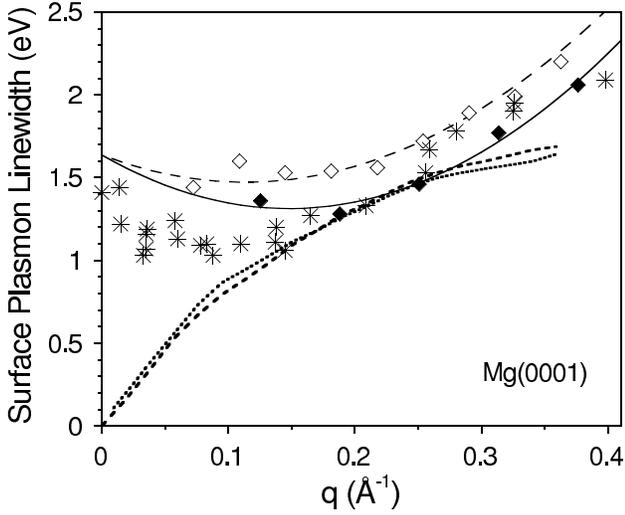} \caption{Linewidth of surface plasmons in Mg(0001), as
a function of the magnitude of the 2D wave vector ${\bf q}$. The
filled (open) diamonds represent {\it ab initio}
calculations~\cite{mg}, as obtained with ${\bf q}$ along the
$\bar\Gamma\bar M$ ($\bar\Gamma\bar K$) direction by employing the
static XC local-field factor of Eq.~(\ref{corradini}) in the
evaluation of the Fourier coefficients of the interacting
density-response function. The solid and long dashed line represent
the best fit of the {\it ab initio} calculations. The short dashed
and dotted lines correspond to jellium and 1D model-potential
calculations, also obtained by employing the static XC local-field
factor of Eq.~(\ref{corradini}). Stars represent the experimental
data reported in Ref.~\cite{plummer5}.}\label{fig19}
\end{figure}

\begin{figure}
\includegraphics[width=0.95\linewidth]
{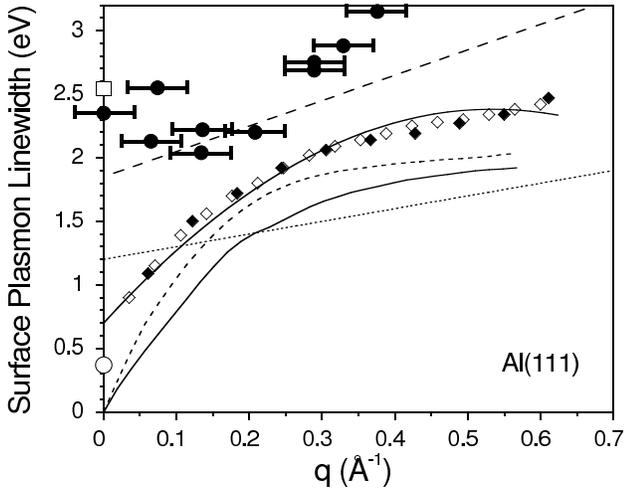} \caption{Linewidth of surface plasmons in Al(111), as a
function of the magnitude of the 2D wave vectors ${\bf q}$. The
filled (open) diamonds represent {\it ab initio} ALDA calculations,
as obtained with ${\bf q}$ along the $\bar\Gamma\bar M$
($\bar\Gamma\bar K$)~\cite{vac}. The solid and long dashed line
represent the best fit of the {\it ab initio} calculations. The
dotted line corresponds to 1D model-potential calculations, also
obtained in the ALDA. Stars represent the experimental data reported
in Ref.~\cite{leed2}. The measured values from Ref.~\cite{alnew} are
represented by dots with error bars.}\label{fig20}
\end{figure}

{\it Ab initio} calculations also show that the band structure is of
paramount importance for a correct description of the
surface-plasmon linewidth. First-principles TDDFT calculations of
the Mg(0001) and Al(111) surface-plasmon linewidth dispersions along
various symmetry directions are shown in Figs.~\ref{fig19} and
~\ref{fig20}, respectively. Also shown in this figures are the
experimental measurements reported in Refs.~\cite{plummer5}
and~\cite{leed2} (stars) and the corresponding jellium calculations
(dashed lines). For small 2D wave vectors, the agreement between
theory and experiment is not as good as in the case of the
surface-plasmon energy dispersion (which can be attributed to
finite-size effects of the supercell geometry). Nevertheless, {\it
ab initio} calculations for Mg(0001) (see Fig.~\ref{fig19}) yield a
negative slope for the linewidth dispersion at small ${\bf q}$, in
agreement with experiment, and properly account for the experimental
linewidth dispersion at intermediate and large wave vectors.
Fig.~\ref{fig19} also shows that the Mg(0001) surface-plasmon
linewidth dispersion depends considerably on the direction of the 2D
wave vector, and that the use of the parameterized one-dimensional
model potential reported in Ref.~\cite{chulkov0} (dotted line) does
not improve the jellium calculations. As for Al(111), we note that
band-structure effects bring the jellium calculations closer to
experiment; however, {\it ab initio} calculations are still in
considerable disagreement with the measurements reported in
Refs.~\cite{leed2} and~\cite{alnew}, especially at the lowest wave
vectors.
At the moment it is not clear the origin of so large discrepancy
between the theory and experiment. From the other hand, considerable
disagreement is also found between the measured surface-plasmon
linewidth at $q\sim 0$ reported in Refs.~\cite{leed2}
($\Delta\omega_s\sim 1.9\,{\rm eV}$) and ~\cite{alnew}
($\Delta\omega_s\sim 2.3\,{\rm eV}$) and the value derived (see
Table~\ref{table2}) from the imaginary part of the surface-response
function of Eq.~(\ref{long2}) with measured values of the bulk
dielectric function ($\Delta\omega_s=0.38\,{\rm eV}$).

\subsection{Acoustic surface plasmons}

\begin{figure}
\includegraphics[width=0.75\linewidth]{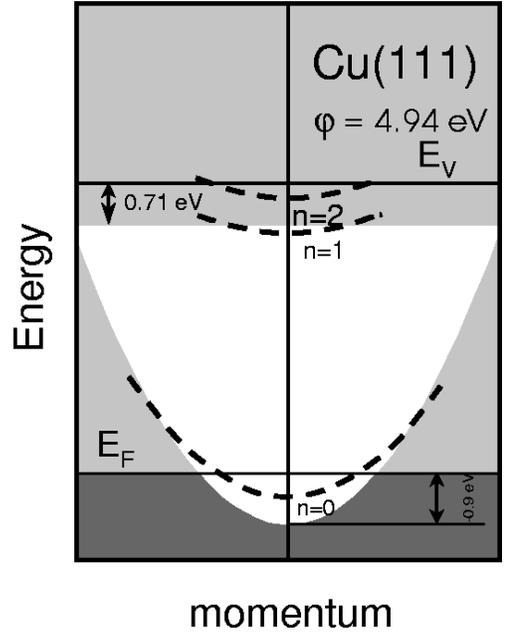}\\
\caption{Schematic representation of the surface band structure on Cu(111) near
the $\bar\Gamma$ point. The shaded region represents the projection of the
bulk bands.}\label{fig8p}
\end{figure}

\begin{table}
\caption{2D Fermi energy ($\varepsilon_F^{2D}$) of surface states at the
$\bar\Gamma$ point of Be(0001) and the (111) surfaces of the noble metals Cu,
Ag, and Au. $v_F^{2D}$ and $m^{2D}$ represent the corresponding 2D Fermi
velocity
and effective mass, respectively. $v_F^{2D}$ is expressed in units of the Bohr
velocity
$v_0=e^2/\hbar$.}
\begin{ruledtabular} \begin{tabular}{lcccccc}
&$\varepsilon_F^{2D}$(eV)&$v_F^{2D}/v_0$&$m^{2D}$\\ \hline
Be(0001)&2.75&0.41&1.18\\
Cu(111)&0.44&0.28&0.42\\
Ag(111)&0.065&0.11&0.44\\
Au(111)&0.48&0.35&0.28\\
\end{tabular}
\end{ruledtabular}\label{tableg}
\end{table}

A variety of metal surfaces, such as Be(0001) and the (111) surfaces of the
noble metals Cu, Ag, and Au, are known to support a partially occupied band of
Shockley surface states within a wide energy gap around the Fermi level
(see Fig.~\ref{fig8p})~\cite{chul87,inglesg}. Since these states are strongly
localized near the surface and disperse with momentum parallel to the surface,
they can be considered to form a quasi 2D surface-state band
with a 2D Fermi energy $\varepsilon_F^{2D}$ equal to the surface-state binding
energy at the
$\bar\Gamma$ point (see Table~\ref{tableg}).

In the absence of the 3D substrate, a Shockley surface state would support a 2D
collective oscillation, the energy of the corresponding plasmon being given
by~\cite{stern}
\begin{equation}\label{2d}
\omega_{2D}=(2\pi n^{2D}q)^{1/2},
\end{equation}
with $n^{2D}$ being the 2D density of occupied surface states, i.e,
\begin{equation}
n^{2D}=\varepsilon_F^{2D}/\pi.
\end{equation}
Eq.~(\ref{2d}) shows that at very long
wavelengths plasmons in a 2D electron gas have low energies; however, they do
not affect e-ph interaction and phonon dynamics near the Fermi level, due to
their
square-root dependence on the wave vector. Much more effective than ordinary 2D
plasmons in mediating, e.g., superconductivity would be the so-called acoustic
plasmons with sound-like long-wavelength dispersion.

Recently, it has been shown that in the presence of the 3D substrate
the dynamical screening at the surface provides a mechanism for the
existence of a {\it new} acoustic collective mode, the so-called
acoustic surface plasmon, whose energy exhibits a linear dependence
on the 2D wave number~\cite{sigael04,note3,pitarke,sipiprb05}. This
{\it novel} surface-plasmon mode has been observed at the (0001)
surface of Be, showing a linear energy dispersion that is in very
good agreement with first-principles calculations~\cite{acexp}.

\subsubsection{A simple model}

\begin{figure}
\includegraphics[width=0.55\linewidth,angle=270]{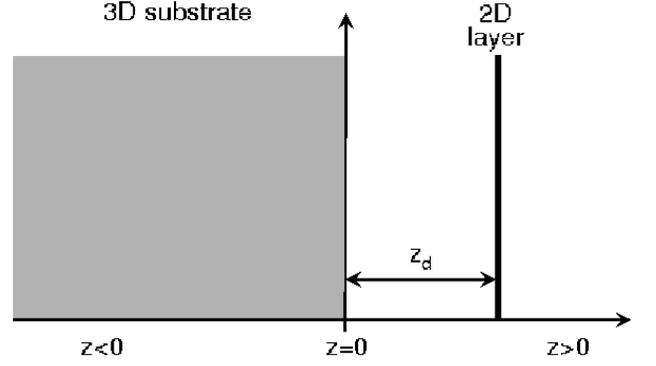}\\
\caption{Surface-state electrons comprise a 2D sheet of interacting free
electrons at $z=z_d$. All other states of the semi-infinite metal comprise a
plane-bounded 3D electron gas at $z\leq 0$. The metal surface is located at
$z=0$.}\label{fig9p}
\end{figure}

First of all, we consider a simplified model in which surface-state electrons
comprise a 2D electron gas at $z=z_d$ (see Fig.~\ref{fig9p}), while all other
states of the semi-infinite metal comprise a 3D substrate at $z\leq 0$
represented by the Drude dielectric function of Eq.~(\ref{drude}). Within this
model, one finds that both e-h and collective excitations occurring within the
2D gas can be described with the use of an effective 2D dielectric function,
which in the RPA takes the form~\cite{pitarke}
\begin{equation}\label{eff}
\epsilon_{eff}^{2D}(q,\omega)=1-W(z_d,z_d;q,\omega)\,\chi_{2D}^0(q,\omega),
\end{equation}
$W(z,z';q,\omega)$ being the screened interaction of Eq.~(\ref{wsurf}), and
$\chi_{2D}^0(q,\omega)$ being the noninteracting density-response function of
a homogeneous 2D electron gas~\cite{stern}.

In the absence of the 3D substrate, $W(z,z';q,\omega)$ reduces to the bare
Coulomb interaction $v(z,z';q)$, and $\epsilon_{eff}^{2D}(q,\omega)$
coincides, therefore, with the RPA
dielectric function of a 2D electron gas, which in the long-wavelength ($q\to
0$) limit has one single zero corresponding to collective excitations at
$\omega=\omega_{2D}$.

In the presence of a 3D substrate, the long-wavelength limit of
$\epsilon_{eff}^{2D}(q,\omega)$ is found to have two zeros. One zero
corresponds to a high-frequency ($\omega>>v_Fq$) oscillation of energy
\begin{equation}
\omega^2=\omega_s^2+\omega_{2D}^2,
\end{equation}
in which 2D and 3D electrons oscillate in phase with one another. The other
zero corresponds to a low-frequency {\it acoustic} oscillation in which both 2D
and 3D electrons oscillate out of phase; for this zero to be present, the
long-wavelength ($q\to 0$) limit of the low-frequency ($\omega\to 0$) screened
interaction $W(z_d,z_d;q,\omega)$,
\begin{equation}
I(z_d)=\lim_{q\to 0}W(z_d,z_d;q,\omega\to 0),
\end{equation}
must be different from zero. The energy of this low-frequency mode is then
found to be of the form~\cite{pitarke}
\begin{equation}\label{acous}
\omega=\alpha\,v_F^{2D}\,q,
\end{equation}
where
\begin{equation} \label{alpha2}
\alpha=\sqrt{1+{\left[I(z_d)\right]^2\over\pi\left[\pi+2\,I(z_d)\right]}}.
\end{equation}

\paragraph{Local 3D response.} If one characterizes the 3D substrate by a Drude
dielectric function of the form of Eq.~(\ref{drude}), the 3D screened
interaction $W(z_d,z_d;q,\alpha v_Fq)$ is easily found to be
\begin{equation}\label{lh}
W(z_d,z_d;q,\alpha v_Fq)=\cases{0&$z_d\leq 0$\cr\cr
4\,\pi\,z_d&$z_d>0$}.
\end{equation}

Hence, in the presence of a 3D substrate that is {\it spatially separated} from
the 2D sheet ($z_d>0$), introduction of $I(z_d)=4\,\pi\,z_d$ into
Eq.~(\ref{alpha2}) yields at $z_d>>1$:
\begin{equation}\label{chapli}
\alpha=\sqrt{2z_d},
\end{equation}
which is the result first obtained by Chaplik in his study of charge-carrier
crystallization in low-density inversion layers~\cite{chaplik}.

If the 2D sheet is located inside the 3D substrate ($z_d\leq 0$),
$I(z_d)=0$, which means that the effective dielectric function of
Eq.~(\ref{eff}) has no zero at low energies ($\omega<\omega_s$).
Hence, in a local picture of the 3D response the characteristic
collective oscillations of the 2D electron gas would be completely
screened by the surrounding 3D substrate, and no low-energy acoustic
mode would exist. This result had suggested over the years that
acoustic plasmons should only exist in the case of {\it spatially
separated} plasmas, as pointed out by Das Sarma and
Madhukar~\cite{sarma}.

Nevertheless, Silkin {\it et al.}~\cite{sigael04} have demonstrated
that metal surfaces where a partially occupied quasi-2D
surface-state band {\it coexists} in the same region of space with
the underlying 3D continuum can indeed support a well-defined
acoustic surface plasmon. This acoustic collective oscillation has
been found to appear as the result of a combination of the
nonlocality of the 3D dynamical screening and the spill out of the
3D electron density into the vacuum~\cite{pitarke}.

\paragraph{Self-consistent 3D response.}

\begin{figure}
\includegraphics[width=0.45\textwidth,height=0.3375\textwidth]{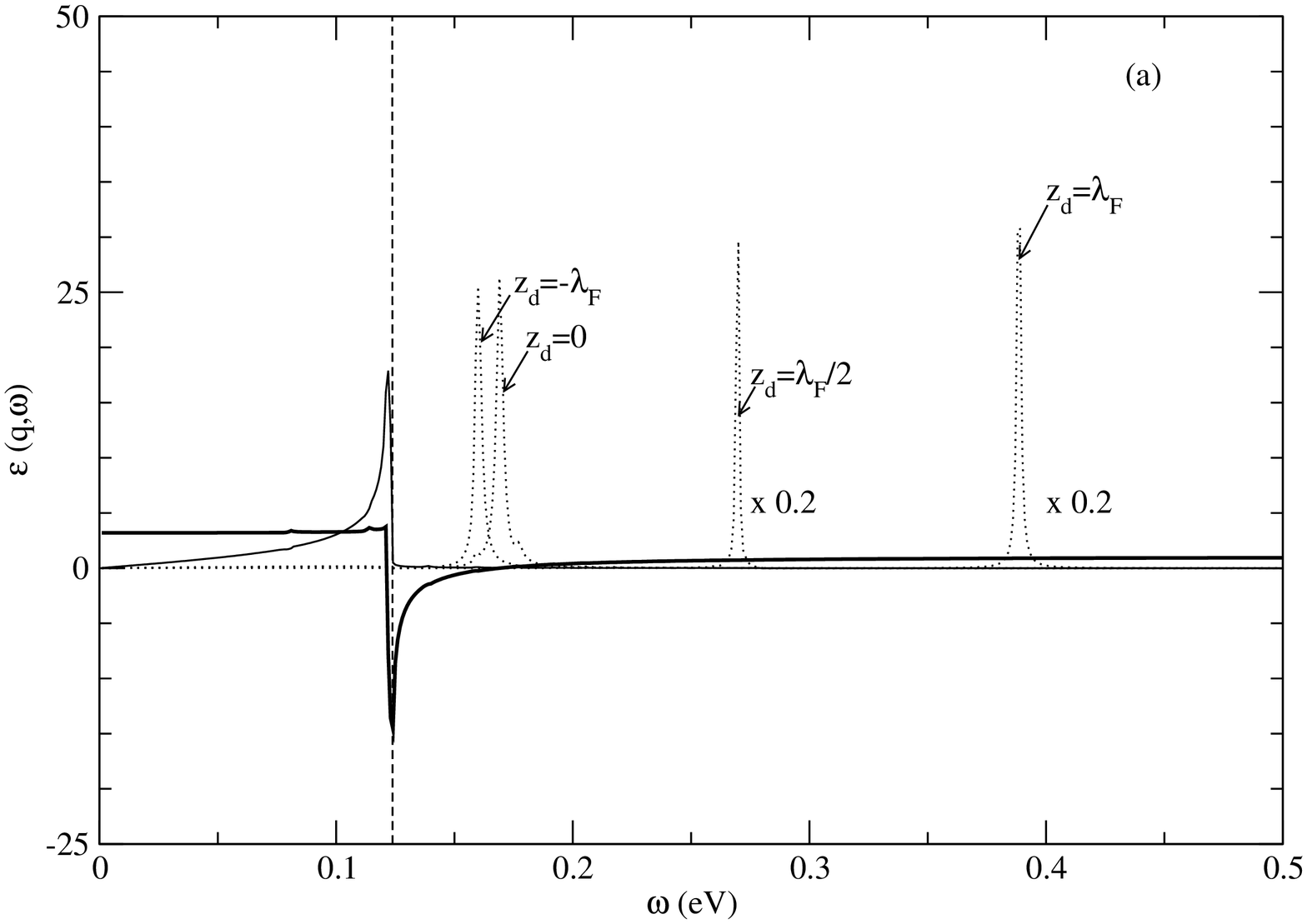}\\
\vspace*{1.2cm}
\includegraphics[width=0.45\textwidth,height=0.3375\textwidth]{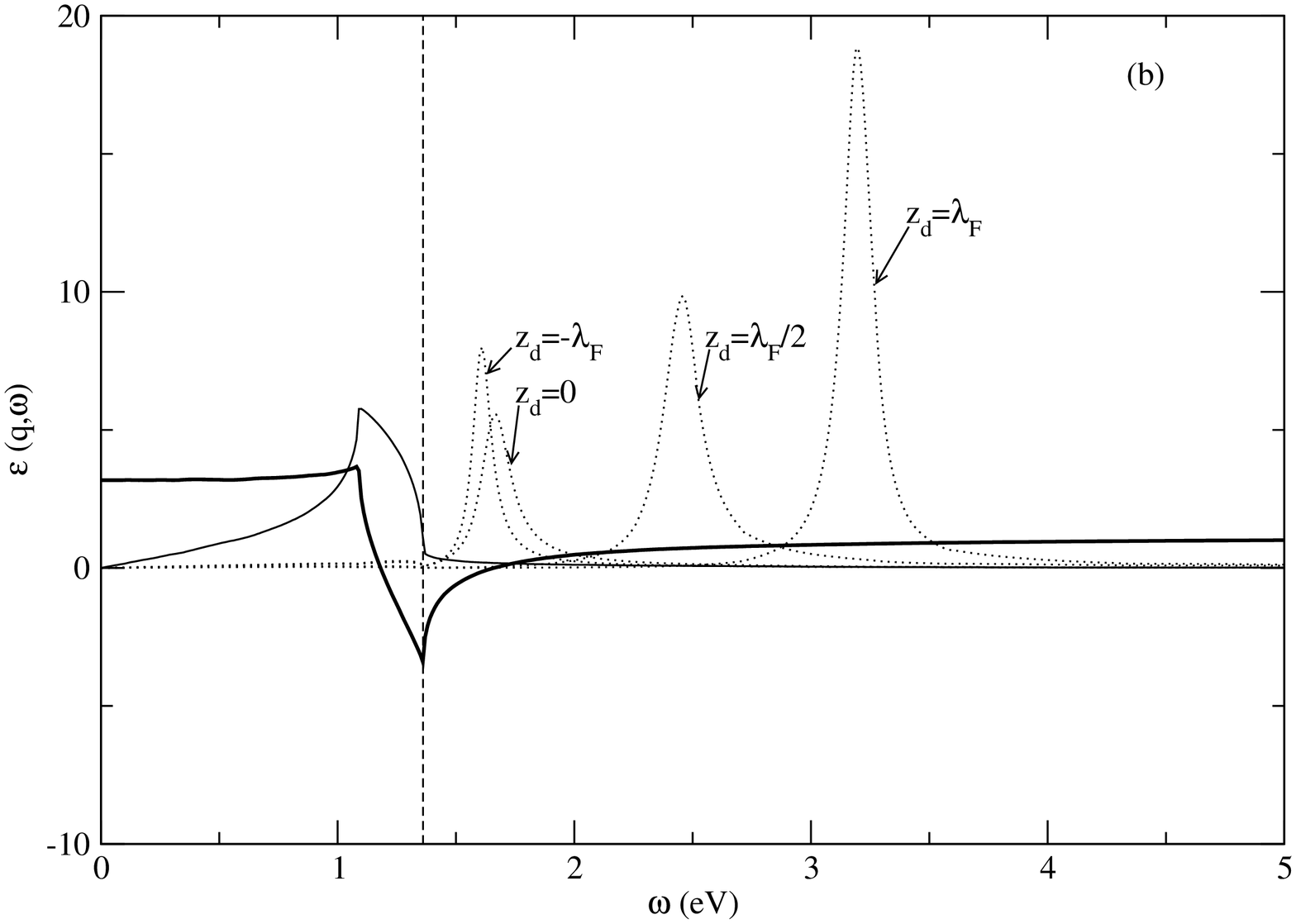}\\
\caption{Effective dielectric function of a 2D sheet that is located
at the jellium edge ($z_d=0$), as obtained from Eq.~(\ref{eff}) and
self-consistent RPA calculations of the 3D screened interaction
$W(z_d,z_d;q,\omega)$ and the 2D density-response function
$\chi^{2D}(q,\omega)$ with (a) $q=0.01$ and (b)
$q=0.1$~\cite{pitarke}. The real and imaginary parts of
$\epsilon_{eff}(q,\omega)$ are represented by thick and thin solid
lines, respectively. The dotted lines represent the effective 2D
energy-loss function ${\rm
Im}\left[-\epsilon_{eff}^{-1}(q,\omega)\right]$ for
$z_d=-\lambda_F$, $z_d=\lambda_F/2$, and $z_d=\lambda_F$. The
vertical dashed line represents the upper edge
$\omega_u=v_Fq+q^2/2m$ of the 2D e-h pair continuum, where 2D e-h
pairs can be excited. The calculations presented here for
$z_d=\lambda_F$ and $q=0.01\,a_0^{-1}$ have been carried out by
replacing the energy $\omega$ by a complex quantity $\omega+i\eta$
with $\eta=0.05\,{\rm eV}$. All remaining calculations have been
carried out for real frequencies, i.e., with $\eta=0$. The 2D and 3D
electron-density parameters have been taken to be $r_s^{2D}=3.14$
and $r_s=1.87$, corresponding to Be(0001).}\label{fig4p}
\end{figure}

Figures~\ref{fig4p}a and \ref{fig4p}b exhibit self-consistent RPA calculations
of the effective dielectric function of Eq.~(\ref{eff}) corresponding to the
(0001) surface of Be, with $q=0.01\,a_0^{-1}$ (Fig.~\ref{fig4p}a) and
$q=0.1\,a_0^{-1}$ (Fig.~\ref{fig4p}b). The real and imaginary parts of the
effective dielectric function (thick and thin solid lines, respectively) have
been displayed for $z_d=0$, as approximately occurs with the quasi-2D
surface-state band in Be(0001). Also shown in these figures (dotted lines) is
the energy-loss function ${\rm Im}\left[-1/\epsilon_{eff}^{2D}(q
\omega)\right]$ for the 2D sheet located inside the metal at $z_d=-\lambda_F$,
at the jellium edge ($z_d=0$), and outside the metal at $z_d=\lambda_f/2$ and
$z_d=\lambda_F$.

Collective excitations are related to a zero of
${\rm Re}\,\epsilon_{eff}^{2D}(q,\omega)$ in a region where
${\rm Im}\,\epsilon_{eff}^{2D}(q,\omega)$ is small, which yields a maximum in
the energy-loss function
${\rm Im}\left[-1/\epsilon_{eff}^{2D}(q,\omega)\right]$. For the 2D electron
density under study ($r_s^{2D}=3.12$), in the absence of the 3D substrate a 2D
plasmon would occur at $\omega_{2D}=1.22\,{\rm eV}$ for $q=0.01a_0^{-1}$ and
$\omega_{2D}=3.99\,{\rm eV}$ for $q=0.1a_0^{-1}$. However, in the presence of
the 3D substrate (see Fig.~\ref{fig4p}) a well-defined low-energy
{\it acoustic} plasmon occurs at energies (above the upper edge
$\omega_u=v_Fq+q^2/2$ of the 2D e-h pair continuum) dictated by
Eq.~(\ref{acous}) with the sound velocity $\alpha v_F$ being just over the 2D
Fermi velocity $v_F$ ($\alpha>1$). When the 2D sheet is located far inside the
metal surface, the sound velocity is found to approach the Fermi velocity
($\alpha\to 1$). When the 2D sheet is located far outside the metal surface,
the coefficient $\alpha$ approaches the classical limit
($\alpha\to\sqrt{2 z_d}$) of Eq.~(\ref{chapli}).

Finally, we note that apart from the limiting case $z_d=\lambda_F$ and
$q=0.01\,a_0^{-1}$ (which yields a plasmon linewidth negligibly small) the
small width of the plasmon peak exhibited in Fig.~\ref{fig4p} is entirely due
to plasmon decay into e-h pairs of the 3D substrate.

\subsubsection{1D model calculations}

For a more realistic (but still simplified) description of
electronic excitations at the (0001) surface of Be and the (111)
surface of the noble metals Cu, Ag, and Au, the 1D model potential
$v_{MP}(z)$ of Ref.~\cite{chulkov0} was employed in
Refs.~\cite{sigael04,sipiprb05}. The use of this model potential
allows to assume translational invariance in the plane of the
surface and to trace, therefore, the presence of collective
excitations to the peaks of the imaginary part of the jellium-like
surface response function of Eq.~(\ref{g}). The important difference
between the screened interaction $W(z,z';q,\omega)$ used here to
evaluate $g(q,\omega)$ and the screened interaction used in
Eq.~(\ref{eff}) in the framework of the simple model described above
lies in the fact that the single-particle orbitals $\psi_i(z)$ and
energies $\epsilon_i$ are now obtained by solving Eq.~(\ref{ks1})
with the jellium Kohn-Sham potential $v_{KS}[n_0](z)$ replaced by
the model potential $v_{MP}(z)$. Since this model potential has been
shown to reproduce the key features of the surface band structure
and, in particular, the presence of a Shockley surface state within
an energy gap around the Fermi level of the materials under study,
it provides a realistic description of surface-state electrons
moving in the presence of the 3D substrate.

\begin{figure}
\includegraphics[width=0.75\linewidth]{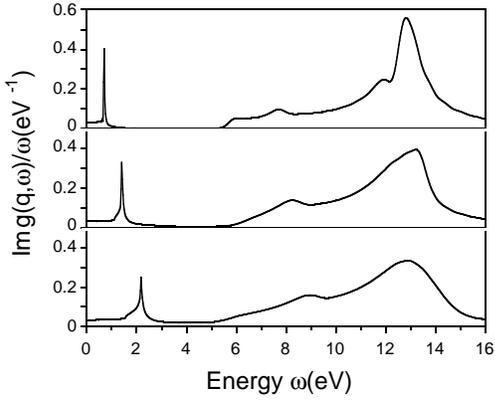}\\
\caption{Energy loss function ${\rm Im}[g(q,\omega)]/\omega$ of
Be(0001) versus the excitation energy $\omega$, obtained from
Eq.~(\ref{g}) with the use of the 1D model potential of
Ref.~\cite{chulkov0}, as reported in Ref.~\cite{sigael04}. The
magnitude of the wave vector ${\bf q}$ has been taken to be 0.05
(top panel), 0.1 (middle panel), and 0.15 (bottom panel), in units
of the inverse Bohr radius $a_0^{-1}$. In the long-wavelength limit
($q\to 0$), $g(q,\omega)$ is simply the total electron density
induced by the potential of Eq.~(\ref{ext2}).}\label{be}
\end{figure}

Figure~\ref{be} shows the imaginary part of the surface-response function of
Eq.~(\ref{g}), as obtained for Be(0001) and for increasing values of $q$ by
using the 1D model potential $v_{MP}(z)$ of Ref.~\cite{chulkov0}. As follows
from the figure, the excitation spectra is clearly dominated by two distinct
features: (i) the conventional surface plasmon at
$\hbar\omega_s\sim 13\,{\rm eV}$, which can be traced to the characteristic
pole that the surface-response function $g(q,\omega)$ of a bounded 3D free
electron gas with $r_s^{3D}=1.87$ exhibits at this energy~\cite{noteg2}, and
(ii) a well-defined low-energy peak with {\it linear} dispersion.

\begin{figure}
\includegraphics[width=0.95\linewidth]{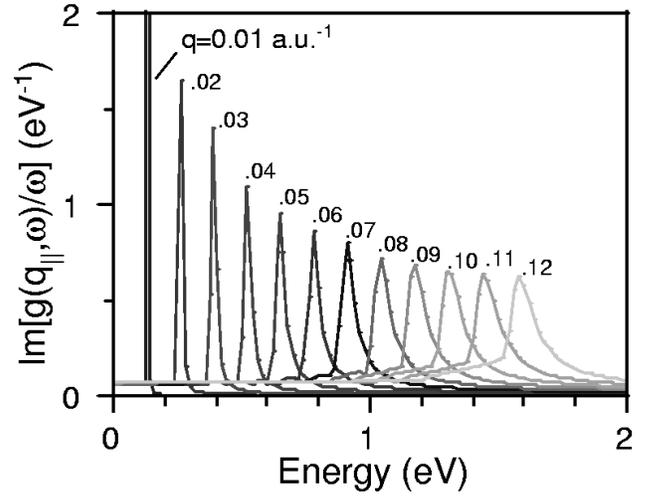}\\
\caption{Energy loss function ${\rm Im}[g(q,\omega)]/\omega$ of
Be(0001) versus the excitation energy $\omega$ obtained from
Eq.~(\ref{g}) with the use of the 1D model potential of
Ref.~\cite{chulkov0} and for various values of
$q$~\cite{opt}.}\label{fig13p}
\end{figure}

That the low-frequency mode that is visible in Fig.~\ref{be} has linear
dispersion is clearly shown in Fig.~\ref{fig13p}, where the imaginary part of
the surface-response function $g(q,\omega)$ of Be(0001) is displayed at low
energies for increasing values of the magnitude of the wave vector in the
range $q=0.01-0.12\,a_0^{-1}$. The excitation spectra is indeed dominated at
low energies by a well-defined {\it acoustic} peak at energies of the form of
Eq.~(\ref{acous}) with an $\alpha$ coefficient that is close to unity, i.e.,
the sound velocity being at long wavelengths very close to the 2D Fermi
velocity $v_F^{2D}$ (see Table~\ref{tableg}).

\begin{figure}
\includegraphics[width=0.75\linewidth]{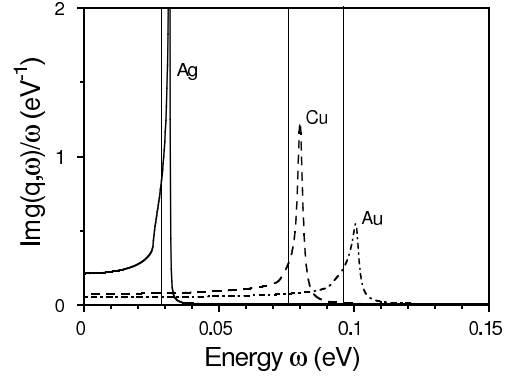}\\
\caption{Energy-loss function ${\rm Im}[g(q,\omega)]/\omega$ of the
(111) surfaces of the noble metals Cu, Ag, and Au~\cite{sipiprb05},
shown by solid, dashed, and dashed-dotted lines, respectively,
versus the excitation energy $\omega$, as obtained from
Eq.~(\ref{g}) with the use of the 1D model potential of
Ref.~\cite{chulkov0} and for $q=0.01\,a_0^{-1}$ and $\eta=1\,{\rm
meV}$. The vertical solid lines are located at the energies
$\omega=v_F^{2D}\,q$, which would correspond to Eq.~(\ref{acous})
with $\alpha=1$.}\label{fig-g}
\end{figure}

The energy-loss function ${\rm Im}\,g(q,\omega)$ of the (111) surfaces of the
noble metals Cu, Ag, and Au is displayed in Fig.~\ref{fig-g} for
$q=0.01\,a_0^{-1}$. This figure shows again the presence of a well-defined
low-energy collective excitation whose energy is of the form of
Eq.~(\ref{acous}) with $\alpha\sim 1$.

\begin{figure}
\includegraphics[width=0.65\linewidth,angle=-90]{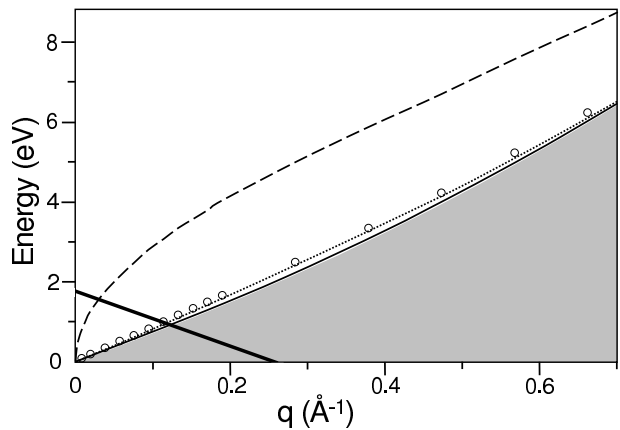}\\
\caption{The solid line shows the energy of the acoustic surface
plasmon of Be(0001)~\cite{sigael04}, as obtained from the maxima of
the calculated surface-loss function ${\rm Im}\,g(q,\omega)$ shown
in Fig.~\ref{fig13p}. The thick dotted line and the open circles
represent the maxima of the energy-loss function ${\rm
Im}\left[-1/\epsilon_{eff}(q,\omega)\right]$ obtained from
Eq.~(\ref{eff}) with $z_d$ far inside the solid (thick dotted line)
and with $z_d=0$ (open circles). The dashed line is the plasmon
dispersion of a 2D electron gas in the absence of the 3D system. The
gray area indicates the region of the ($q$,$\omega$) plane (with the
upper limit at $\omega_{2D}^{up}=v_F^{2D}q+q^2/2m_{2D}$) where e-h
pairs can be created within the 2D Shockley band of Be(0001). The
area below the thick solid line corresponds to the region of
momentum space where transitions between 3D and 2D states cannot
occur. The quantities $\omega_{inter}^{min}$ and $q^{min}$ are
determined from the surface band structure of Be(0001). 2D and 3D
electron densities have been taken to be those corresponding to the
Wigner radii $r_s^{2D}=3.14$ and $r_s^{3D}=1.87$,
respectively.}\label{fig27}
\end{figure}

\begin{figure}
\includegraphics[width=0.75\linewidth]{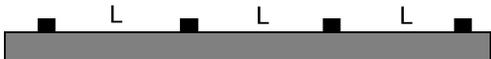}\\
\caption{A periodic grating of constant $L$. The grating periodic
structure can provide an impigning free electromagnetic radiation
with additional momentum $2\pi/L$.}\label{fig16p}
\end{figure}

Figure~\ref{fig27} shows the energy of the acoustic surface plasmon of Be(0001)
versus $q$ (solid line), as derived from the maxima of the calculated
${\rm Im}\,g(q,\omega)$ of Fig.~\ref{fig13p} (solid line) and from the maxima
of the effective energy-loss function
${\rm Im}\left[-1/\epsilon_{eff}(q,\omega)\right]$ (see Fig.~\ref{fig4p})
obtained from Eq.~(\ref{eff}) for $z_d=0$ (dashed line). Little discrepancies
between these two calculations should be originated in (i) the absence in the
simplified model leading to Eq.~(\ref{eff}) of transitions between 2D and 3D
states, and (ii) the nature of the decay and penetration of the surface-state
orbitals, which in the framework of the model leading to Eq.~(\ref{eff}) are
assumed to be fully localized in a 2D sheet at $z=z_d$.

\subsubsection{First-principles calculations}

First-principles calculations of the imaginary part of the
surface-response function $g_{{\bf g}=0,{\bf g}=0}({\bf q},\omega)$
of Be(0001) have been carried out recently~\cite{acexp}, and it has
been found that this metal surface is indeed expected to support an
acoustic surface plasmon whose energy dispersion agrees with the
solid line represented in Fig.~\ref{fig27} (if the dispersion of
Fig.~\ref{fig27} calculated for surface state effective mass $m=1$
is scaled according to the {\it ab initio} value $m=1.2$~\cite{sibaprb01}).
Furthermore, these calculations have been found to agree
closely with recent high-resolution EELS measurements on the (0001)
surface of Be~\cite{acexp} (under grazing incidence), which
represent the first evidence of the existence of acoustic surface
plasmons.

\subsubsection{Excitation of acoustic surface plasmons}

As in the case of the conventional surface plasmon at the Ritchie's frequency
$\omega_s$, acoustic surface plasmons should be expected to be excited not
only by moving electrons (as occurs in the EELS experiments reported
recently~\cite{acexp}) but also by light. Now we focus on a possible mechanism
that would lead to the excitation of acoustic surface plasmons by light in,
e.g., vicinal surfaces with high indices~\cite{opt}.

At long wavelengths ($q\to 0$), the acoustic surface-plasmon
dispersion curve is of the form of Eq.~(\ref{acous}) with $\alpha\sim
1$. As the 2D Fermi velocity $v_F^{2D}$ is typically about three
orders of magnitude smaller than the velocity of light, there is, in
principle, no way that incident light can provide an ideal surface
with the correct amount of momentum and energy for the excitation of
an acoustic surface plasmon to occur. As in the case of conventional surface
plasmons, however, a periodic corrugation or grating in the metal
surface should be able to provide the missing momentum.

\begin{figure}
\includegraphics[width=0.65\linewidth,angle=-90]{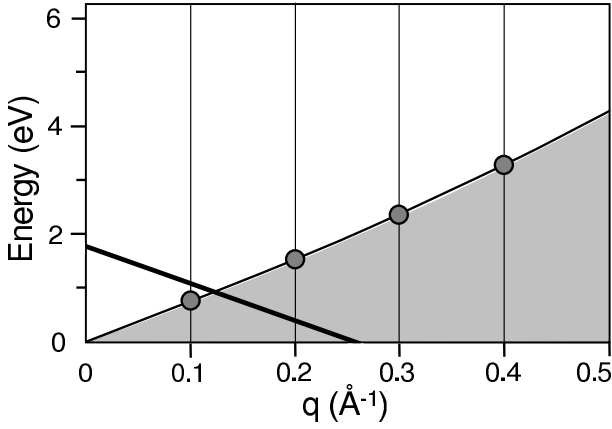}\\
\caption{A schematic representation of the dispersion relation of
acoustic surface plasmons (solid line) and free light impigning on a
periodic grating of constant $L$ (essentially vertical dotted
lines).}\label{fig30}
\end{figure}

Let us consider a periodic grating of constant $L$ (see
Fig.~\ref{fig16p}). If light hits such a surface, the grating
periodic structure can provide the impigning free electromagnetic
waves with additional momentum arising from the grating periodic
structure. If free electromagnetic radiation hits the grating at an
angle $\theta$, its wave vector along the grating surface has
magnitude
\begin{equation}\label{grat}
q={\omega\over c}\,\sin\theta\pm{2\pi\over L}\,n,
\end{equation}
where $L$ represents the grating constant, and $n=1,2,\dots$. Hence, the linear
(nearly vertical) dispersion relation of free light changes into a set of
parallel straight lines, which can match the acoustic-plasmon dispersion
relation as shown in Fig.~\ref{fig30}.

For a well-defined acoustic surface plasmon in Be(0001) to be
observed, the wave number $q$ needs to be smaller than $q\sim
0.06\,a_0^{-1}$ (see Fig.~\ref{fig27})~\cite{noteslava}. For
$q=0.05\,a_0^{-1}$, Eq.~(\ref{grat}) with $n=1$ yields a grating constant
$L=66\,{\rm\AA}$. Acoustic surface plasmons of energy $\omega\sim
0.6\,{\rm eV}$ could be excited in this way. Although a grating
period of a few nanometers sounds unrealistic with present
technology, the possible control of vicinal surfaces with high
indices could provide appropriate grating periods in the near
future.

\section{Applications}

Surface plasmons have been employed over the years in a wide spectrum of
studies ranging from condensed matter and surface physics~
\cite{inglesvw,zk,ser,feibelmani,ritchie2,mahan1,lucas1,gadzuk,lucas2,ingles,langreth,
chabal,persson,pendry1,inkson,ueba} to electrochemistry~\cite{knoll0},
wetting~\cite{wetting}, biosensing~\cite{bio0,bio1,bio2}, scanning tunneling
microscopy~\cite{stm}, the ejection of ions from
surfaces~\cite{compton}, nanoparticle growth~\cite{nano1,nano2},
surface-plasmon microscopy~\cite{nature1,science0}, and
surface-plasmon resonance
technology~\cite{spr1,spr2,nature2,spr3,spr4,spr5,spr6}. Renewed interest
in surface plasmons has come from recent advances in the investigation of the
optical properties of nanostructured materials~\cite{pendry2,halas}, one of
the most attractive aspects of these collective excitations now being
their use to concentrate light in subwavelength structures and to
enhance transmission through periodic arrays of subwavelength holes
in optically thick metallic films~\cite{ebbesen1,ebbesen3}, as well
as the possible fabrication of nanoscale photonic circuits operating
at optical frequencies~\cite{ebbesen2} and their use as mediators in
the transfer of energy from donor to acceptor molecules on opposite
sides of metal films~\cite{barnes}.

Here we focus on two distinct applications of collective electronic excitations
at metal surfaces: The role that surface plasmons play in particle-surface
interactions and the new emerging field called plasmonics.

\subsection{Particle-surface interactions: energy loss}

Let us consider a recoilless fast point particle of charge $Z_1$
moving in an arbitrary inhomogeneous many-electron system at a given
impact vector ${\bf b}$ with nonrelativistic velocity ${\bf v}$, for
which retardation effects and radiation losses can be
neglected~\cite{noteps}. Using Fermi's golden rule of time-dependent
perturbation theory, the lowest-order probability for the probe
particle to transfer momentum ${\bf q}$ to the medium is given by
the following expression~\cite{pc}:
\begin{eqnarray}\label{eq51p}
P_{\bf q}&=&-{4\pi\over LA}\,Z_1^2\int_0^\infty d\omega
\int{d{\bf q}'\over(2\pi)^3}\,{\rm e}^{i{\bf
b}\cdot({\bf q}+{\bf q}')}\cr\cr
&\times& {\rm Im}W({\bf q},{\bf q}';\omega)\,\delta(\omega-{\bf q}\cdot{\bf
v})\,\delta(\omega+{\bf q}'\cdot{\bf v}),
\end{eqnarray}
where $L$ and $A$ represent the normalization length and area, respectively,
and $W({\bf q},{\bf q}';\omega)$ is the double Fourier transform of the
screened interaction $W({\bf r},{\bf r}';\omega)$ of Eq.~(\ref{screened}):
\begin{equation}
W({\bf q},{\bf q}';\omega)=\int d{\bf r}\int d{\bf r}'\,{\rm e}^{-i({\bf
q}\cdot{\bf r}+{\bf q}'\cdot{\bf r}')}\,W({\bf r},{\bf r}';\omega).
\end{equation}

Alternatively, the total decay rate $\tau^{-1}$ of the probe
particle can be obtained from the knowledge of the imaginary part of
the self-energy. In the $GW$ approximation of many-body
theory~\cite{hedin}, and replacing the probe-particle Green function
by that of a non-interacting recoilless particle, one
finds~\cite{chemphys}:
\begin{eqnarray}\label{auto4}
\tau^{-1}=&&-2\,Z_{1}^{2}\sum_f\int d{\bf r}\int d{\bf r'}
\,\phi_i^\ast({\bf r})\,\phi_{f}^\ast({\bf r'})\cr\cr
&&\times {\rm Im}W({\bf r},{\bf r'},\varepsilon_i-\varepsilon_f)\,
\phi_{i}({\bf r'})\,\phi_{f}({\bf r}),
\end{eqnarray}
where $\phi_i({\bf r})$ represents the probe-particle initial state of energy
$\varepsilon_i$, and the sum is extended over a complete set of final states
$\phi_f({\bf r})$ of energy $\varepsilon_f$. Describing the probe-particle
initial and final states by plane waves in the direction of motion and
a Dirac $\delta$ function in the transverse direction, i.e.,
\begin{equation}\label{auto8}
\phi({\bf r})={1\over\sqrt A}\,e^{i{\bf v}\cdot{\bf
r}}\,\sqrt{\delta({\bf r}_\perp-{\bf b})},
\end{equation}
where ${\bf r}_\perp$ represents the position vector perpendicular to the
projectile velocity, one finds that the decay rate of Eq.~(\ref{auto4}) reduces
indeed to a sum over the probability $P_{\bf q}$ of Eq.~(\ref{eq51p}), i.e.:
\begin{equation}\label{tau}
\tau^{-1}={1\over T}\sum_{\bf q}\,P_{\bf q},
\end{equation}
$T$ being a normalization time.

For a description of the total energy $\Delta E$ that the moving probe particle
loses due to electronic excitations in the medium, one can first define the
time-dependent probe-particle charge density
\begin{equation}\label{rho}
\rho^{ext}({\bf r},t)=Z_1\,\delta({\bf r}-{\bf b}-{\bf v}\,t),
\end{equation}
and one then obtains the energy that this classical particle loses per unit
time as follows~\cite{flores}
\begin{equation}\label{flores}
-{dE\over dt}=-\int d{\bf r}\,\rho^{ext}({\bf r},t)\,{\partial V^{ind}({\bf
r},t)\over\partial t},
\end{equation}
where $V^{ind}({\bf r},t)$ is the potential induced by the probe particle at
position ${\bf r}$ and time $t$, which to first order in the external
perturbation yields [see Eq.~(\ref{tpotnew})]:
\begin{eqnarray}\label{vi}
V^{ind}({\bf r},t)&=&\int d{\bf
r}'\int_{-\infty}^{+\infty} dt'\int_{-\infty}^{+\infty}{d\omega\over
2\pi}\,{\rm e}^{-i\omega(t-t')}\cr\cr\cr
&\times&\tilde W({\bf r},{\bf r}';\omega)\,\rho^{ext}({\bf r}',t')
\end{eqnarray}
with
\begin{equation}
\tilde W({\bf r},{\bf r}';\omega)=W({\bf r},{\bf r}';\omega)-v({\bf r},{\bf
r}'),
\end{equation}
Finally, one writes:
\begin{equation}\label{loss1}
-\Delta E=\int_{-\infty}^{+\infty} dt\left(-{dE\over dt}\right).
\end{equation}
Introducing Eqs.~(\ref{rho}) and (\ref{vi}) into Eq.~(\ref{flores}), and
Eq.~(\ref{flores}) into Eq.~(\ref{loss1}), one finds that the total energy loss
$\Delta E$ can indeed be written as a sum over the probability $P_{\bf q}$ of
Eq.~(\ref{eq51p}), i.e.:
\begin{equation}\label{loss2}
-\Delta E=\sum_{\bf q}\left({\bf q}\cdot{\bf v}\right)P_{\bf q},
\end{equation}
where ${\bf q}\cdot{\bf v}$ is simply the energy transferred by our
recoilless probe particle to the medium.

\subsubsection{Planar surface}

In the case of a plane-bounded electron gas that is translationally invariant
in two directions, which we take to be normal to the $z$ axis,
Eqs.~(\ref{rho})-(\ref{vi})
yield the following expression for the energy that the probe particle loses per
unit time:
\begin{eqnarray}\label{general_1}
-{dE\over dt}&=&i\,{Z_1^2\over\pi}\int {d^2{\bf
q}\over(2\pi)^2}\int_{-\infty}^{+\infty}
dt' \int_0^\infty d\omega\,\omega\cr\cr\cr
&\times&{\rm e}^{-i(\omega-{\bf
q}\cdot{\bf v}_\parallel)(t-t')}\,\tilde W[z(t),z(t');q,\omega],
\end{eqnarray}
where ${\bf q}$ is a 2D wave vector in the plane of the surface,
${\bf v}_\parallel$ represents the component of the velocity that is parallel
to the surface, $z(t)$ represents the position of the projectile relative to
the surface, and  $\tilde W(z,z';q,\omega)$ is the 2D Fourier transform
of $\tilde W({\bf r},{\bf r}';\omega)$.

In the simplest possible model of a bounded semi-infinite electron gas in
vacuum, in which the screened interaction $W(z,z';q,\omega)$ is given by the
classical expression Eq.~(\ref{clas1}) with $\epsilon_1$ being the Drude
dielectric function of Eq.~(\ref{drude}) and $\epsilon_2=1$, explicit
expressions can be found for the energy lost per unit path length by probe
particles that move along a trajectory that is either parallel or normal to
the surface.

\begin{figure}
\includegraphics[width=0.5\linewidth]{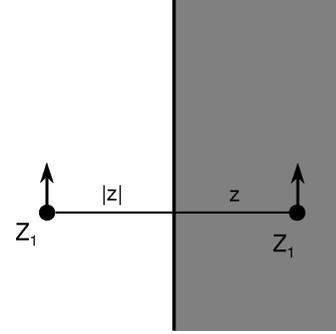}\\
\caption{Particle of charge $Z_1$ moving with constant velocity at a
fixed distance $z$ from the surface of a plane-bounded electron
gas. Inside the solid, the presence of the surface causes (i) a
decrease of loss at the bulk-plasmon frequency $\omega_p$ varying
with $z$ as $\sim K_0(2\omega_pz/v)$ and (ii) an additional loss at
the surface-plasmon frequency $\omega_s$ varying with $z$ as
$\sim K_0(2\omega_sz/v)$.
Outside the solid, energy losses are dominated by
a surface-plasmon excitation at $\omega_s$.}\label{fig2g}
\end{figure}

\paragraph{Parallel trajectory.} In the case of a probe particle moving with
constant velocity at a fixed distance $z$ from the surface (see
Fig.~\ref{fig2g}), introduction of Eq.~(\ref{clas1}) into
Eq.~(\ref{general_1}) yields~\cite{opt}
\begin{widetext}
\begin{equation}\label{parallel}
-{dE\over dx}={Z_1^2\over v^2}\cases{\omega_p^2\left[{\rm
ln}(k_cv/\omega_p)-K_0(2\omega_pz/v)\right]+\omega_s^2
\,K_0(2\omega_sz/v),&$z<0$\cr\cr
\omega_s^2\,K_0(2\omega_s|z|/v),&$z>0$,}
\end{equation}
\end{widetext}
where $K_0(\alpha)$ is the zero-order modified Bessel function~\cite{abra}, and
$k_c$ denotes the magnitude of a wave vector above which long-lived bulk
plasmons are not sustainable.

For particle trajectories outside the solid ($z>0$), Eq.~(\ref{parallel})
reproduces the classical expression of Echenique and Pendry~\cite{pendry1},
which was found to describe correctly EELS experiments~\cite{howie1} and which
was extended to include relativistic corrections~\cite{milne}. For
particle trajectories inside the solid ($z>0$), Eq.~(\ref{parallel})
reproduces the result first obtained by Nu\~nez {\it et al.}~\cite{nunez}.
Outside the solid, the energy loss is dominated by the excitation of surface
plasmons at $\omega_s$. When the particle moves inside the solid, the effect
of the boundary is to cause (i) a decrease in loss at the bulk plasma
frequency $\omega_p$, which in an infinite electron gas would be
$-dE/dx=Z_1^2\omega_p^2{\rm ln}(k_cv/\omega_p)/v^2$ and (ii) an additional loss
at the surface-plasma frequency $\omega_s$.

Nonlocal effects that are absent in the classical Eq.~(\ref{parallel}) were
incorporated approximately by several authors in the framework of the
hydrodynamic approach and the specular-reflection model described in
Sections~\ref{hydrog} and
\ref{srm}~\cite{inaki1,inaki2,inaki3,inaki4,inaki5,inaki6}. More recently,
extensive RPA and ALDA calculations of the energy-loss spectra of charged
particles moving near a jellium surface were carried out~\cite{aran1}
within the self-consistent scheme described in
Section~\ref{self}. At high velocities (of a few Bohr units) and
for charged particles moving far from the surface into the vacuum, the actual
energy loss was found to converge with the classical limit dictated by the
first line of Eq.~(\ref{parallel}). However, at low and intermediate
velocities substantial changes in the energy loss were observed as a realistic
description of the surface response was considered.

Corrections to the energy loss of charged particles (moving far from the
surface into the vacuum) due to the finite width of the surface-plasmon
resonance that is not present, in principle, in jellium self-consistent
calculations, have been discussed recently~\cite{toketsi}. These corrections
have been included to investigate the energy loss of highly charged ions
undergoing distant collisions at grazing incidence angles with the internal
surface of microcapillary materials, and it has been suggested that the
correlation between the angular distribution and the energy loss of
transmitted ions can be used to probe the dielectric properties of the
capillary material.

For a more realistic description of the energy loss of charged particles moving
near a Cu(111) surface, the Kohn-Sham potential $v_{KS}(z)$ used in the
self-consistent jellium calculations of Ref.~\cite{aran1} was replaced in
Ref.~\cite{inaki7} by the 1D model potential $v_{MP}$ of Ref.~\cite{chulkov0}.
It was shown, however, that although the Cu(111) surface exhibits a wide band
gap around the Fermi level and a well-defined Shockley surface state the
energy loss expected from this model does not differ significantly from its
jellium counterpart. This is due to the fact that the presence of the surface
state compensates the reduction of the energy loss due to the band gap.

Existing first-principles calculations of the interaction of charged
particles with solids invoke periodicity of the solid in all
directions and neglect, therefore, surface effects and, in
particular, the excitation of surfaces
plasmons~\cite{campillo,trickey}. An exception is a recent
first-principles calculation of the energy loss of ions moving
parallel with a Mg(0001) surface~\cite{maia}, which accounts
naturally for the finite width of the surface-plasmon resonance that
is present neither in the self-consistent jellium calculations of
Ref.~\cite{aran1} nor in the 1D model calculations of
Ref.~\cite{inaki7}.

A typical situation in which charged particles can be approximately assumed to
move along a trajectory that is parallel to a solid surface occurs in the
glancing-incidence geometry, where ions penetrate into the solid, they skim
the outermost layer of the solid, and are then specularly repelled by a
repulsive, screened Coulomb potential, as discussed by Gemell~\cite{gemmell}.
By first calculating the ion trajectory under the combined influence of the
repulsive planar potential and the attractive image potential, the total
energy loss can be obtained approximately as follows
\begin{equation}
\Delta E=2v\,\int_{z_{tp}}^\infty{dE\over dx}(z)\left[v_z(z)\right]^{-1}\,dz,
\end{equation}
$z_{tp}$ and $v_z(z)$ denoting the turning point and the value of
the component of the velocity normal to the surface, respectively,
which both depend on the angle of incidence.

Accurate measurements of the energy loss of ions being reflected
from a variety of solid surfaces at grazing incidence have been
reported by several
authors~\cite{kimura1,kimura2,winter1,robin,winter2}. In particular,
Winter {\it et al.}~\cite{winter1} carried out measurements of the
energy loss of protons being reflected from Al(111). From the
analysis of their data at $120\,{\rm keV}$, these authors deduced
the energy loss ${dE/dx}(z)$ and found that at large distances from
the surface the energy loss follows closely the energy loss expected
from the excitation of surface plasmons. Later on, RPA jellium
calculations of the energy loss from the excitation of valence
electrons were combined with a first-Born calculation of the energy
loss due to the excitation of the inner shells and reasonably good
agreement with the experimental data was obtained for all angles of
incidence~\cite{juaristic}.

\paragraph{Normal trajectory.} Let us now consider a situation in which the
probe particle moves along a normal trajectory from the vacuum side of the
surface ($z>0$) and enters the solid at $z=t=0$. The position of the
projectile relative to the surface is then $z(t)=-vt$. Assuming that the
electron gas at $z\leq 0$ can be described by the
Drude dielectric function of Eq.~(\ref{drude}) and introducing
Eq.~(\ref{clas1}) into Eq.~(\ref{general}) yields~\cite{opt}:
\begin{widetext}
\begin{equation}\label{normalt}
-{dE\over dz}={Z_1^2\over v^2}\cases{
\omega_p^2\left[{\rm ln} (k_cv/\omega_p)-h(\omega_pz/v)\right]
+\omega_s^2\,h(\omega_sz/v)
,&$z<0$\cr\cr
\omega_s^2\,f(2\omega_s|z|/v),&$z>0$,}
\end{equation}
\end{widetext}
where
\begin{equation}\label{b2}
h(\alpha)=2\cos(\alpha)\,f(\alpha)-f(2\alpha),
\end{equation}
with $f(\alpha)$ being given by the following expression:
\begin{equation}
f(\alpha)=\int_0^\infty{x\,{\rm e}^{-\alpha x}\over 1+x^2}dx.
\end{equation}
Eq.~(\ref{normalt}) shows that (i) when the probe particle is moving outside
the
solid the effect of the boundary is to cause energy loss at the
surface-plasmon energy $\omega_s$, and (ii) when the probe particle is moving
inside the solid the effect of the boundary is to cause both a decrease in
loss at the bulk-plasmon energy $\omega_p$ and an additional loss at the
surface-plasmon energy $\omega_s$, as predicted by Ritchie~\cite{ritchie1}.

\begin{figure}
\centering
\includegraphics[width=0.5\linewidth]{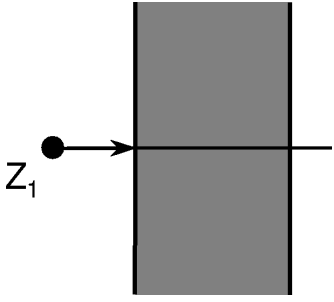}\\
\caption{Particle of charge $Z_1$ passing perpendicularly through
a finite foil of thickness $a$. The presence of the boundaries
leads to a decrease in the energy loss at the bulk-plasmon frequency $\omega_p$
and an additional loss at the surface-plasmon frequency $\omega_s$, the net
boundary effect being an increase in the total energy loss in comparison to
the case of a particle moving in an infinite medium with no boundaries.}
\label{fig1g}
\end{figure}

Now we consider the real situation in which a fast charged particle passes
through a {\it finite} foil of thickness $a$ (see Fig.~\ref{fig1g}). Assuming
that the foil is thick enough
for the effect of each boundary to be the same as in the case of a
semi-infinite medium, and
integrating along the whole trajectory from minus to plus infinity, one finds
the total energy
that the probe particle loses to collective excitations:
\begin{equation}\label{delta}
-\Delta E={Z_1^2\over v^2}\left[a\,\omega_p^2\,{\rm ln}{k_cv\over\omega_p}
-{\pi\over 2}\omega_p+\pi\omega_s\right].
\end{equation}

This is the result first derived by Ritchie in a different way~\cite{ritchie1},
which brought him to the realization that surface collective excitations exist
at the lowered frequency $\omega_s$. The first term of Eq.~(\ref{delta}),
which is proportional to the thickness of the film represents the bulk
contribution, which would also be present in the absence of the
boundaries. The second and third terms, which are both due to the presence of
the boundaries and become more important as the foil thickness decreases,
represent the decrease in the energy loss at the plasma frequency $\omega_p$
and the energy loss at the lowered frequency $\omega_s$, respectively.
Eq.~(\ref{delta}) also
shows that the net boundary effect is an increase in the total energy loss
above the value which would exist in its absence, as noted by
Ritchie~\cite{ritchie1}. A more accurate jellium self-consistent description
of the energy loss of charged particles passing through thin foils has been
performed recently in the RPA and ALDA~\cite{arann}.

\subsection{STEM: Valence EELS}

The excitation of both surface plasmons on solid surfaces and localized Mie
plasmons on small particles has attracted great interest over the years in the
fields of scanning transmission electron
microscopy~\cite{batson,howie1,ferrell,rivacoba1,howie2,rivacoba2} and
near-field optical spectroscopy~\cite{klar}.

EELS of fast electrons in STEM shows two types of losses, depending on the
nature of the excitations that are produced in the sample: atomically defined
core-electron excitations at energies $\omega>100\,{\rm eV}$ and
valence-electron (mainly collective) excitations at energies up to
$\sim 50\,{\rm eV}$. Core-electron excitations occur when the probe moves
across the target, and provide chemical information about atomic-size regions
of the target~\cite{core}. Conversely, valence-electron excitations provide
information about the surface structure with a resolution of the order of
several nanometers. One advantage of valence EELS is that it provides a strong
signal, even for non-penetrating trajectories (the so-called {\it aloof beam
energy loss spectroscopy}~\cite{fw1,fw2}), and generates less specimen
damage~\cite{howie1}.

The central quantity in the interpretation of valence EELS experiments is the
total probability $P(\omega)$ for the STEM beam to exchange energy $\omega$
with the sample. In terms of the screened interaction
$W({\bf r},{\bf r}';\omega)$ and for a probe electron in the state
$\phi_0({\bf r})$ with energy $\varepsilon_0$, first-order perturbation theory
yields:
\begin{eqnarray}\label{pomega}
P(\omega)&=&-2\sum_f\int d{\bf r}\int d{\bf r}'\,\phi_f^*({\bf r}')\,
\phi_0({\bf r}')\,\phi_f({\bf r})\,\phi_0^*({\bf r})\cr\cr
&\times&{\rm Im}W({\bf r},{\bf r}';\omega)\
\delta[\omega-\varepsilon_0-\varepsilon_F],
\end{eqnarray}
where the sum is extended over a complete set of final states
$\phi_f({\bf r})$ of energy $\varepsilon_f$. For probe electrons moving on a
definite trajectory (and having, therefore, a charge density of the form of
Eq.~(\ref{rho}) with no beam recoil), the total energy loss $\Delta E$ of
Eqs.~(\ref{loss1}) and (\ref{loss2}) can be expressed in the expected form
\begin{equation}\label{deltae}
\Delta E=\int_0^\infty d\omega\,\omega\,P(\omega).
\end{equation}
Ritchie and Howie~\cite{rh} showed that in EELS experiments where all of the
inelastic scattering is collected treating the fast electrons as a classical
charge of the form of Eq.~(\ref{rho}) is indeed adequate. Nonetheless, quantal
effects due to the spatial extension of the beam have been addressed by
several authors~\cite{rivacoba1,kohl,cohen}.

\subsubsection{Planar surface}

In the case of a classical beam of electrons moving with constant velocity at a
fixed distance $z$ from a planar surface (see Fig.~\ref{fig2g}), the initial
and
final states can be described by taking a $\delta$ function in the transverse
direction and plane waves in the direction of motion. Neglecting recoil, the
probability $P(\omega)$ of Eq.~(\ref{pomega}) then takes the following form:
\begin{equation}\label{planar}
P(\omega)=-{1\over\pi^2 v}\int_0^\infty dq_x\,{\rm Im}W(z,z;q,\omega),
\end{equation}
with $q=\sqrt{q_x^2+(\omega/v)^2}$.

In a classical model in which the screened interaction $W(z,z';q,\omega)$ is
given by the classical expression Eq.~(\ref{clas1}), the probability
$P(\omega)$ is easy to calculate. In particular, if the beam of electrons is
moving outside the sample, one finds:
\begin{equation}\label{par}
W(z,z;q,\omega)={2\pi\over q}\,\left[1-g\,{\rm e}^{-2qz}/\epsilon_2\right],
\end{equation}
with the surface-response function $g$ given by Eq.~(\ref{g1}). Introduction of
Eq.~(\ref{par}) into Eq.~(\ref{planar}) yields the classical probability
\begin{equation}
P(\omega)={2L\over\pi v^2}\,K_0(2\omega z/v)\,{\rm Im}g(\omega),
\end{equation}
which in the case of a Drude metal in vacuum yields
\begin{equation}\label{pomegas}
P(\omega)=L\,{\omega_s\over v^2}\,\delta(\omega-\omega_s),
\end{equation}
and, therefore [see Eq.~(\ref{deltae})], the energy loss per unit path length
given by Eq.~(\ref{parallel}) with $Z_1=-1$ and $z>0$.

The classical Eq.~(\ref{par}) can be easily extended to the case in which the
sample is formed by a semi-infinite medium characterized by $\epsilon_1$ and
covered by a layer of dielectric function $\epsilon_3$ and thickness $a$. One
finds:
\begin{equation}\label{onel}
W(z,z;q,\omega)={2\pi\over q}\,\left[1-{\rm e}^{-2qz}
{\xi_{32}+\xi_{13}\,{\rm e}^{2qa}\over
\xi_{32}\xi_{13}+{\rm e}^{2qa}}\right],
\end{equation}
where
\begin{equation}\label{twol}
\xi_{32}={\epsilon_3-\epsilon_2\over\epsilon_3+\epsilon_2}
\end{equation}
and
\begin{equation}\label{threel}
\xi_{13}={\epsilon_1-\epsilon_3\over\epsilon_1+\epsilon_3}.
\end{equation}
Eqs.~(\ref{onel})-(\ref{threel}) show that while for a clean surface ($a=0$)
the
energy-loss function $P(\omega)$ is dominated by the excitation of surface
plasmons at $\omega=\omega_s$  [see Eq.~(\ref{pomegas})], EELS should be
sensitive to
the presence of sub-surface structures. This effect was observed by
Batson~\cite{batson} in the energy-loss spectra corresponding to an Al surface
coated with an Al$_2$O$_3$ layer of increasing thickness.

\subsubsection{Spheres}

\paragraph{Definite trajectories.} In the case of a classical beam of electrons
moving with constant velocity at a
fixed distance $b$ from the center of a single sphere of radius $a$ and
dielectric function $\epsilon_1(\omega)$ that is immersed in a host medium of
dielectric function $\epsilon_2(\omega)$, the initial and final states can be
described (as in the case of the planar surface) by taking a $\delta$ function
in the transverse direction and plane waves in the direction of motion. If
recoil is neglected and the classical screened interaction of
Eqs.~(\ref{scr1})-(\ref{scr4}) is used, then Eq.~(\ref{pomega}) yields the
following expression for the energy-loss probability~\cite{ferrell}:
\begin{eqnarray}\label{pspheres1}
P(\omega)&=&{4 a\over\pi v^2}\,\sum_{l=1}^\infty\,\sum_{m=0}^l\,
{\mu_m\over(l-m)(l+m)}\left[\omega a\over v\right]^{2l}\cr\cr
&\times&K_m^2(\omega_b/v)\,{\rm Im} g_l(\omega),
\end{eqnarray}
for outside trajectories ($a\leq b$), and~\cite{rivacoban}
\begin{eqnarray}\label{pspheres2}
&&P(\omega)={4 a\over\pi v^2}\,\sum_{l=1}^\infty\,\sum_{m=0}^l\,
\mu_m{(l-m)\over(l+m)}\left\{\left[A_{lm}^0+A_{lm}^i\right]^2
\right.\cr\cr
\times&&{\rm Im}g_l(\omega)+\left.A_{lm}^i\left[A_{lm}^0+A_{lm}^i\right]\,
{\rm Im}\left[\epsilon^{-1}(\omega)\right]\right\},
\end{eqnarray}
for inside trajectories ($a\geq b$), with
\begin{eqnarray}
A_{lm}^0(\omega)&=&{1\over a}\int_c^\infty dx
\left[{\sqrt{b^2+x^2}\over a}\right]^{l+1}
P_l^m\left[{x\over\sqrt{b^2+x^2}}\right]\cr\cr
&\times&c_{lm}(\omega x/v)
\end{eqnarray}
and
\begin{eqnarray}
A_{lm}^i(\omega)&=&{1\over a}\int_0^c dx
\left[{\sqrt{b^2+x^2}\over a}\right]^l
P_l^m\left[{x\over\sqrt{b^2+x^2}}\right]\cr\cr
&\times&c_{lm}(\omega x/v)
\end{eqnarray}
Here, $c=\sqrt{a^2-b^2}$, $P_l^m(x)$ denote Legendre functions,
$c_{lm}(x)=\cos(x)$, if $(l+m)$ is even, os $\sin(x)$, if $(l+m)$ is odd, and
$g_l(\omega)$ denotes the classical function of Eq.~(\ref{scr4}), which
has poles at the classical surface-plasmon condition of Eq.~(\ref{spheres}).

Equations~(\ref{pspheres1}) and (\ref{pspheres2}) show that an infinite number
of multipolar modes can be excited, in general, which contribute to the energy
loss of moving electrons. It is known, however, that for small spheres with
$a<<v/\omega_s$ the dipolar mode ($l=1$) dominates, which in the case of a
Drude sphere in vacuum occurs at $\omega_1=\omega_p/\sqrt{3}$. For spheres with
$a\sim v/\omega_s$, many multipoles contribute with similar weight. For very
large spheres ($a>>v/\omega_s$) and $a<b$, the main contribution arises from
high multipolar modes occurring approximately at the planar surface plasmon
energy $\omega_s$, since in the limit of large $l$ $g_l(\omega)$ of
Eq.~(\ref{scr4}) reduces to the energy-loss function of Eq.~(\ref{g1}). Indeed,
this is an expected result, due to the fact that for very large spheres the
probe electron effectively interacts with an almost planar surface.

\paragraph{Broad beam.} We now consider a broad beam geometry, and we therefore
describe the probe electron states by plane waves of the form:
\begin{equation}\label{plane1}
\phi_0({\bf r})={1\over\sqrt\Omega}\,{\rm e}^{{\rm i}{\bf k}_0\cdot{\bf r}}
\end{equation}
and
\begin{equation}\label{plane22}
\phi_f({\bf r})={1\over\sqrt\Omega}\,{\rm e}^{{\rm i}{\bf k}_f\cdot{\bf r}},
\end{equation}
where $\Omega$ represents the normalization volume. Then, introducing
Eqs.~(\ref{plane1}), (\ref{plane22}) and (\ref{scr1})-(\ref{scr4}) into
Eq.~(\ref{pomega}), and neglecting recoil, one finds
\begin{equation}\label{phomo}
P(\omega)={1\over\pi^2}\int{d{\bf Q}\over Q^2}\,{\rm Im}
\left[-\epsilon_{eff}^{-1}(Q,\omega)\right]\delta(\omega-{\bf Q}\cdot{\bf v}).
\end{equation}
This is precisely the energy-loss probability corresponding to a probe electron
moving in a 3D homogeneous electron gas but with the inverse dielectric
function $\epsilon^{-1}(Q,\omega)$ replaced by the effective inverse dielectric
function $\epsilon_{eff}^{-1}(Q,\omega)$ of Eq.~(\ref{effsphere}).

\subsubsection{Cylinders}

\paragraph{Definite trajectory.} In the case of a classical beam of electrons
moving with constant velocity at a fixed distance $b$ from the axis of a single
cylinder of radius $a$ and dielectric function $\epsilon_1(\omega)$ that is
immersed in a host medium of dielectric function $\epsilon_2(\omega)$, the
initial and final states can be described (as in the case of the planar
surface) by taking a $\delta$ function in the transverse direction and plane
waves in the direction of motion. If recoil is neglected and the classical
screened interaction of Eqs.~(\ref{cyl1})-(\ref{cyl4}) is used, then
Eq.~(\ref{pomega}) yields the following expression for the energy-loss
probability~\cite{rivacoba2}:
\begin{widetext}
\begin{eqnarray}
&&P(\omega)=-{2\over\pi v}\cr\cr&&\times{\rm
Im}\left\{(\epsilon_1^{-1}-\epsilon_2^{-1})\sum_{m=0}^\infty
\mu_m\left[I_m\left({\omega
b\over v}\right)K_m\left({\omega b\over v}\right)+
{\epsilon_2K_m({\omega a\over v})K_m'({\omega a\over v})I_m^2\left({\omega
b\over v}\right)\over\epsilon_1 I_m'({\omega a\over v})K_m({\omega
a\over v})-\epsilon_2I_m({\omega a\over v})K_m'({\omega
a\over v})}\right]\right\},
\end{eqnarray}
for inside trajectories ($a\leq b$), and
\begin{equation}
P(\omega)=-{2\over\pi v}{\rm
Im}\left\{(\epsilon_1^{-1}-\epsilon_2^{-1})
\sum_{m=0}^\infty\mu_m
{\epsilon_1 I_m({\omega a\over v})I_m'({\omega a\over
v})K_m^2\left({\omega b\over v}\right)\over\epsilon_1 I_m'({\omega a\over
v})K_m({\omega a\over v})-\epsilon_2I_m({\omega a\over v})K_m'({\omega
a\over v})}\right\},
\end{equation}
\end{widetext}
for outside trajectories ($a\geq b$). In particular, for axial trajectories
($b=0$), the energy-loss probability $P(\omega)$ is due exclusively to the
$m=0$ mode.

\paragraph{Broad beam.} For a broad beam geometry the probe electron states can
be described by plane waves of the form of Eqs.~(\ref{plane1}) and
(\ref{plane22}), which after introduction into Eq.~(\ref{pomega}), neglecting
recoil, and using the classical screened interaction of
Eqs.~(\ref{cyl1})-(\ref{cyl4}) yield an energy-loss probability of the form of
Eq.~(\ref{phomo}) but now with the inverse dielectric function
$\epsilon^{-1}(Q,\omega)$ replaced by the effective inverse dielectric
function $\epsilon_{eff}^{-1}(Q,\omega)$ of Eq.~(\ref{effcyl1}).

\subsection{Plasmonics}

Renewed interest in surface plasmons has come from recent advances in the
investigation of the electromagnetic properties of nanostructured
materials~\cite{pendry2,halas}, one of the most attractive aspects of these
collective excitations now being their use to concentrate light in
subwavelength
structures and to enhance transmission through periodic arrays of subwavelength
holes in optically thick metallic films~\cite{ebbesen1,ebbesen3}.
Surface-plasmon polaritons are tightly bound to metal-dielectric interfaces
penetrating around 10 nm into the metal (the so-called skin-depth) and
typically more than 100 nm into the dielectric (depending on the wavelength),
as shown in Fig.~\ref{fig10}. Indeed, surface plasmons of an optical wavelength
concentrate light in a region that is considerably smaller than their
wavelength, a feature that suggests the possibility of using surface-plasmon
polaritons for the fabrication of nanoscale photonic circuits operating
at optical frequencies~\cite{ozsc06,ebbesen2}.

Here we discuss only a few aspects of the most recent research that has been
carried out in the so-called field of plasmonics. This constitutes an important
area of research, since surface-plasmon based circuits are known to merge the
fields of photonics and electronics at the nanoscale, thereby
enabling to overcome the existing difficulties related to the large size
mismatch between the micrometer-scale bulky components of photonics and the
nanometer-scale electronic chips. Indeed, the surface-plasmon polariton
described in Sec.~\ref{spp1} can serve as a base for constructing nano-circuits
that will be able to carry optical signals and electric currents. These
optoelectronic circuits would consist of various components such as couples,
waveguides, switches, and modulators.

In order to transmit optical signals to nanophotonic devices and to efficiently
increase the optical far-field to near-field conversion, a nanodot coupler
(fabricated from a linear array of closely spaced metallic nanoparticles) has
been combined recently with a surface-plasmon polariton condenser (working as a
phase array) fabricated from hemispherical metallic
nanoparticles~\cite{noohapl05}. By focusing surface-plasmon polaritons with a
spot size as small as 400 nm at $\lambda=785\,{\rm nm}$, their transmission
length through the nanodot coupler was confirmed to be three times longer than
that of a metallic-core waveguide, owing to the efficient near-field coupling
between the localized surface plasmon of neighboring nanoparticles.

Achieving control of the light-solid interactions involved in nanophotonic
devices requires structures that guide electromagnetic energy with a lateral
mode confinement below the diffraction limit of light. It was suggested that
this so-called subwavelength-sized wave guiding can occur along chains of
closely spaced metal nanoparticles that convert the optical mode into surface
(Mie) plasmons~\cite{quleol98}. The existence of guided long-range
surface-plasmon waves was observed experimentally by using thin metal
films~\cite{chbeol00,lakrapl01,nileapl03}, nanowires~\cite{krlael02,krwe04},
closely spaced silver rods~\cite{makinm03}, and metal
nanoparticles~\cite{muasprb04}.

The propagation of guided surface plasmons is subject to significant ohmic
losses that limit the maximum propagation length. In order to avoid these
losses, various geometries have been devised using arrays of features of
nanosize dimensions~\cite{mabaapl04,mafrapl05}. The longest propagation length
(13.6 mm) has been achieved with a structure consisting of a thin lossy metal
film lying on a dielectric substrate and covered by a different dielectric
superstrate~\cite{bechjap05}. The main issue in this context is to strongly
confine the surface-plasmon field in the cross section perpendicular to the
surface-plasmon propagation direction, while keeping relatively low propagation
losses. Recently, it has been pointed out that strongly localized channel
plasmon polaritons (radiation waves guided by a channel cut into a planar
surface of a solid characterized by a negative dielectric
function~\cite{nomaprb02}) exhibit relatively low propagation
losses~\cite{piogapl05}. The first realization and characterization of the
propagation of channel plasmon polaritons along straight subwavelength metal
grooves was reported by Bozhevilnyi {\it et al.}~\cite{bovoprl05}. More
recently, the design, fabrication, and
characterization of channel plasmon polariton based subwavelength
waveguide components have been reported; these are Y-splitters, Mach-Zehnder
interferometers, and waveguide-ring resonators operating at telecom
wavelengths~\cite{bovon06}.

In the framework of plasmonics, modulators and switches have also been
investigated. Switches should serve as an active element to control
surface-plasmon polariton waves~\cite{krzhapl04,krzajo05}. This approach
takes advantage of the strong dependence of the propagation of surface-plasmon
polaritons on the dielectric properties of the metal in a thin surface
layer that may be manipulated using light. This idea was realized introducing
a few-micron long gallium switching section to a gold-on-silica
waveguide~\cite{krzajo05}. An example of an active plasmonic device
has been demonstrated by using a thin silver film covered from
both sides by thin polymer films with molecular chromophores~\cite{barnes}.
In this case, coupled surface-plasmon polaritons provide an effective transfer
of excitation energy from donor molecules to acceptor molecules on the opposite
sides of a metal film up to 120 nm thick.

Another emerging area of active research in the field of plasmonics is based on
the generation and manipulation of electromagnetic radiation of various
wavelengths from microwave to optical frequencies. For instance, the coating of
semiconductor quantum wells by nanometer metal films results in an increased
spontaneous emission rate in the semiconductor that leads to the enhancement of
light emission. This enhancement is due to an efficient energy transfer from
electron-hole pair recombination in the quantum well to surface-plasmon
polaritons at the surface of semiconductor heterostructures coated by metal.
Recently, a 32-fold increase in the spontaneous
emission rate at 440 nm in an InGaN/GaN quantum well has been probed by
time-resolved photoluminescence spectroscopy~\cite{okniapl05}. Also probed has
been the enhancement of photoluminescence up to an order of magnitude through a
thin metal film from organic light emitting diodes, by removing the
surface-plasmon polariton quenching with the use of a periodic
nanostructure~\cite{wewaapl04}.

Recent theoretical and experimental work also suggest that surface-plasmon
polaritons play a key role both in the transmission of electromagnetic waves
through a single aperture and the enhanced transmission of light through
subwavelength hole arrays~\cite{ebbesen1,thpeol01,pemas04,bamuprl04}. This
enhanced transmission has also been observed at millimeter-waves and
micro-waves~\cite{besool04,akbuapl04} and at THz-waves~\cite{canaoe04}.

Finally, we note that surface-plasmon polaritons have been used in
the field of nanolithography. This surface-plasmon based
nanolithography can produce subwavelength structures at surfaces,
such as sub-100 nm lines with visible
light~\cite{kimaprb04,luisapl04,shchapl05,waupnl06}. On the other
hand, the enhancement of evanescent waves through the excitation
of surface plasmons led Pendry~\cite{peprl00} to the concept of
the so-called superlens~\cite{rainer}.

\section{Acknowledgments}

Partial support by the University of the Basque Country, the Basque
Unibertsitate eta Ikerketa Saila, the Spanish Ministerio de Educaci\'on
y Ciencia, and the EC 6th framework Network of Excellence NANOQUANTA (Grant No.
NMP4-CT-2004-500198) are acknowledged.


\begin{thebibliography}{1}
\bibitem{optical} The electron density $n$ is usually
characterized by the density parameter $r_s=(3/4\pi n)^{1/3}/a_0$,
$a_0$ being the Bohr radius, $a_0=0.529\,{\rm \AA}$. In metals the
electron-density parameter of valence electrons is typically in the
range $2<r_s<6$, which corresponds to plasmon energies on the order
of $10\,{\rm eV}$ and frequencies that lie in the optical regime.
\bibitem{note1} Since the finite angular acceptance of typical
energy-loss spectrometers guarantees that the momentum transfer is
associated with a plasmon wavelength larger than the wavelength of
light at $\omega=\omega_s$ ($2\pi c/\omega_s\sim 10^3\,{\rm\AA}$),
only the nonretarded region is observed in these experiments.
\bibitem{note3} The sound velocity of this acoustic mode is, however,
close to the Fermi velocity of the 2D surface-state band, which is
typically a few orders of magnitude larger than the sound velocity
of acoustic phonons in metals but still about three orders of
magnitude smaller than the velocity of light.
\bibitem{eta} Although in the case of an ideal damping-free electron
gas the quantity $\eta$ entering the Drude dielectric function of
Eq.~(\ref{drude}) should be a positive infinitesimal, a
phenomenological finite parameter $\eta$ is usually introduced in
order to account for the actual electron damping occurring in real
solids.
\bibitem{noteh} While the use of the Thomas-Fermi functional $G[n]$ of
Eq.~(\ref{tf}), which assumes static screening, predicts
$\beta=\sqrt{1/3}\,(3\pi^2n_0)^{1/3}$, the value $\beta=\sqrt{3/5}\
(3\pi^2n_0)^{1/3}$ should be more appropriate when high-frequencies of the
order of the plasma frequency are involved; see, e.g., Ref.~\cite{lundqvist}.
\bibitem{p} In the case of $p$-polarized electromagnetic waves,
infinitely long cylinders can be represented by 2D circular
inclusions.
\bibitem{long} As we are considering longitudinal fields, where the
electric field ${\bf E}$ and the wave vector ${\bf q}$ have the same
direction, in the limit as $x=q_z a\to0$ the electric field lies in
the plane perpendicular to the axis of the cylinder.
\bibitem{notep} This mode, however, does not contribute to the
effective inverse dielectric function and, therefore, to the energy
loss of moving charged particles, since when $x=q_za=0$ ($p$
polarization) the strength of this mode is equal to zero for all
values of the total wave
vector $Qa$ (see Ref.~\cite{pitarke3}).
\bibitem{inside} $\omega_Q^B$ denotes the energy of bulk plasmons with
momentum $Q$, and $Q_c$ denotes the critical momentum for which the
bulk-plasmon dispersion $\omega_Q^B$ enters the electron-hole pair
excitation spectrum. For $r_s=2.07$ and $Q=0.4q_F$, the RPA values of
$\omega_Q^B$ and $\omega_{Q_c}^B$ are $17.6$ and $23.6\,{\rm eV}$,
respectively.
\bibitem{lambdaf} The Fermi wavelength $\lambda_F$ is defined as follows:
$\lambda_F=2\pi/q_F$, $q_F$ being the magnitude of the Fermi wave vector.
\bibitem{drudenote} For an ideal free-electron gas, $\epsilon(\omega)$
is given by the Drude dielectric function of Eq.~(\ref{drude}).
\bibitem{dper} In general, the $d$-parameter $d_\perp(\omega)$
entering Eq.~(\ref{expansion}) should be replaced by
$d_\perp(\omega)-d_\parallel(\omega)$. However, as for neutral jellium
surfaces $d_\parallel(\omega)$ coincides with the jellium edge,
$d_\perp(\omega)-d_\parallel(\omega)$ reduces to the centroid
$d_\perp(\omega)$ of the induced electron density with respect to the
jellium edge.
\bibitem{noteg2} This surface plasmon would also be present in the absence of
surface states, i.e., if the 1D model potential $v_{MP}$ of
Ref.~\cite{chulkov0} were replaced by the jellium Kohn-Sham potential of
Eq.~(\ref{vks}).
\bibitem{noteslava} Acoustic surface plasmons in Be(0001) have been observed
for energies up to $2\,{\rm eV}$~\cite{acexp}.
\bibitem{noteps} This approximation is valid for heavy charged particles, e.g.,
ions, and also for swift electrons moving with nonrelativistic velocities that
are large compared to the velocity of target electrons. In the case of
electrons, $Z_1=-1$.
\end{thebibliography}
\end{document}